\documentclass[a4paper,fleqn,usenatbib]{mnras}

\usepackage{graphicx}
\usepackage[percent]{overpic}
\usepackage{lipsum}
\usepackage{multirow}
\usepackage{color}
\usepackage{rotating}
\usepackage{mwe,tikz}
\usepackage[percent]{overpic}
\usepackage{mathtools}
\usepackage{physics}
\usepackage{amsmath}
\usepackage[utf8x]{inputenc}
\usepackage{hyperref}
\usepackage{longtable}
\usepackage{wasysym}
\usepackage[normalem]{ulem}
\hypersetup{
    colorlinks=true,
    linkcolor=blue,
    filecolor=magenta,      
    urlcolor=blue,
}
\usepackage{soul}
\usepackage{pdflscape}
\usepackage{longtable}
\usepackage{booktabs,tabularx}

\usepackage{graphicx}	
\usepackage{amsmath}	
\usepackage{amssymb}	
\usepackage{latexsym}

\newcommand{\lta}{$\; \buildrel < \over \sim \;$}
\newcommand{\simlt}{\lower.5ex\hbox{\ltsima}}
\newcommand{\gta}{$\; \buildrel > \over \sim \;$}
\newcommand{\simgt}{\lower.5ex\hbox{\gtsima}}

\def\FeH{{\rm[Fe/H]}}

\def\masyr{{\rm\,mas/yr}}
\def\kpc{{\rm\,kpc}}

\def\pc{{\rm\,pc}}

\def\deg{^\circ}

\def\s{\ifmmode \widetilde \else \~\fi}
\def\={\overline}

\def\spose#1{\hbox to 0pt{#1\hss}}

\def\eg{{e.g.,\ }}
\def\ie{{i.e.,\ }}
\def\lta{\mathrel{\spose{\lower 3pt\hbox{$\mathchar"218$}}
     \raise 2.0pt\hbox{$\mathchar"13C$}}}
\def\gta{\mathrel{\spose{\lower 3pt\hbox{$\mathchar"218$}}
     \raise 2.0pt\hbox{$\mathchar"13E$}}}
\def\Dt{\spose{\raise 1.5ex\hbox{\hskip3pt$\mathchar"201$}}}    
\def\dt{\spose{\raise 1.0ex\hbox{\hskip2pt$\mathchar"201$}}}    

\def\dotsfill{\leaders\hbox to 1em{\hss.\hss}\hfill}

\def\FeH{{\rm[Fe/H]}}

\newcommand{\Pris}{{\emph{Pristine}}}
\newcommand{\Gaia}{{\emph{Gaia}}}
\newcommand{\SF}{\texttt{STREAMFINDER}}
\def\FeHPr{{\rm\FeH_\mathrm{Pr}}}
\def\FeH{{\rm[Fe/H]}}
\def\FeHPrave{{\rm\langle\FeH_\mathrm{Pr}\rangle}}

\title[\Pris\ metallicities of stellar streams]{The \Pris\ survey -- XVI. The metallicity of 26 stellar streams around the Milky Way detected with the \SF\ in \Gaia\ EDR3}

\author[N. F. Martin et al.]{Nicolas F. Martin$^{1,2}$\thanks{E-mail: nicolas.martin@astro.unistra.fr}, Rodrigo A. Ibata$^1$, Else Starkenburg$^{3}$, Zhen Yuan$^1$, Khyati Malhan$^{4,2}$,
\newauthor Michele Bellazzini$^5$, Akshara Viswanathan$^3$, David Aguado$^6$, Anke Arentsen$^1$, Piercarlo Bonifacio$^7$,
\newauthor Ray Carlberg$^8$, Jonay I. Gonz\'{a}lez Hern\'{a}ndez$^{9,10}$, Vanessa Hill$^{11}$, Pascale Jablonka$^{12,7}$,
\newauthor Georges Kordopatis$^{11}$, Carmela Lardo$^{13}$, Alan W. McConnachie$^{14}$, Julio Navarro$^{15}$,
\newauthor Rub\'en S\'anchez-Janssen$^{16}$, Federico Sestito$^{15}$, Guillaume F. Thomas$^{9,10}$, Kim A. Venn$^{15}$ \& 
\newauthor Sara Vitali$^{17}$ \& Karina T. Voggel$^1$\\
$^1$Universit\'e de Strasbourg, CNRS, Observatoire astronomique de Strasbourg, UMR 7550, F-67000 Strasbourg, France\\
$^2$Max-Planck-Institut f\"{u}r Astronomie, K\"{o}nigstuhl 17, D-69117 Heidelberg, Germany \\
$^3$Kapteyn Astronomical Institute, University of Groningen, Landleven 12, 9747 AD Groningen, The Netherlands\\
$^4$The Oskar Klein Centre, Department of Physics, Stockholm University, AlbaNova, SE-10691 Stockholm, Sweden\\
$^5$INAF - Osservatorio di Astrofisica e Scienza dello Spazio, via Gobetti 93/3, I-40129 Bologna, Italy\\
$^6$Dipartimento di Fisica e Astronomia, Universit\'a degli Studi di Firenze, Via G. Sansone 1, I-50019 Sesto Fiorentino, Italy\\
$^7$GEPI, Observatoire de Paris, Universit\'e PSL, CNRS, 5 Place Jules Janssen, 92195, Meudon, France\\
$^8$Department of Astronomy \& Astrophysics, University of Toronto, Toronto, ON M5S 3H4, Canada \\
$^9$Instituto de Astrof\'isica de Canarias, E-38205 La Laguna, Tenerife, Spain\\
$^{10}$Universidad de La Laguna, Dpto. Astrof\'isica, E-38206 La Laguna, Tenerife, Spain\\
$^{11}$Universit\'e C\^ote d'Azur, Observatoire de la C\^ote d'Azur, CNRS, Laboratoire Lagrange, Nice, France\\
$^{12}$Laboratoire d'astrophysique, \'Ecole Polytechnique F\'ed\'erale de Lausanne (EPFL), Observatoire, 1290 Versoix, Switzerland\\
${^13}$Dipartimento di Fisica e Astronomia, Universit\`a degli Studi di Bologna, Via Gobetti 93/2, I-40129 Bologna, Italy\\
$^{14}$NRC Herzberg Astronomy and Astrophysics, 5071 West Saanich Road, Victoria, BC V9E 2E7, Canada \\
$^{15}$Dept. of Physics and Astronomy, University of Victoria, P.O. Box 3055, STN CSC, Victoria BC V8W 3P6, Canada \\
$^{16}$UK Astronomy Technology Centre, Royal Observatory, Blackford Hill, Edinburgh, EH9 3HJ, UK\\
$^{17}$N\'ucleo de Astronom\'ia, Facultad de Ingenier\'ia y Ciencias Universidad Diego Portales, Ej\'ercito 441, Santiago, Chile\\
}

\date{Accepted XXX. Received YYY; in original form ZZZ}

\pubyear{2022}

\begin{document}
\label{firstpage}
\pagerange{\pageref{firstpage}--\pageref{lastpage}}
\maketitle

\begin{abstract}
We use the photometric metallicities provided by the panoramic \Pris\ survey to study the veracity and derive the metallicities of the numerous stellar streams found by the application of the \SF\ algorithm to the \Gaia\ EDR3 data. All 26~streams present in \Pris\ show a clear metallicity distribution function, which provides an independent check of the reality of these structures, supporting the reliability of \SF\ in finding streams and the power of \Pris\ to measure precise metallicities. We further present 6~candidate structures with coherent phase-space and metallicity signals that are very likely streams. The majority of studied streams are very metal-poor (14 structures with $\FeH<-2.0$) and include 3~systems with $\FeH<-2.9$ (C-11, C-19, and C-20). These streams could be the closest debris of low-luminosity dwarf galaxies or may have originated from globular clusters of significantly lower metallicity than any known current Milky Way globular cluster. Our study shows that the promise of the \Gaia\ data for Galactic Archeology studies can be substantially strengthened by quality photometric metallicities, allowing us to peer back into the earliest epochs of the formation of our Galaxy and its stellar halo constituents.
\end{abstract}




\section{Introduction}

Since its launch in  2013, the \Gaia\ satellite has been charting the motions and brightnesses of more than a billion stars throughout the Galaxy and Local Group, in an effort to build up an unprecedented six-dimensional map of the positions and velocities of stars in the nearby universe \citep{Gaia16a,Gaia18a,Gaia21a}. One of the aims of the \Gaia\ mission is to identify signs of the ancient accretions that built up our Milky Way (MW) through the use of conserved dynamical quantities derived from the kinematic observables. In the inner regions of the MW halo ($\lta30\kpc$), the sought-for accretion events may be quite difficult to uncover, due to the complex interactions that have taken place over cosmic time, especially with time-dependent non-axisymmetric components such as spiral arms and the Galactic bar \citep[e.g.,][]{pearson17}. There, our best hopes for using dynamics for Galactic Archeology is via action variables, which are conserved during adiabatic changes of the potential \citep{binney08}. In the more distant reaches of our Galaxy, or when the dynamical interactions have been less violent, the accretions may yet be detectable as dissolving overdensities in position and velocity space.

The revolution brought about by the deep astrometric (and photometric) survey of the \Gaia\ satellite is truly staggering. With each new data release, the MW's phase space is revealed in a wealth of ever more complex and intricate detail \citep[e.g.,][]{malhan18,antoja18,haywood18,helmi18,myeong18,ibata19,myeong19,belokurov20,yuan20a,yuan20b}. Our view of the stellar halo, in particular, is being completely transformed and revealed to be, at least in part, a nest of stellar streams of all shapes and forms, at all distances, and over the whole sky \citep{ibata21}. 

Prior to the advent of \Gaia, most Galactic Archeology studies relied on the chemical make-up of stars, as the composition of their atmospheres is generally conserved during stellar evolution \citep[e.g.,][]{freeman02,venn04,hogg16,martell17,fernandez-alvar19,recio-blanco21,li22}. Combining the power of dynamics with the discriminatory power of chemistry promises to be very powerful, and is indeed the main scientific motivation behind the planned Galactic Archaeology halo projects of the future multi-object spectrographs WEAVE and 4MOST \citep{dalton18,christlieb19,helmi19}. These surveys build on very successful past and current surveys that slowly help build a chemodynamical portait of the MW (\eg SEGUE, \citealt{yanny09}; APOGEE, \citealt{majewski17}; GALAH, \citealt{martell17}; RAVE, \citealt{steinmetz06}). Nevertheless, while these surveys will provide very precise chemical abundances, the targeted nature of the spectroscopic observations means that they will provide this information only for a small fraction of stars in the Galactic halo. There is therefore huge scope for major contributions to the field from photometric metallicity surveys, which can in principle provide a metallicity estimate for all stars in \Gaia, and many more fainter counterparts as well. The key requirement is that the photometric metallicities need to be accurate enough, especially in the low metallicity regime of the halo. Numerous efforts have been undertaken to ensure a good photometric metallicity mapping of the MW's sky \citep[e.g.,][]{ivezic08,ibata17,thomas19,an21}. More recently, the observation of narrow-band photometry focused on metallicity-sensitive lines  over very large swaths of the sky by the SkyMapper \citep{onken19,chiti21} and \Pris\ \citep{starkenburg17b} surveys are revealing more and more of the metallicity portait of our Galaxy on a global scale \citep[e.g.,][]{youakim20}.

On a smaller scale, the dozens of cold stellar structures found in the MW halo are particularly powerful tools for Galactic Archeology because they reveal (some of) the past accretions that made up our Galaxy \citep[\eg][]{bonaca21}; their track on the sky is a sensitive probe of the Galactic potential \citep[\eg][]{johnston99}; and, because the streams themselves may take billions of years to form, they place constraints on how much the potential may have changed since their formation \citep[\eg][]{erkal19}. While stellar streams are thus very useful, they can be challenging to detect because of their exceedingly low surface brightness. It was for this reason that \citet{malhan18} developed the \SF\ algorithm to make use of the astrometric and photometric information expected from the \Gaia\ Data Release~2 (DR2) data. For any star in a \Gaia\ data release, \SF\ determines the significance of that star being part of a stream, i.e. of a group of stars along a specific orbit \citep{ibata19}. While computationally very intensive, this technique has proven incredibly powerful to isolate unambiguous stellar streams and stellar-stream candidates \citep{ibata21}.

One of the outcomes of \SF\ is the expected radial velocity of the stream stars. These can therefore be used as predictions to confirm (or not) that a stream is real when these are confronted against measured radial velocities \citep{malhan18}. That has been shown to be a very successful approach and the exploration of data from publicly accessible spectroscopic surveys as well as specific spectroscopic follow-up programs \citep{ibata21} have confirmed many of the \SF \ streams. But this is a cumbersome approach and requires either luck for stream stars to be already observed in surveys or an extensive follow-up program. As such, only 11.5\% of \SF\ stars selected from \Gaia\ DR2 have a spectroscopic observation \citep{ibata21}. In addition, the spectroscopic sampling is often very low, with only a handful of stars observed at high signal-to-noise, thereby limiting the constraints on the global properties of a stream and, particularly, on its metallicity or metallicity dispersion.

Here, we present a new approach, based on the photometric metallicity survey \Pris\ \citep{starkenburg17b}, to both confirm the validity of a stream and derive its metallicity without spectroscopic follow-up. For any stream, the expectation of a coherent metallicity signal in the \Pris\ survey can be used to confirm that \SF\ did not simply group random MW stars that happen to have apparently similar motion. This is especially efficient for very low metallicity streams ($\FeH\lta-1.8$) since these stars are rare in the MW and, if \SF\ strings together many such low metallicity stars, it is very unlikely to be the product of contamination \citep{ibata21}. \Pris\ can further be used to effectively clean that stream of stars with clearly conflicting metallicities, such as foreground disk stars with solar-like metallicities in front of a very metal-poor stream, for a more efficient spectroscopic follow up.

The paper is structured as follows: in Section~2, we present the two datasets we combine in this paper, from \SF\ and \Pris; Section~3 provides a detailed description of the 26 streams in common between the two data sets; and we discuss our results in a global context and conclude in Section~4.

\section{Data}
\subsection{The \SF\ catalogue}
\begin{figure*}
\begin{center}
\includegraphics[width=1\hsize,angle=270]{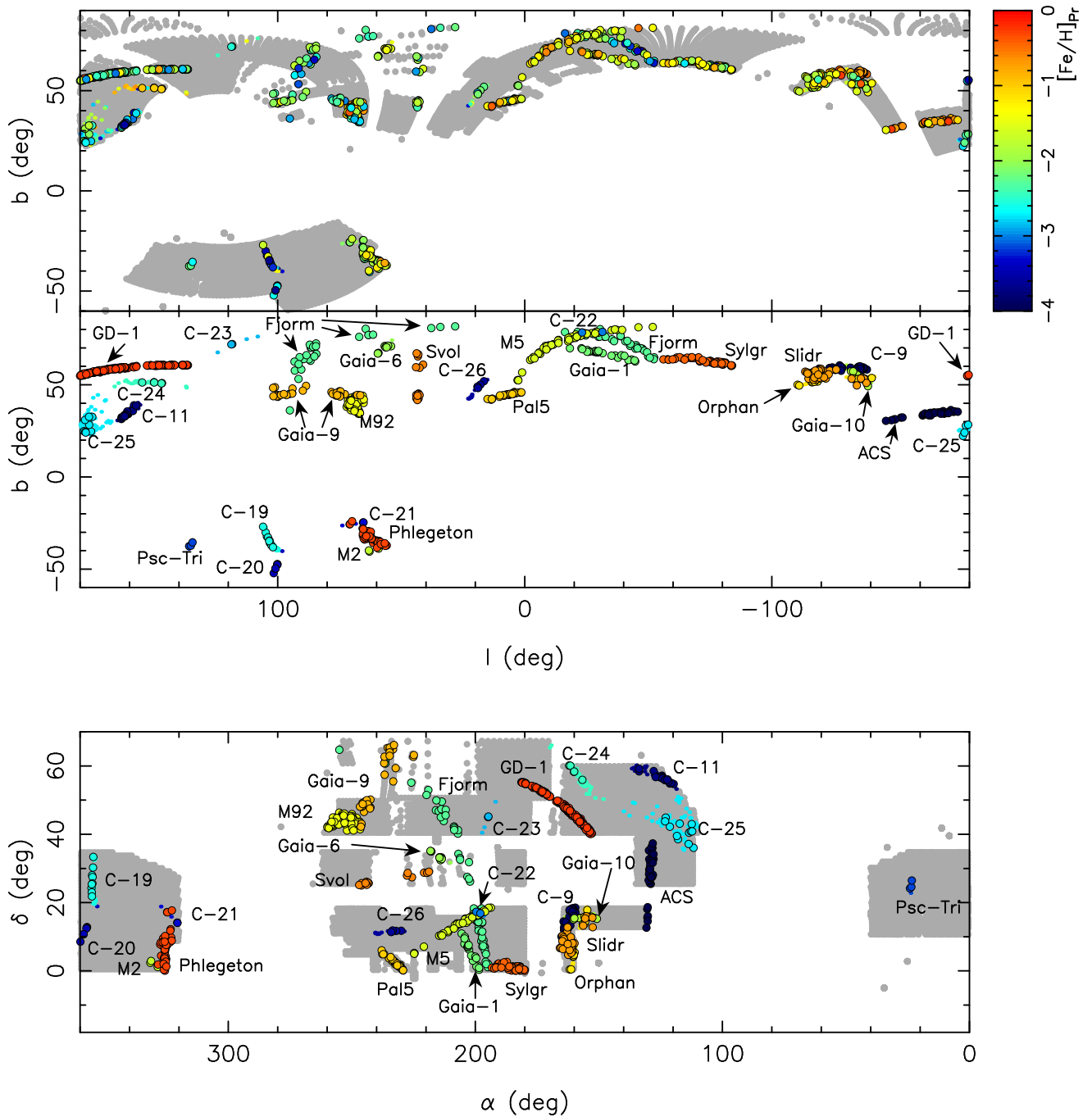}
\caption{\emph{Top panel:} Galactic coordinate map of stars in common between the sample of \SF\ stars significantly detected to be part of a stellar stream and the \Pris\ survey. Large colored symbols correspond to stars from our sample of very high \SF\ significance ($>10\sigma$) while small symbols correspond to slightly lower significance detections ($>8\sigma$) that help us populated sparse streams at the cost of a little more contamination. All stars are color-coded by the \Pris\ metallicity of the star ($\FeHPr$), as shown in the color bar on the right-hand side of the panel. The \Pris\ footprint is represented by the complex grey area, with each large dot corresponding to the central pointing of a MegaCam field. \emph{Middle panel:} The same stream stars color-coded by their membership to the 26 streams discussed in this paper. \emph{Bottom panel:} The same as the middle panel, but in equatorial coordinates. \label{fig:map}}
\end{center}
\end{figure*}

We refer the reader to the papers describing the \SF\ algorithm \citep{malhan18}, its application to the \Gaia\ DR2 data \citep{ibata19}, and to the \Gaia\ Early Data Release~3 (EDR3) data \citep{ibata21} for a detailed description of the process followed by \SF\ to find stars that are part of stellar streams at high significance. As a quick summary, for any star in a \Gaia\ DR, \SF\ tests a large set of orbits, assuming radial velocity values and other orbital properties consistent with the observed properties of the star (proper motion, parallax, and photometry). The algorithm then calculates the likelihood that this star is part of a localized stellar stream in addition to the expected field contamination, given all neighboring stars in a region of $\sim10\deg$. Since the distance to the star is often poorly known and its radial velocities very rarely known, this usually means testing dozens of potential orbits for every star in \Gaia, and calculating a very complex likelihood function for each of these orbits.

The complexity of \SF\ means that it becomes exceedingly costly to run it close to the MW plane because of the increasing stellar density. This process has nevertheless been extremely successful in uncovering previously unknown streams for $|b|>10\deg$. Moreover, relying on a well-defined algorithm means that the search is as objective as possible. It will also eventually allow, beyond a simple tally of stellar streams, for a determination of the detection limits of stellar streams in the MW halo.

The version of the \SF\ catalogue that this paper is based on was produced by the application of the algorithm to the recently released \Gaia\ EDR3 data and is presented in \citet{ibata21}. A likelihood value is calculated for every star processed by the algorithm and for a series of stellar population templates with varying metallicity and distance. Here, we consider only the template that maximizes the likelihood of a given star to be part of a stream and we purposefully restrict ourselves only to those stars that are most significantly found to be part of a stream. Following  \citet{ibata21}, we rely mainly on candidate stream-star detections at more than $10\sigma$, i.e. with $\textrm{ln}(\mathcal{L})>52.3$, where $\mathcal{L}$ is the likelihood of a star to be part of a stellar stream according to \SF. Those stars are the ones displayed in the maps of Figures~10 and 12 of \citet{ibata21}. So far, all thin structures found by \SF\ with at least a handful of stars with an individual significance higher than $10\sigma$ have been confirmed to be genuine through follow-up spectroscopic observations \citep{ibata21}. When the (candidate) streams are poorly sampled or only partly overlap with \Pris, we also increase the size of this sample with slightly lower significance detections and retain stars detected at more than $8\sigma$ instead ($\textrm{ln}(\mathcal{L})>34.1$). These lower significance stars are represented with smaller symbols in all figures. When included, these stars are not treated differently from the more significant stars in the samples. The analysis presented below confirms that including these stars does not significantly increase the contamination but can be essential to properly characterize some low density structures.

We note that cluttering stars that belong to the very prominent Sagittarius stream are removed from this sample since we focus on thin structures with this particular application of \SF. Following \citet{ibata21}, we also do not consider stars with a total proper motion $<1\masyr$. This region of the proper motion space is particularly confused and difficult to disentangle.

\subsection{The \Pris\ data}
The \Pris\ survey \citep{starkenburg17b} is a narrow-band photometric survey that observes a significant fraction of the northern sky of the MW halo with the MegaCam imager \citep{boulade03} on the Canada-France-Hawaii Telescope with its CaHK filter. Focussing on the Calcium H\&K lines, this narrow-band photometry is combined with broadband photometry provided by the SDSS \citep{blanton17} to infer the photometric metallicity $\FeHPr$ of any observed star. The survey is tailored to the depth of the \Gaia\ data ($G\simeq20.5$) and has been shown to be significantly more efficient than broadband photometry to derive the metallicity of stars in the metal-poor regime, especially below $\FeH\lta-1.5$ \citep{starkenburg17b}. The survey is particularly efficient at finding extremely metal-poor (EMP) star candidates \citep[$\FeH<-3.0$;][]{youakim17,aguado19,venn20}.

For this analysis, we use the \Pris\ internal Data Release 2, whose metallicity model is summarized by \citet[][their section 2.2]{fernandez-alvar21}. Compared to the version of the Pristine survey presented by \citet{starkenburg17b}, the version used here has a significantly larger footprint. But the processing of the \Pris\ data, the calibration of the $CaHK$ magnitudes, and the model to go from $(CaHK,g,i)_0$ to $\FeH_\mathrm{Pr}$ are all very similar.

In this paper, we rely on the \Pris\ photometric metallicities, $\FeHPr$, inferred using the SDSS $(g-i)_0$ color as a proxy for temperature and the $CaHK_0-g_0 -1.5(g-i)_0$ color as an estimate of the amount of absorption exhibited by a star in the region of the metallicity-sensitive Ca H\&K lines. The location of stars in this ``Pristine color-color space,'' generated by these two quantities, is a direct test of the clumping of stars in metallicity, even without any consideration for the inferred $\FeHPr$ and are provided below for all the streams considered in this paper. We refrain at this stage from replacing the SDSS $g$ and $i$ broadband magnitudes by the \Gaia\ magnitudes as the width of the \Gaia\ filters makes them at the same time less sensitive to temperature and very sensitive to the extinction correction, requiring more assumptions. The publication of reliable individual stellar parameters in \Gaia's full Data Release 3 \citep{andrae18} should eventually circumvent this issue. We note that, in this version of the \Pris\ catalogue, we have put a particular focus on removing as many sources of contamination as possible in order to limit the number of spurious EMP stars in the sample \citep{fernandez-alvar21}. The updated model that converts broadband and $CaHK$ magnitudes into $\FeHPr$ now provides generic metallicities as well as specific metallicities under the assumption that the star is a dwarf or a giant. In the case of the stellar streams studied here, it is easy to determine a rough location for the main-sequence turnoff of the structures and we therefore use the relevant dwarf or giant metallicities.

It is worth noting that $CaHK$ magnitude uncertainties are difficult to map into uncertainties in $\FeHPr$ because of the non-linear transformations required to translate $(CaHK,g,i)_0$ into a metallicity. As a consequence, the uncertainties are often non-Gaussian. The metallicity uncertainties we present throughout this paper are therefore obtained through a Monte Carlo scheme by drawing from the uncertainties on the magnitudes $(\delta CaHK,\delta g,\delta i)$ and reporting the central 68\% confidence interval. Building on the comparison of \Pris\ metallicities with external spectroscopic catalogues performed by \citet{starkenburg17b}, we add an uncertainty of 0.25\,dex in quadrature to the formal \Pris, internal uncertainties on $\FeHPr$.

After more than 5 years of data gathering at CFHT, the \Pris\ survey has accumulated the equivalent of more than 700\,h of MegaCam observations and now covers  $>5,000 \textrm{\,deg}^2$. Most of these observations were conducted as part of a bad-weather, filler program. While this has little impact on our inference of photometric metallicities over the \Gaia\ magnitude range, it does mean that our footprint is rather patchy (see the grey area in the top panel of Figure~\ref{fig:map}). So far, the survey has been purposefully focussed on the Galactic caps to maximize follow-up observations of EMP star candidates with the WEAVE spectroscopic survey. The targeted extension of \Pris\ that focusses on the inner galaxy, the \Pris\ Inner Galaxy Survey or PIGS \citep{arentsen20b}, is not considered in this paper as the \SF\ catalogue only includes regions with $|b|>20\deg$.

\subsection{The \SF+\Pris\ sample}
The cross-match of the \SF\ list of stars significantly in a stream and the \Pris\ survey is shown in Figure~\ref{fig:map}, projected on Galactic coordinates. As mentioned by \citet{ibata21}, \SF\ does not, for computational reasons, currently keep the information of which other stars specifically contributed to increase the likelihood of a studied star. In other words, the algorithm only provides a list of stars that are likely to be part of streams, but not which star is part of which stream. This can easily be circumvented at the post-processing stage by manually extracting the different star associations using their location on the sky and their proper motions. At this stage, we do not yet use the \Pris\ photometric metallicities for the selection. One of our goals here is to confirm the veracity of these streams from the metallicity distributions, so using the metallicity information in this step would be detrimental to the whole analysis\footnote{Eventually, it would be ideal to include the photometric metallicities directly into the calculation of the \SF\ likelihoods. The algorithm would then straightforwardly upweight or downweight the likelihood of a star to be part of a stream based on whether or not nearby stars with similar proper motions, parallaxes, and the appropriate CMD distribution also display similar \Pris\ metallicities. However, our empirical modeling of the contamination from the data themselves means that sharp feature in any dimension used is very difficult to handle. The patchy \Pris\ footprint is therefore the current limiting factor but, with the ever increasing \Pris\ coverage, there is hope that this can be achieved in the future.}. While it may be considered sub-optimal to manually isolate candidate members from the various streams instead of relying on an algorithm, the vastness of phase-space actually makes this process very straightforward as it is extremely rare for two streams to cross in 6 dimensions.

In total, 1083 stars can be linked to a structure, including 6 new candidate streams described later in this paper (C-21, C-22, C-23, C-24, C-25, and C-26). All of these structures are shown and labeled in the bottom panel of Figure~\ref{fig:map}, in which stars of a given stream are all displayed with the same color. Some of the streams are cut off by the edges of the complex \Pris\ footprint but that is of little consequence for their metallicity characterization, as we will see below. Stars that could not be associated to a specific structure are mainly located at the edge of the MW disk and are likely part of the complex stellar structures that are known to exist in this region \citep[\eg][]{slater14}.

\section{The streams}

\begin{table}
\begin{center}
  \caption{Average \Pris\ metallicities after iterative $2.5\sigma$ clipping calculated for all 26 streams described in the paper. We indicate the number of stars, $N$, used in the calculation for the final iteration of the sigma-clipping procedure. The final two columns report the mean orbit poles of the structures, obtained by averaging the pole of the favored \SF\ orbit of sample stars. \label{table}}
  \begin{tabular}{@{}l|c|c|cc@{}}
  \hline
   Stream & $\FeHPrave$ & $N$ & $\langle\ell_\mathrm{pole}\rangle$ ($\deg$) & $\langle b_\mathrm{pole}\rangle$ ($\deg$)\\
     \hline
M~5 stream & $-1.35 \pm 0.05$ &  51 & $60.3$ & $-5.7$\\
M~92 stream & $-2.07 \pm 0.04$ &  86 & $-6.3$ & $-16.8$\\
Pal~5 stream & $-1.07 \pm 0.05$ &  26 & $-7.6$ & $-44.3$\\
ACS & $-0.77 \pm 0.05$ &  32 & $162.7$ & $-50.9$\\
Fj\"orm & $-2.33 \pm 0.05$ &  60 & $-129.9$ & $5.3$\\
GD-1 & $-2.49 \pm 0.03$ & 190 & $-20.1$ & $31.0$\\
Orphan & $-1.65 \pm 0.06$ & 28 & $-72.7$ & $-23.3$\\
Psc-Tri & $-2.36\pm0.11$ & 5 & $109.0$ & $49.4$\\
Phlegeton & $-1.98 \pm 0.05$ &  45 & $-175.3$ & $-39.5$\\
Slidr & $-1.70 \pm 0.05$ &  42 & $-48.5$ & $-11.6$\\
Sylgr ($G_0<18.0$)  & $-2.17 \pm 0.10$ &  10 & $-14.5$ & $-14.3$\\
Sv\"ol & $-1.98 \pm 0.10$ &  13 & $82.2$ & $-29.0$\\
Gaia-1 & $-1.43 \pm 0.07$ &  26 & $-0.2$ & $-19.6$\\
Gaia-6 & $-1.53 \pm 0.12$ & 10 & $-4.6$ & $-10.0$\\
Gaia-9 & $-2.21 \pm 0.07$ &  30 & $-44.1$ & $30.4$\\
Gaia-10/300S & $-1.40 \pm 0.06$ &  24 & $-31.9$ & $9.3$\\
C-9 & $-0.72 \pm 0.04$ &   36 & $-39.3$ & $1.7$\\
C-11 & $-2.91 \pm 0.10$ &   15 & $60.0$ & $16.1$\\
C-19 & $-3.58 \pm 0.08$ &   17 & $-168.6$ & $1.4$\\
C-20 & $-2.93 \pm 0.14$ &   9 & $7.8$ & $-2.4$\\
C-21 & $-2.03 \pm 0.13$ &   5 & $142.9$ & $29.9$\\
C-22 & $-2.18 \pm 0.09$ &   12 & $64.9$ & $0.7$\\
C-23 & $-2.36 \pm 0.14$ &   6 & $-57.9$ & $17.5$\\
C-24 & $-0.93 \pm 0.05$ &   37 & $74.2$ & $-9.0$\\
C-25 & $-2.30 \pm 0.04$ &   75 & $71.3$ & $25.0$\\ 
C-26 & $-2.04 \pm 0.08$ &   15 & $86.0$ & $-20.3$\\
    \hline
\end{tabular}
\end{center}
\end{table}

In this section, we show the properties of the 26 streams or candidate streams found in the joint \SF+\Pris\ catalogue. Diagnostic plots are shown in Figures~\ref{fig:M5stream} to \ref{fig:C-26} and include, for each stream: the \Gaia\ and SDSS color-magnitude diagram of a given stream's stars (left), their distribution in the \Pris\ color-color space (top-middle), the location of those stars in the right ascension and declination proper motion plane (top-right), the distribution of metallicities along the stream in the Galactic longitude and latitude directions (the two bottom-middle panels), and their metallicity distribution function (MDF; bottom-right). In these figures, large and small symbols correspond to stars detected by \SF\ as belonging to a stream with a significance above $10\sigma$ and $8\sigma$, respectively. For all streams, we determine their mean metallicities, $\FeHPrave$, weighted by the individual $\FeHPr$ uncertainties of members, through an iterative $2.5\sigma$-clipping procedure. For this $\sigma$-clipping procedure, we use the individual metallicity uncertainties as "$\sigma$" as these are usually larger than the dispersion of the metallicity distribution. This more conservative approach is found to mainly remove clear outliers while using the dispersion of the MDF further truncates the wings of the (non-Gaussian) distributions. The resulting values are summarized in Table~\ref{table}. The table also lists the mean orbital poles of the resulting sample of stars for each structure.

In the following, we first discuss our conclusions for known GC streams, whose progenitors have well-studied understood in the literature. These serve as a confirmation that the joint analysis of the \SF\ and \Pris\ yields reliable results. We then present our results for other known streams, candidate streams from \citet{ibata21} that we confirm here, and detections of new candidate stellar streams that were not reported previously.
 
\subsection{Globular cluster streams}
The comparison of the mean \Pris\ photometric metallicities of the three following GC streams for which the GCs have well measured spectroscopic metallicities shows that the \Pris\ metallicities are reliable and that they can be used to infer the mean metallicity of a stellar stream within $\sim0.25$\,dex for metal-poor to very metal-poor structures. This is not surprising \citep{starkenburg17b} but is reassuring and gives us confidence that our procedure to derive the mean metallicity of \SF\ structures is adequate.

\subsubsection{The M~5 stream}
\begin{figure*}
\begin{center}
\includegraphics[width=9cm,angle=270]{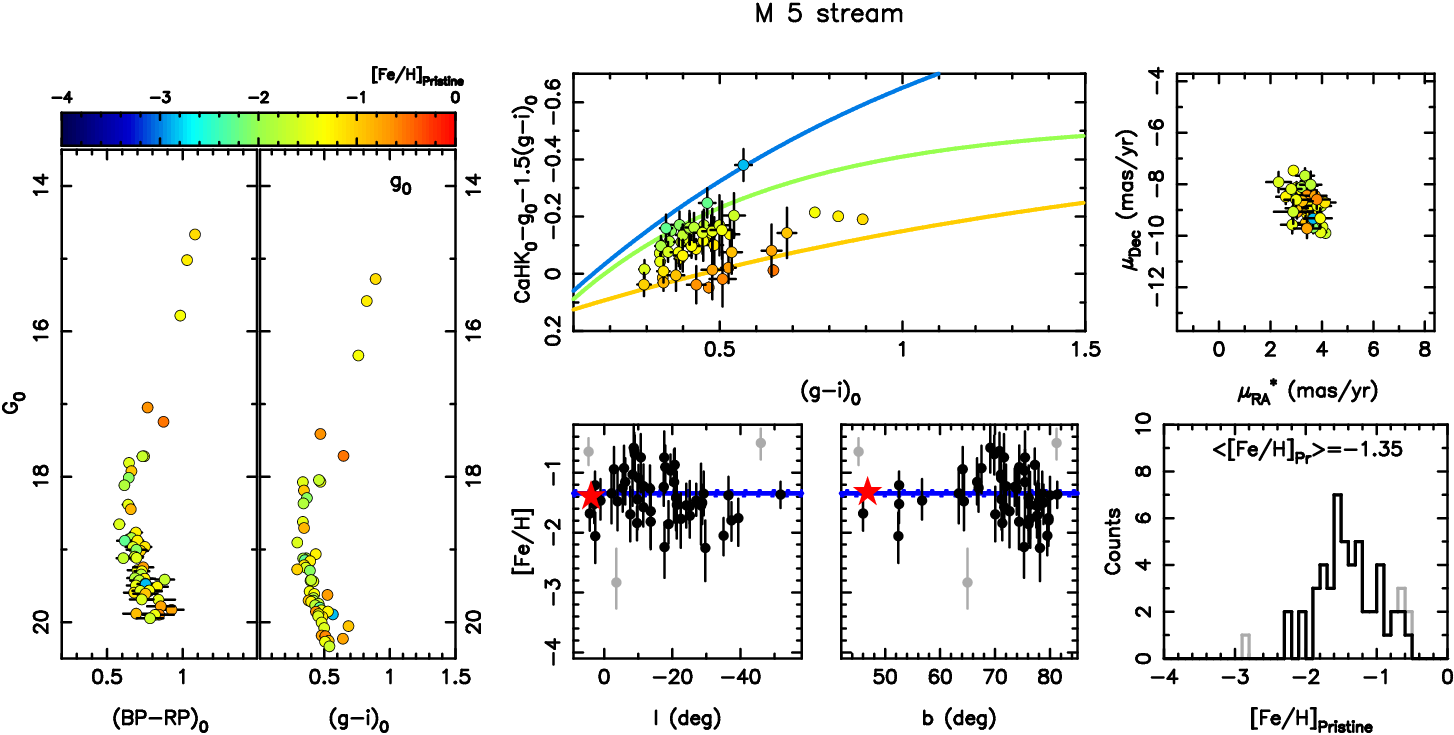}
\caption{\SF+\Pris\ diagnostics for the M~5 stream. \emph{Top-left:} Color-magnitude diagram (CMD) of the stars in the sample in \Gaia\ and the SDSS photometries. Colors in this panel and others represent the \Pris\ metallicity of a star, $\FeHPr$, as represented on the color bar above this panel. \emph{Top-middle:} Distribution of the stream stars in the \Pris\ color-color space. The colored lines correspond to iso-metallicity values of $\FeHPr=-1.0$ (orange), $-2.0$ (light green), and $-3.0$ (blue). \emph{Top-right:} Distribution of stream stars in the \Gaia\ proper motion space $\mu^*_\mathrm{\alpha}$ (i.e. $\mu_\mathrm{\alpha}\cos(\delta)$) and $\mu_\mathrm{\delta}$. \emph{Bottom-middle and bottom-right:} Metallicity of stream stars along the stream. The red star corresponds to the location and metallicity of the progenitor GC when known. The blue line is the mean metallicity of the stream stars as calculated through our $2.5\sigma$-clipping procedure and the dotted blue lines correspond to the uncertainty on this value. In these panels, the grey points were rejected by the clipping procedure and not taken into account to calculate the mean. \emph{Bottom-right:} Metallicity distribution function (MDF) of high likelihood stream stars in the sample. The grey histogram corresponds to the full sample while the black distribution includes only stars retained after the $2.5\sigma$-clipping procedure. The weighted mean metallicity, calculated through a $2.5\sigma$-clipping procedure is also reported at the top of the panel.\label{fig:M5stream}}
\end{center}
\end{figure*}

\citet{grillmair19} reported the stream produced by the tidal disruption of the M~5 globular cluster, which was clearly mapped by \SF\ \citep{ibata21} as a long thin stream that extends to the Galactic north of M~5. Even though the cluster itself is not within the \Pris\ footprint, most of the stream is located over a large contiguous block of the survey and we isolate 51 members in common between the $10\sigma$ \SF\ detections and \Pris. The properties of these stars are shown in Figure~\ref{fig:M5stream} and display a very clumped metallicity distribution, from which we calculate a mean metallicity $\FeHPrave=-1.35\pm 0.05$. This metallicity is in excellent agreement with the mean metallicity of the cluster, measured to be $\FeH=-1.33\pm0.02$ \citep{carretta09}.

\subsubsection{The M~92 stream}
\label{M92}
\begin{figure*}
\begin{center}
\includegraphics[width=9cm,angle=270]{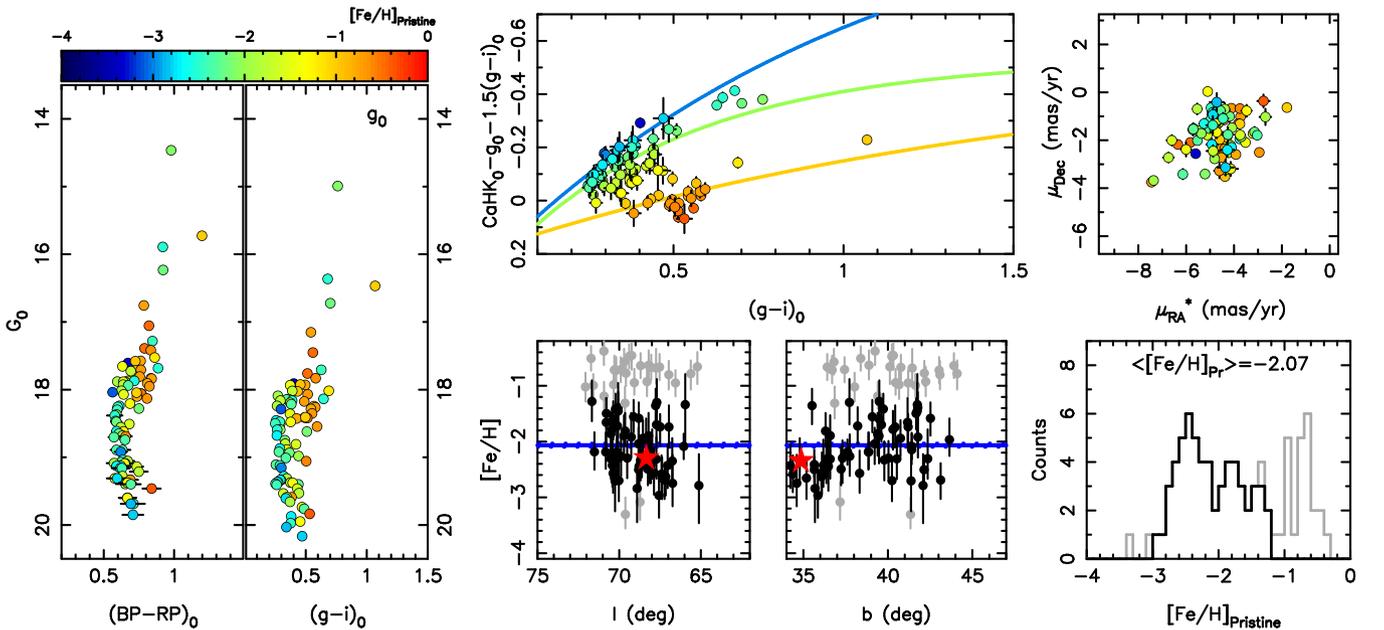}
\caption{Same as Figure~\ref{fig:M5stream} but for the M~92 stream.\label{fig:M92stream}}
\end{center}
\end{figure*}

The stream of the cluster M~92 was independently reported by \citet[][based on \Gaia\ DR2 data]{sollima20} and \citet{thomas20}, who revealed a longer extent of the stream thanks to the addition of the Canada-France-Imaging Survey (CFIS) $u$ band \citep{ibata17} and Pan-STARRS1 $3\pi$ broadband data. \SF\ easily extracts M~92 stream members from \Gaia\ EDR3, despite the cluster being near apocenter and, therefore, producing a scattered stream. This cluster is known to be one of the most metal-poor in the MW, with $\FeH=-2.35\pm0.05$ \citep{carretta09}, and this is indeed what we find for most \SF-selected stars that fall in the \Pris\ footprint\footnote{We remove all stars within $0.8\deg$ of the center of M~92 so we are not biased by stars in the clear outskirts of the cluster and properly track the properties of the stream itself.} (Figure~\ref{fig:M92stream}, $\FeHPrave=-2.07\pm0.04$). We note, however, that the metallicity we derive, while similar, is significantly more metal-rich than the literature value. We suspect that this is because \SF\ picks up a large number of significantly more metal-rich field stars that, while they separate themselves distinctly in the \Pris\ color-color space (top-middle panel of the Figure), drive the average metallicity upwards. The presnce of these contaminants is likely due to the scattered energies of the orbits near apocenter and to a higher foreground density than for most of the others streams (M~92 being at quite low latitude) but, thanks to the \Pris\ metallicities, these contaminating stars can be easily tagged.

\subsubsection{The Pal~5 stream}
\begin{figure*}
\begin{center}
\includegraphics[width=9cm,angle=270]{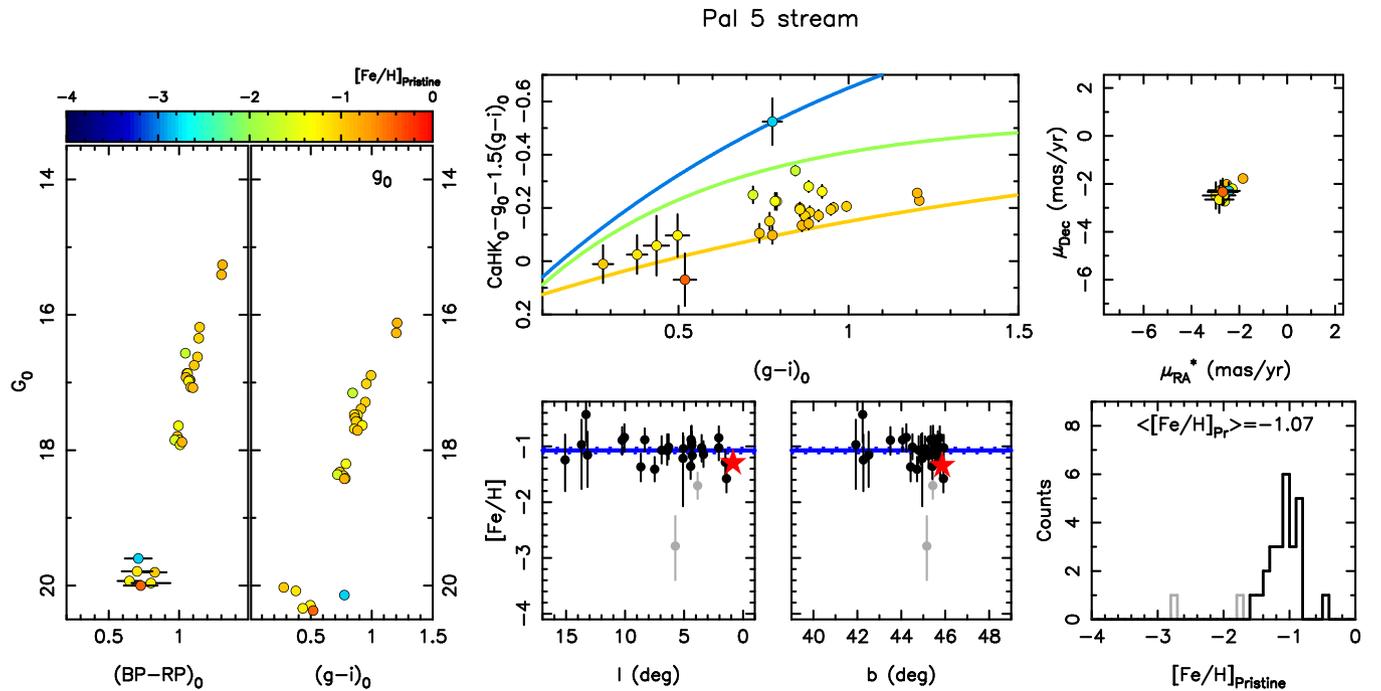}
\caption{Same as Figure~\ref{fig:M5stream} but for the Pal~5 stream.\label{fig:Pal5stream}}
\end{center}
\end{figure*}

The stream of stars that are currently escaping the Pal~5 GC is the prototpyical examples of GC streams in the MW halo since it was discovered by \citet{odenkirchen01a}, before extensive study in the literature \citep[e.g.,][]{rockosi02,carlberg12,ibata16,thomas16,bonaca20b}. Not surprisingly, the stream is detected with very high confidence by \SF, despite being beyond the nominal $10\kpc$ for which the \Gaia\ EDR3 data is the most constraining \citep{ibata21}. The north (or Galactic west) tail of the stream is within the \Pris\ footprint and, from the stars in common, we measure $\FeHPrave=-1.07\pm0.05$ (Figure~\ref{fig:Pal5stream}), in reasonable agreement with the spectroscopic metallicity of the cluster's stars \citep[$\FeH=-1.35\pm0.06$;][]{ishigaki16}, even though some other spectroscopic studies reach more metal-poor value \citep[$\FeH=-1.56\pm0.02\textrm{ (stat.)}\pm0.06\textrm{ (syst.)}$;][]{koch17}.

\subsection{Known streams}
\subsubsection{The Anti-Center Stream}
\begin{figure*}
\begin{center}
\includegraphics[width=9cm,angle=270]{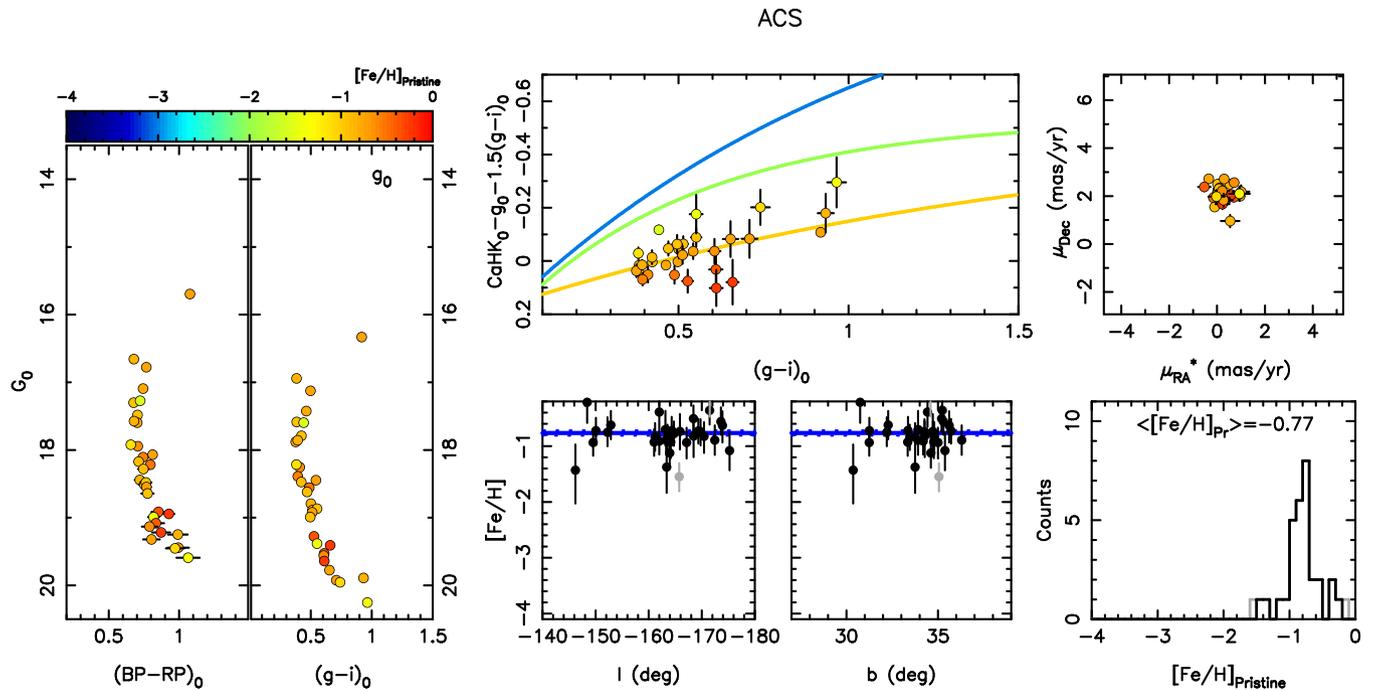}
\caption{Same as Figure~\ref{fig:M5stream} but for the Anti-Center Stream.\label{fig:ACS}}
\end{center}
\end{figure*}

The Anti-Center Stream (or ACS) was first highlighted by \citet{grillmair09} as part of the intricate suite of stellar structures just above the MW plane in the direction of the anticenter \citep[e.g.,][]{slater14,ramos21}. \citet{laporte20} argue that the ACS is in fact made of disk stars perturbed away from the Galactic plane by the interaction of the MW with a massive satellite. They measure a mean metallicity $\langle\FeH\rangle=-0.73$ (and a spread $\sigma_\FeH=0.26$) based on SEGUE metallicities. The analysis of the \Pris\ metallicities does show that this is one of the most metal-rich structures in the sample, with $\FeHPrave=-0.77\pm0.05$ (Figure~\ref{fig:ACS}), in perfect agreement with the spectroscopic measurement.

\subsubsection{Fj\"orm}
\begin{figure*}
\begin{center}
\includegraphics[width=9cm,angle=270]{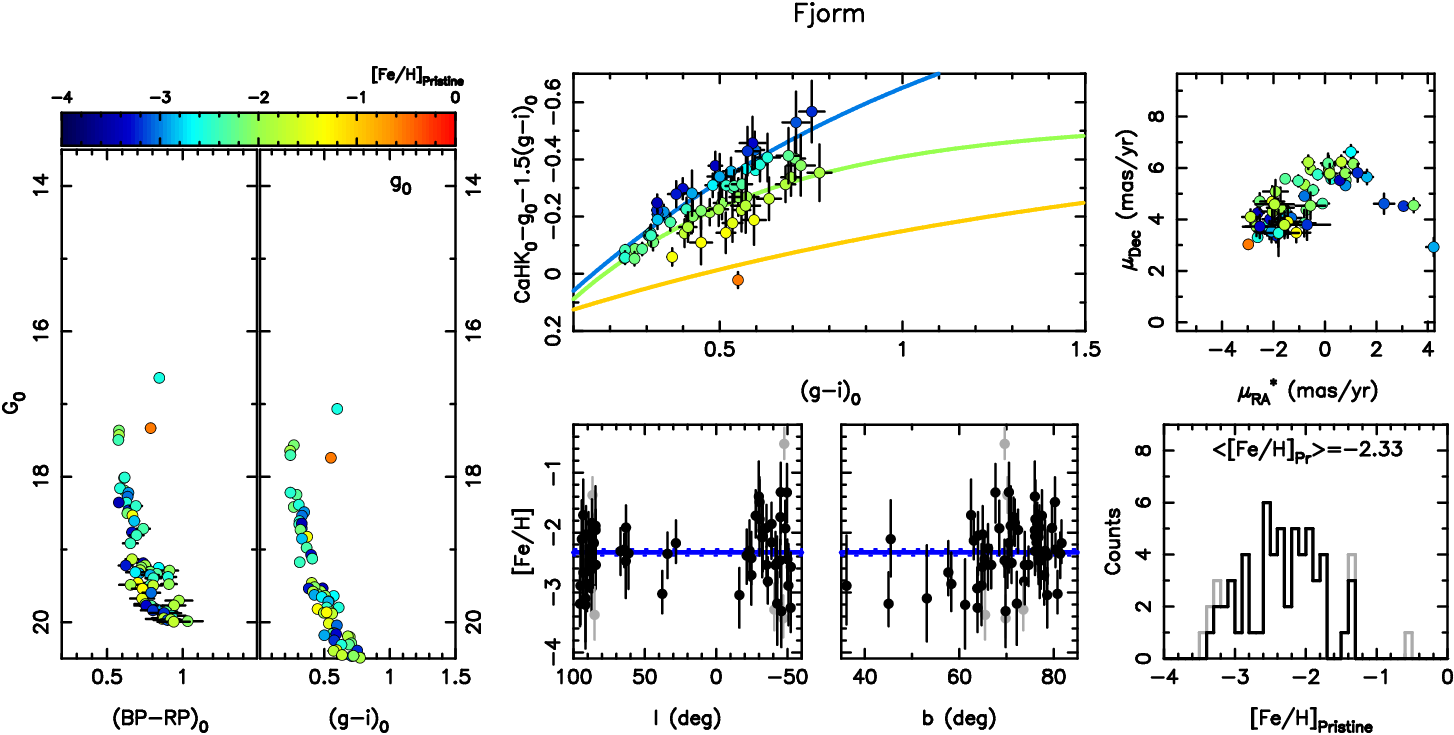}
\caption{Same as Figure~\ref{fig:M5stream} but for the Fj\"orm stream.\label{fig:Fjorm}}
\end{center}
\end{figure*}

This stellar stream was discovered by the application of \SF\ to the \Gaia\ DR2 data \citep{ibata19}. It is a nearby, very long stream that passes near the north Galactic pole. A large fraction of the stream is within the \Pris\ footprint, with 62 stars in common with the golden sample of \SF\ stars. Figure~\ref{fig:Fjorm} shows that Fj\"orm is, on average, very metal-poor ($\FeHPrave=-2.33\pm0.05$) and that its stars nicely cluster along the $\FeH=-2$ and $-3$ line in the \Pris\ color-color space. This is in good agreement with the spectroscopic metallicities of 8 stream stars found by \citet{ibata19} in the SDSS or the LAMOST survey ($\langle\FeH\rangle=-2.2$). \citet{palau19} further suggests a connection between Fj\"orm and the GC M~68. Its metallicity, $\FeH=-2.27\pm0.04$ \citep{carretta09}, is certainly compatible with the metallicity of the stream.

The MDF of Fj\"orm appears significantly broader than that of most of the other structures in the sample, despite \SF\ only picking up a single obvious, high-metallicity contaminant. While some of the metallicity uncertainties are large at the faint end, \Pris\ shows metallicities that range from $\FeHPr\sim-1.5$ down to $\sim-3.5$. This could be the sign that Fj\"orm is the left-over of a now defunct dwarf galaxy instead of a GC and shall be confirmed by future spectroscopic studies of the chemical abundances of these stars.

\subsubsection{Gaia-1}
\begin{figure*}
\begin{center}
\includegraphics[width=9cm,angle=270]{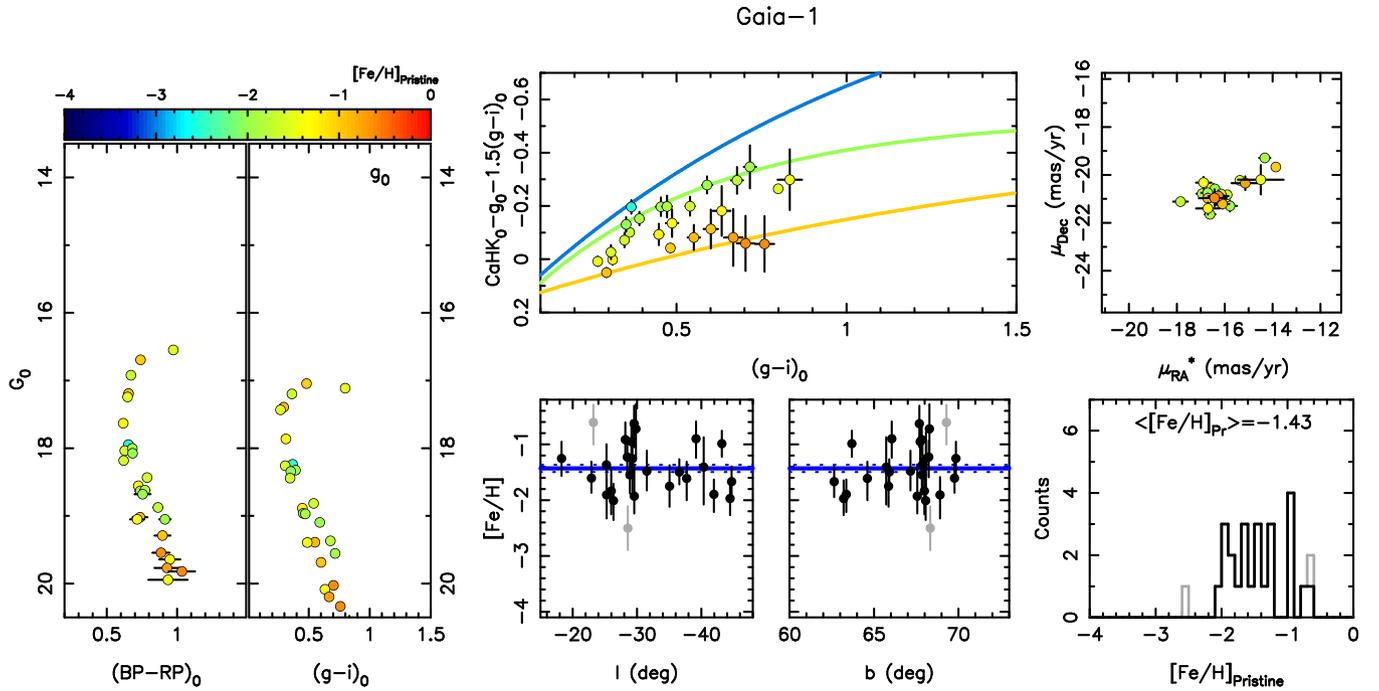}
\caption{Same as Figure~\ref{fig:M5stream} but for the Gaia-1 stream. \label{fig:Gaia-1}}
\end{center}
\end{figure*}

Gaia-1 was revealed on the map generated from the very first application of \SF\ to the \Gaia\ data \citep{malhan18}. Its particularly negative proper motion (in both the right ascension and declination directions) led to a very easy detection. \citet{malhan22} reported two stars in common between this stream and SDSS, yielding an average metallicity of $\langle\FeH\rangle=-1.36$. The 26 stars in common with \Pris\ yield $\FeHPrave=-1.43\pm0.07$ (Figure~\ref{fig:Gaia-1}), which is in good agreement with the spectroscopic metallicity value.

\subsubsection{Gaia-6}
\begin{figure*}
\begin{center}
\includegraphics[width=9cm,angle=270]{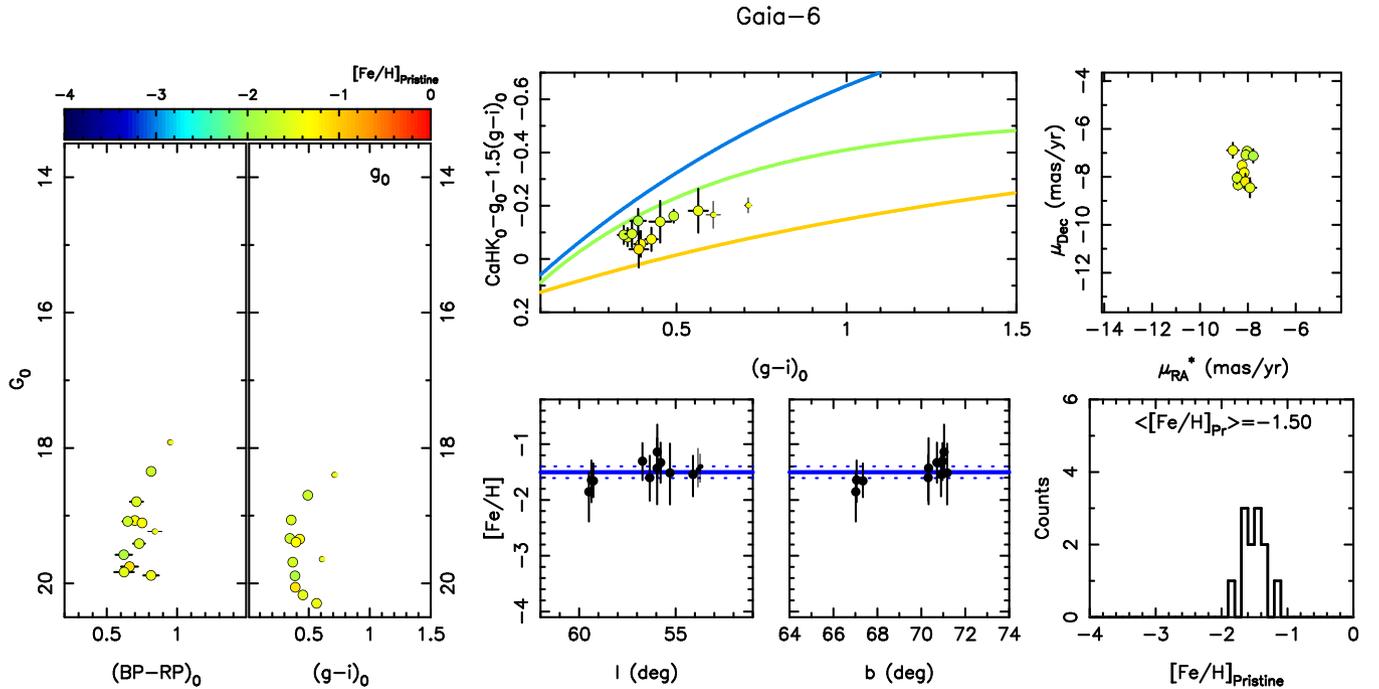}
\caption{Same as Figure~\ref{fig:M5stream} but for the Gaia-6 stream. Small symbols correspond to stars with a \SF\ significance between $8\sigma$ and $10\sigma$.\label{fig:Gaia-6}}
\end{center}
\end{figure*}

Gaia-6 was discovered very recently by \citet{ibata21} through the application of \SF\ to the \Gaia\ EDR3 data and has no star in common with past spectroscopic surveys. The region spanned by Gaia-6 it not well covered by \Pris\ but we can nevertheless measure $\FeHPrave=-1.53\pm0.12$ from 10 stars with a significance higher than $8\sigma$ that are very clumped in metallicity space (Figure~\ref{fig:Gaia-6}).

\subsubsection{Gaia-9}
\begin{figure*}
\begin{center}
\includegraphics[width=9cm,angle=270]{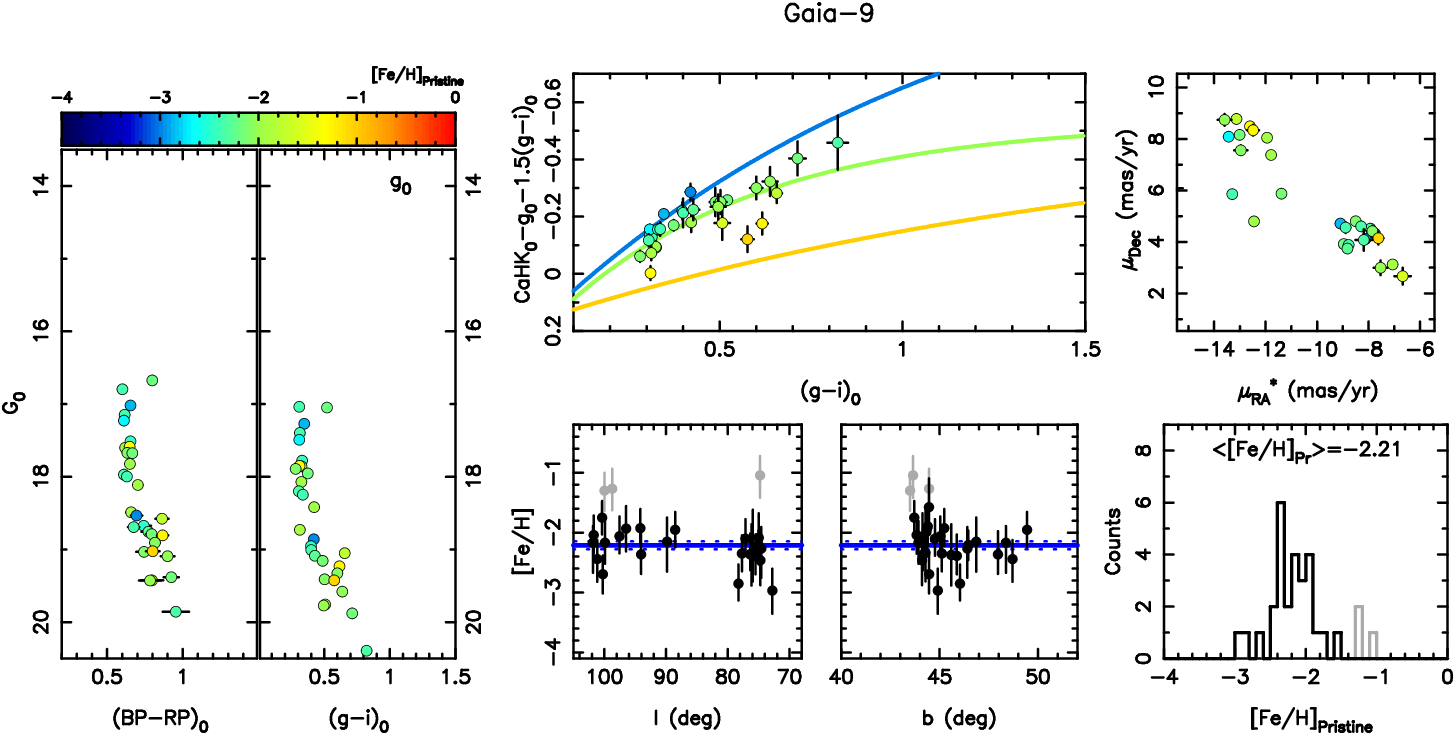}
\caption{Same as Figure~\ref{fig:M5stream} but for the Gaia-9 stream.\label{fig:Gaia-9}}
\end{center}
\end{figure*}

This stream, found by \citet{ibata21} through the application of \SF\ to the \Gaia\ EDR3 data, is located in a busy region that also includes GD-1 and the M~92 stream. We therefore choose to be cautious with our selection of Gaia-9 members in phase-space and avoid any candidate member that could potentially be part of any of the two other streams. The resulting \Pris\ view of Gaia-9 is shown in Figure~\ref{fig:Gaia-9}. Despite a few contaminants, there is a well-defined peak in the metallicity distribution that yields $\FeHPrave=-2.21\pm0.07$ from 30 stars. 

\subsubsection{Gaia-10/300S}
\begin{figure*}
\begin{center}
\includegraphics[width=9cm,angle=270]{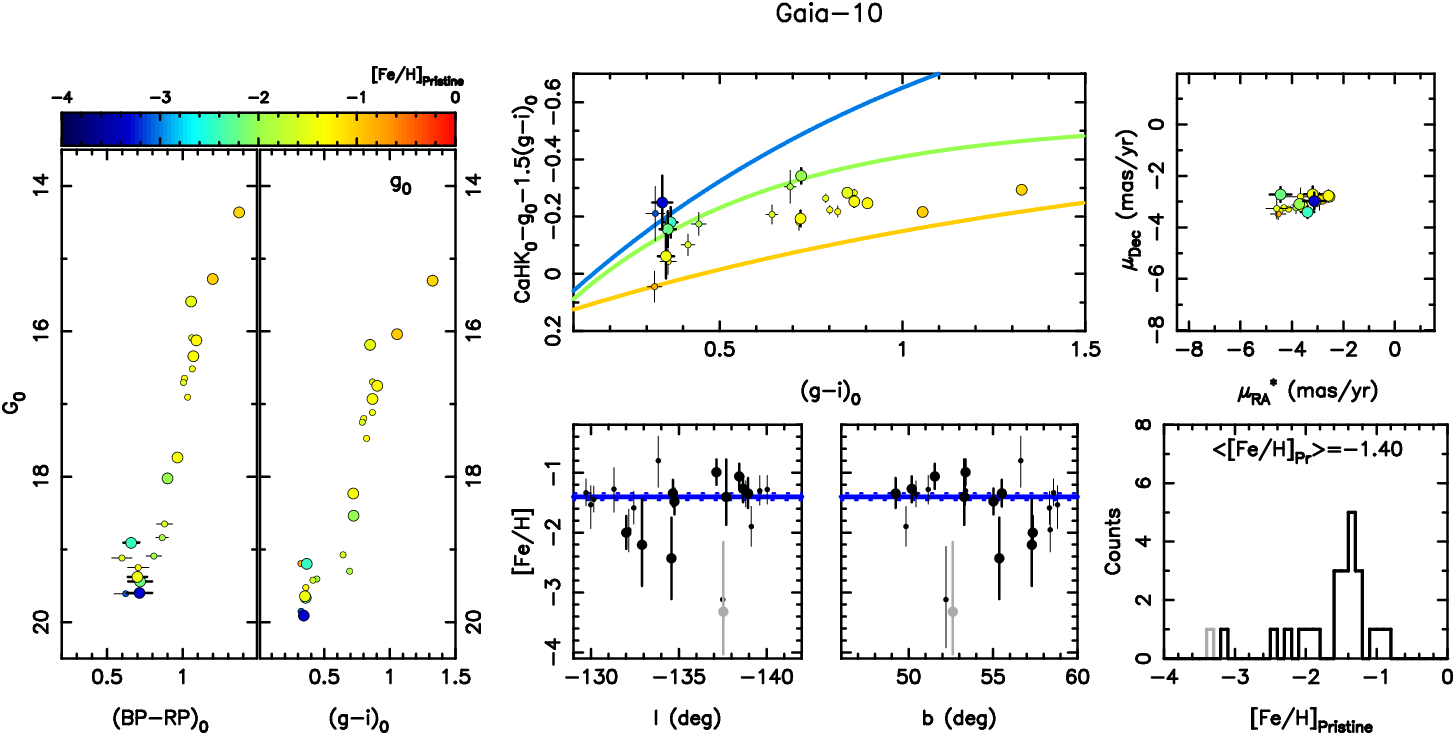}
\caption{Same as Figure~\ref{fig:M5stream} but for the Gaia-10/300S stream. Small symbols correspond to stars with a \SF\ significance between $8\sigma$ and $10\sigma$.\label{fig:Gaia-10}}
\end{center}
\end{figure*}

This stream, which appears in the \SF\ application to the \Gaia\ EDR3 data \citep{ibata21}, is in fact the 300S originally discovered by \citet{simon11} and thoroughly characterized by \citet{li22}. It is well defined in the \SF\ distant (10--30\kpc) map and \SF\ assigns a distance of $\sim15\kpc$ to the stream stars, as expected for the 300S stream. The stream falls entirely in the \Pris\ footprint but shows a complex picture in Figure~\ref{fig:Gaia-10} with both a clear grouping of stars around $\FeHPr\sim-1.3$ and a distribution of very metal-poor stars at the faint end of the CMD. However, the large uncertainties on the metallicity of the faint stars means that they could be spurious. They also do not significantly impact the mean metallicity of the stream that we calculate to be $\FeHPrave=-1.40\pm0.06$, which is in good agreement with the independent, spectroscopic measurement of \citet{li22}.

\subsubsection{GD-1}
\begin{figure*}
\begin{center}
\includegraphics[width=9cm,angle=270]{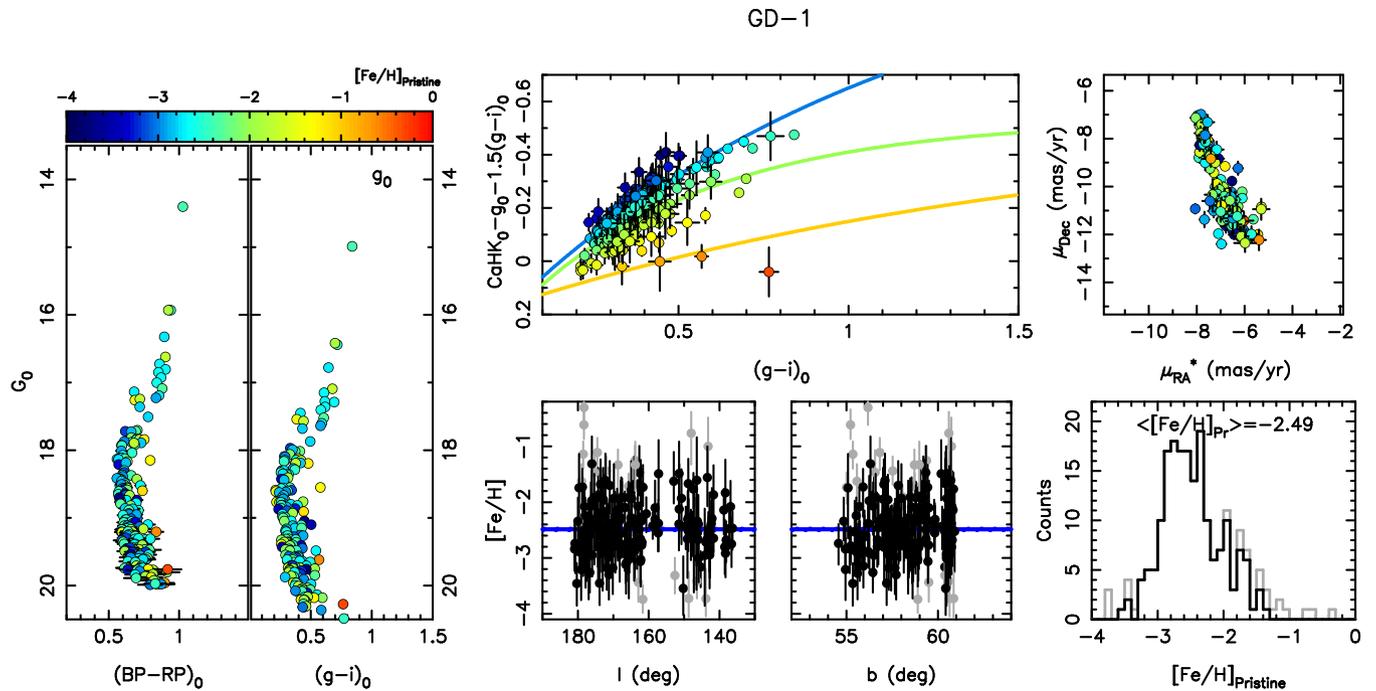}
\caption{Same as Figure~\ref{fig:M5stream} but for the GD-1 stream. In this case, only stars with $\textrm{ln}(\mathcal{L})>500$ are kept in the sample to avoid significant contamination. \label{fig:GD-1}}
\end{center}
\end{figure*}

The GD-1 stream was discovered in the SDSS photometric survey \citep{grillmair06b}. The stream has been used to constrain the shape and mass of the MW's dark matter halo \citep{koposov10,bovy16,malhan19c}. Also, given that it is both a cold stream and very long \citep{price-whelan18}, it has received a lot of attention as it is thought that perturbations in the shape and dynamics of the stream could be related to the recent and nearby passage of dark matter sub-halos \citep[e.g., ][]{carlberg13,bonaca19,banik21}. However, such attempts are made difficult by the complex morphology of the stream's perturbations that could also be related to the dynamics of the progenitor \citep{ibata20} or its orbiting within its own dark matter sub-halo \citep[][]{malhan19b}. Irrespective of the perturbed shape of the stream and possibly associated features, it is usually agreed from the search of public spectroscopic surveys or dedicated spectroscopic follow-up that the stream is very metal-poor, with a mean metallicity $\FeH\simeq-2.3$ \citep{malhan19c,bonaca20}. Recently, \citet{balbinot22} conducted a dedicated spectroscopic analysis of GD-1 member stars that yielded $\FeH=-2.06\pm0.10$.

In the \SF\ sample, the GD-1 stream is quite heavily contaminated. We track this contamination to the \SF\ hypothesis that stellar streams are mere perturbations over a denser local MW background. In the case of the very well populated GD-1 stream, this assumption is not valid anymore and many contaminating stars get their likelihood of being part of a stream boosted by the nearby presence of many GD-1 stars, even when the properties of this star are only marginally close to the properties of the stream stars. It is however easy to remove this contamination by considering only the most significant stars associated with GD-1. Limiting ourselves to stars with $\textrm{ln}(\mathcal{L})>500$ (instead of $\textrm{ln}(\mathcal{L})>52.3$) yields the sample shown in Figure~\ref{fig:GD-1}. Restricting the sample to even more significant stars makes it sparser but does not visibly change the properties of the distributions in the figure.

\Pris\ yields a mean metallicity $\FeHPrave=-2.49\pm0.03$ and indeed confirms that GD-1 is a very metal-poor stream from the sample of 190 very significant stream stars. The MDF is, however, surprisingly wide with clear scatter in the location of stars in the \Pris\ color-color space for even bright, high signal-to-noise member stars (the stars with very small uncertainties in the top-middle panel of Figure~\ref{fig:GD-1}). Given the difficulties to isolate a very pure sample for this structure, it is difficult to unambiguously confirm that this apparent metallicity spread is genuine but we note that GD-1 is known to be complex and also intersects the Kshir stream with which it shares common kinematic properties \citep{malhan19a}.

\subsubsection{Orphan}
\begin{figure*}
\begin{center}
\includegraphics[width=9cm,angle=270]{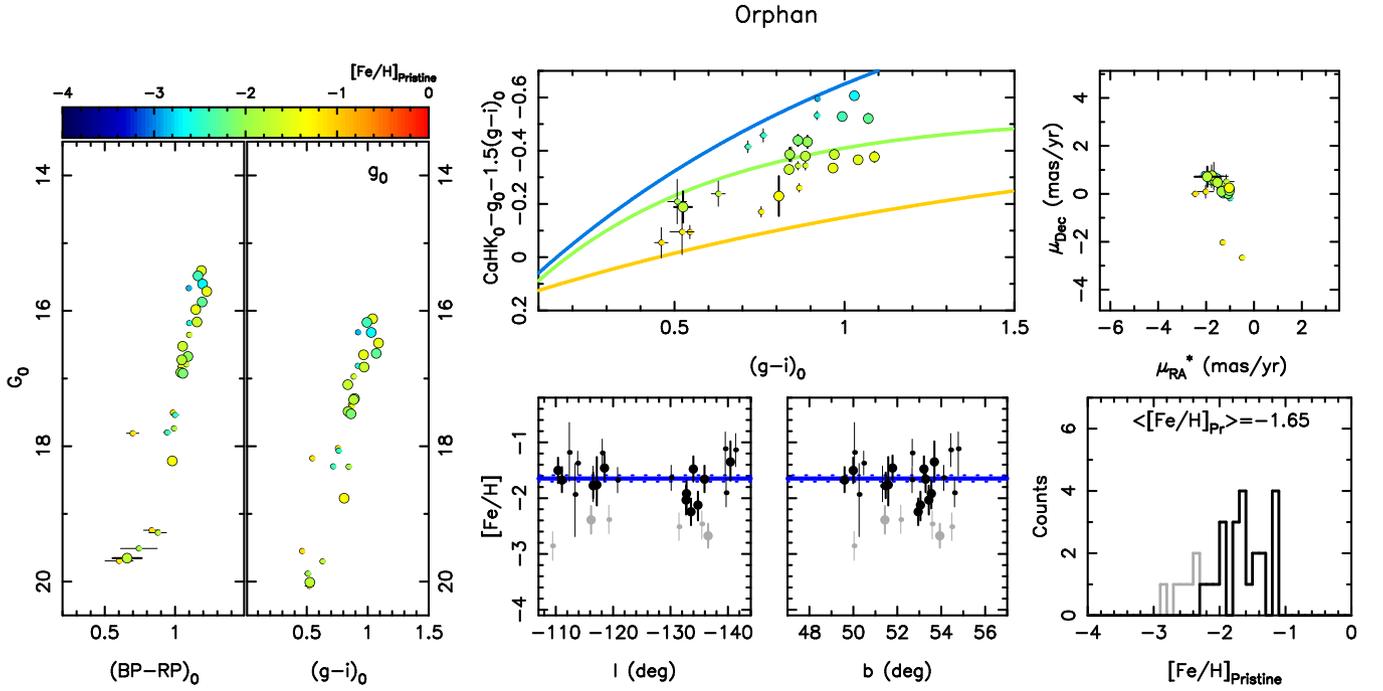}
\caption{Same as Figure~\ref{fig:M5stream} but for the Orphan stream. Small symbols correspond to stars with a \SF\ significance between $8\sigma$ and $10\sigma$.\label{fig:Orphan}}
\end{center}
\end{figure*}

The Orphan stream \citep{grillmair06,belokurov07} is a stream that almost wraps fully around the MW \citep{koposov19} and stems for the tidal disruption of a dwarf galaxy \citep{shelton21}. \SF\ picks up the well-populated red-giant branch of the system and the \Pris\ metallicities for those show a large scatter despite small uncertainties.  We determine $\FeHPrave=-1.65\pm0.06$ but the MDF has a long tail to low metallicity values, down to $\FeH\sim-3.0$ (Figure~\ref{fig:Orphan}). This value is in reasonable agreement with spectroscopic metallicities obtained for dozens of stars by the $S^5$ collaboration \citep[][$\FeH=-1.85\pm0.07$]{li22}, especially given the large dispersion measured in that study ($\sigma_\FeH=0.42^{+0.07}_{-0.06}$).

\subsubsection{Phlegeton}
\begin{figure*}
\begin{center}
\includegraphics[width=9cm,angle=270]{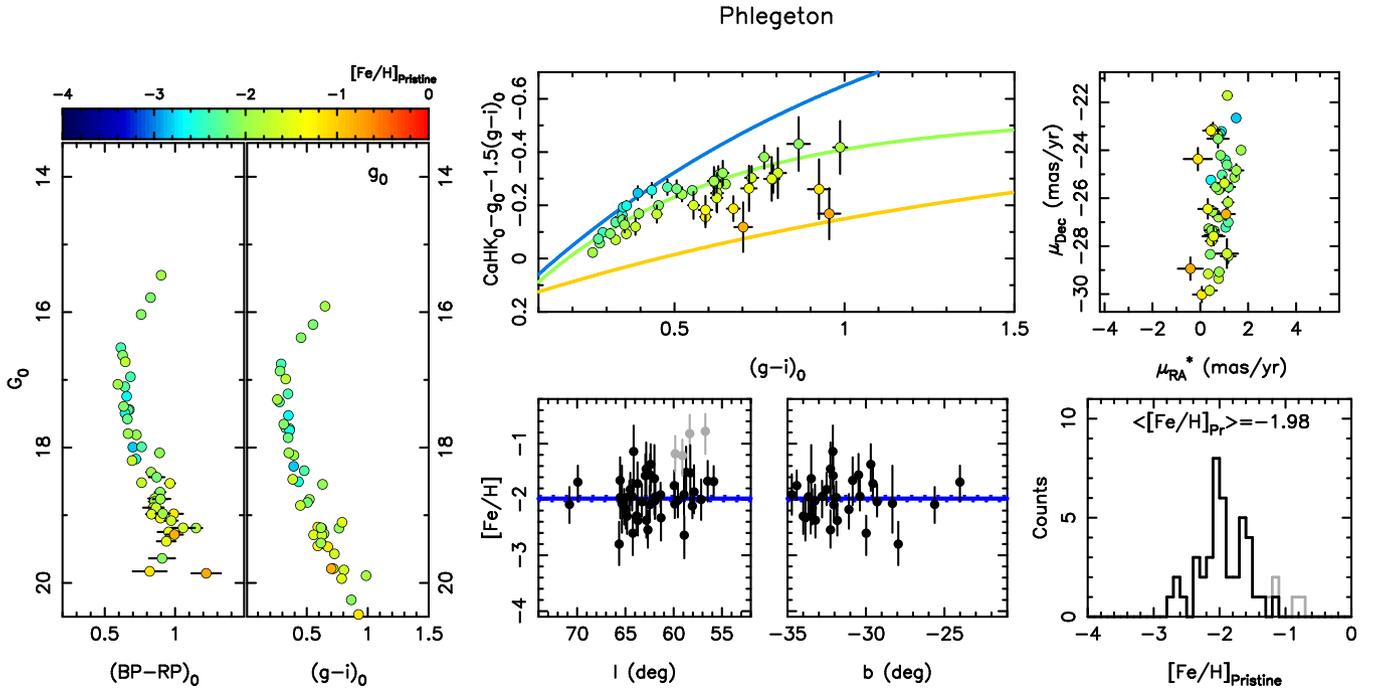}
\caption{Same as Figure~\ref{fig:M5stream} but for the Phlegeton stream.\label{fig:Phlegeton}}
\end{center}
\end{figure*}

Phlegeton was reported by \citet{ibata18} as one of the most significant stream detections with \SF\ in the \Gaia\ DR2 data. It is also one of the closest streams, at a heliocentric distance of $\sim3.8\kpc$. The \Pris\ survey just clips the Galactic eastern edge of the stream but it is so dense that we still find 45 stars in common between the two samples. Based on the \Pris\ metallicities, we find that this stream is very metal-poor, with $\FeHPrave=-1.98\pm0.05$. Its stars nicely track the $\FeH=-2$ iso-metallicity line in the top-middle panel of Figure~\ref{fig:Phlegeton}. \citet{ibata18} mention that one of the possible Phlegeton stars selected by \SF\ has an SDSS metallicity of $\FeH=-1.56\pm0.04$. While this is significantly more metal-rich that the mean \Pris\ metallicity, the metallicity histogram does extend to these values.

\subsubsection{Pisces-Triangulum}
\begin{figure*}
\begin{center}
\includegraphics[width=9cm,angle=270]{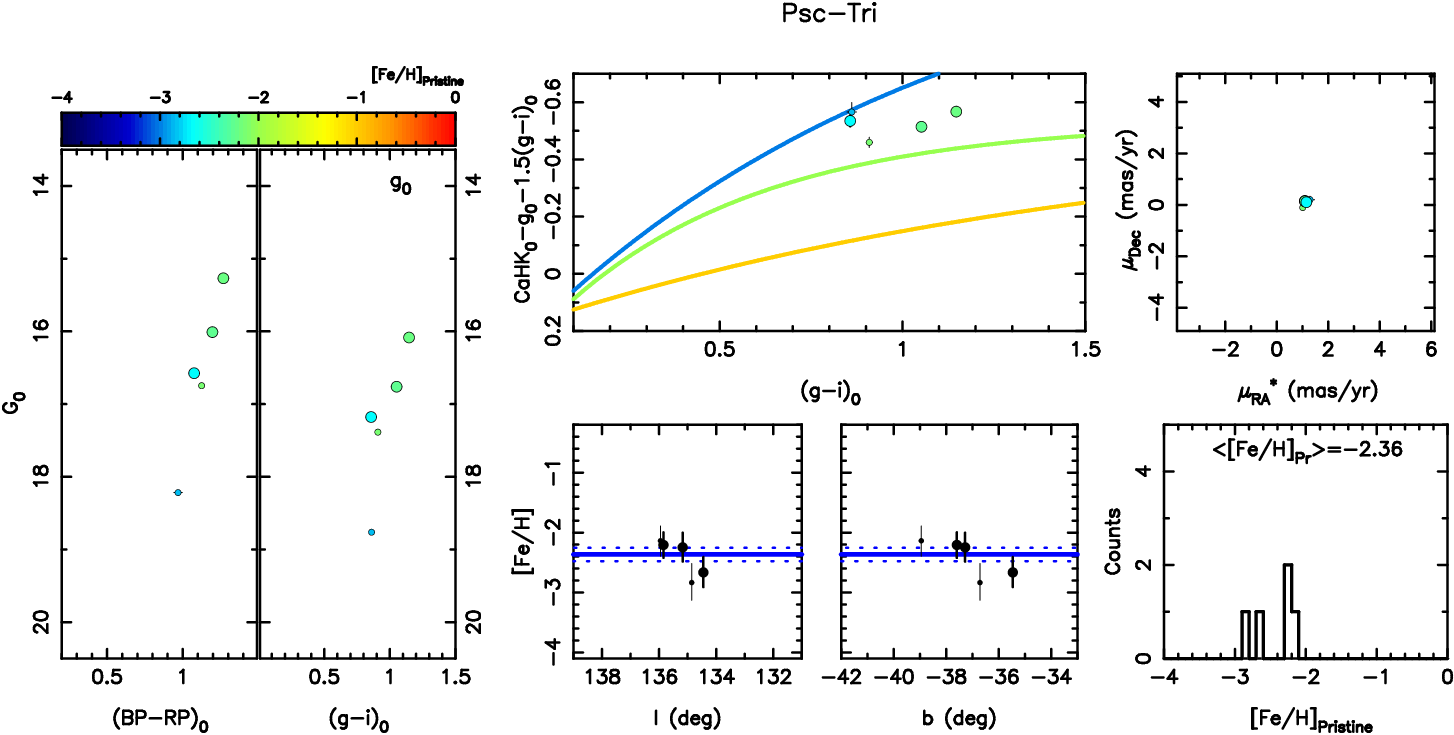}
\caption{Same as Figure~\ref{fig:M5stream} but for the the Psc-Tri stream. Small symbols correspond to stars with a \SF\ significance between $8\sigma$ and $10\sigma$\label{fig:Psc-Tri}}
\end{center}
\end{figure*}

We identify a small group of 5 stars with a \SF\ significance above $8\sigma$ that gave a coherent, metal-poor signal in Pristine with $\FeHPrave=-2.36\pm0.11$ (Figure~\ref{fig:Psc-Tri}). After verification with known structures, it became apparent that this structure corresponds to the Pisces-Triangulum (Psc-Tri) that, at a distance of $\sim30\kpc$ \citep{bonaca12,martinc13,yuan22a}, is located at the very edge of the \SF\ exploration. \citet{martinc13} determine an average metallicity for this stream from 11 stars and find $\langle\FeH\rangle\simeq-2.2$, which is in very good agreement with our measurement using Pristine and further confirms that the few stars we identify in the \SF\ sample belong to Pisces-Triangulum.

\subsubsection{Slidr}
\begin{figure*}
\begin{center}
\includegraphics[width=9cm,angle=270]{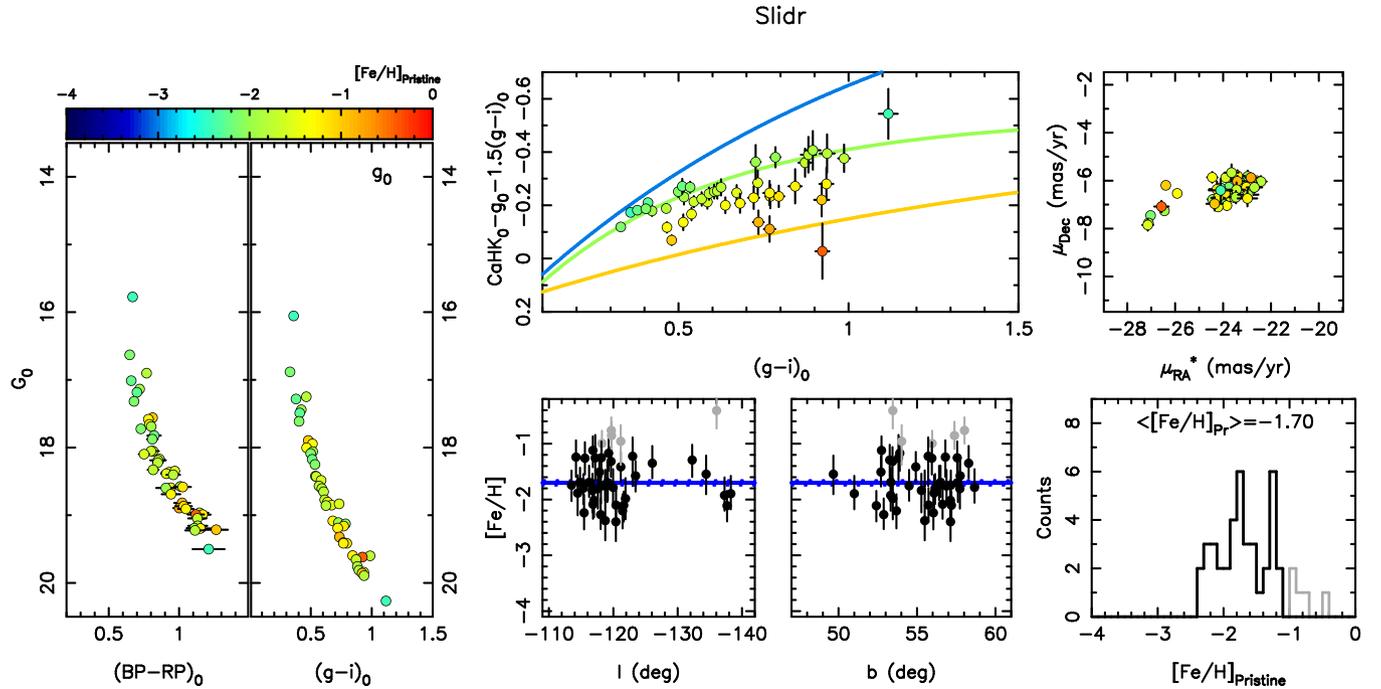}
\caption{Same as Figure~\ref{fig:M5stream} but for the Slidr stream.\label{fig:Slidr}}
\end{center}
\end{figure*}

Slidr is presented in \citet{ibata19}, based on the application of \SF\ to the \Gaia\ DR2 data. A comparison of candidate member stars with the LAMOST and SDSS spectroscopic surveys revealed some contamination in the \SF\ list but, also, that the stream displays a mean spectroscopic metallicity $\langle\FeH\rangle=-1.8$ based on 14 stars. While \Pris\ only includes the Galactic northern edge of the stream, we can confirm this metallicity measurement in Figure~\ref{fig:Slidr}, with $\langle\FeH\rangle=-1.70\pm0.05$ from 42 stars in \Pris. This sample is also very clean for the region that was quite contaminated in the discovery paper (their Figure~9), hinting at the improvement between the \Gaia\ DR2 and EDR3 data sets. We find a low-density extension to the stream westwards of the DR2 detection (with $\mu_\mathrm{RA}^*<-25\masyr$ in the top-right panel of Figure~\ref{fig:Slidr}). These few stars have similar metallicities to those of Slidr stars and, with such negative proper motions, are more easily connected to Slidr than any other MW structure.

\subsubsection{Sv\"ol}
\begin{figure*}
\begin{center}
\includegraphics[width=9cm,angle=270]{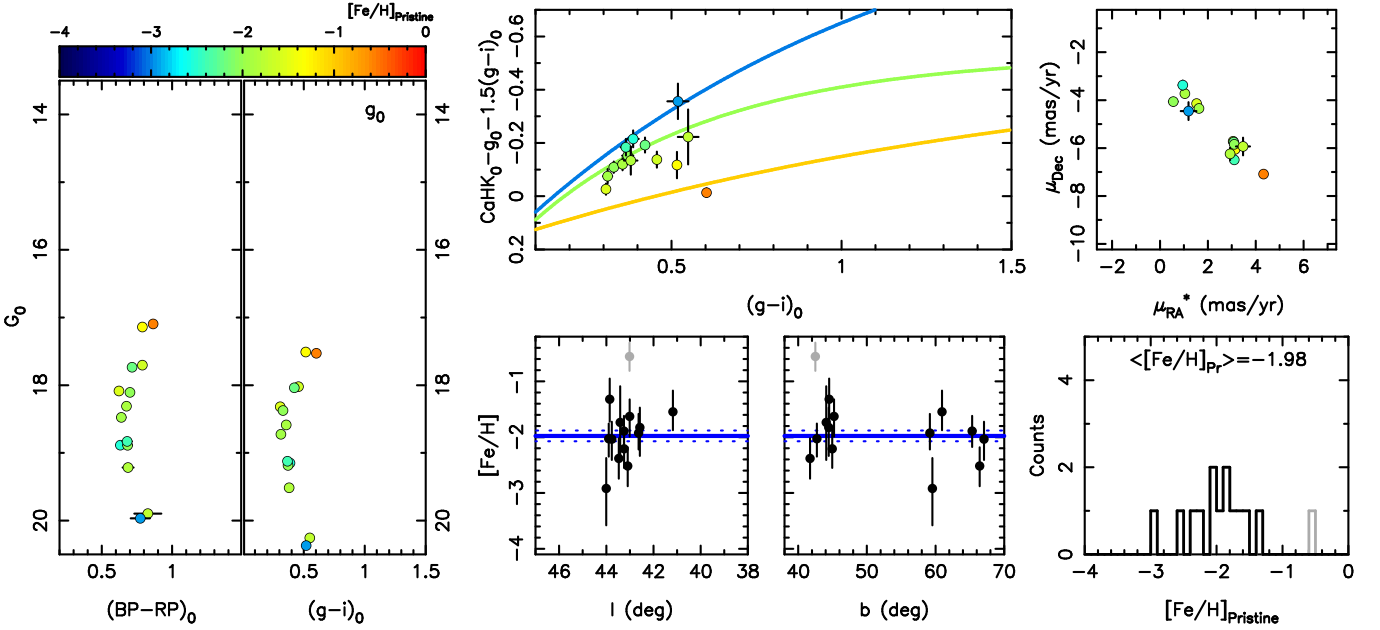}
\caption{Same as Figure~\ref{fig:M5stream} but for the Sv\"ol stream.\label{fig:Svol}}
\end{center}
\end{figure*}

Sv\"ol was also discovered by \citet{ibata19}, based on the \Gaia\ DR2 data. The only star in common between the \SF\ sample and spectroscopic surveys is a LAMOST star with discrepant velocity \citep{ibata19} and likely a contaminant. Despite a very non-contiguous coverage of \Pris\ in this region, there are 13 high-significance Sv\"ol stars in common with \Pris. We unambiguously show that this stream is metal-poor, with $\FeHPrave=-1.98\pm0.10$ (Figure~\ref{fig:Svol}).  

\subsubsection{Sylgr}
\begin{figure*}
\begin{center}
\includegraphics[width=9cm,angle=270]{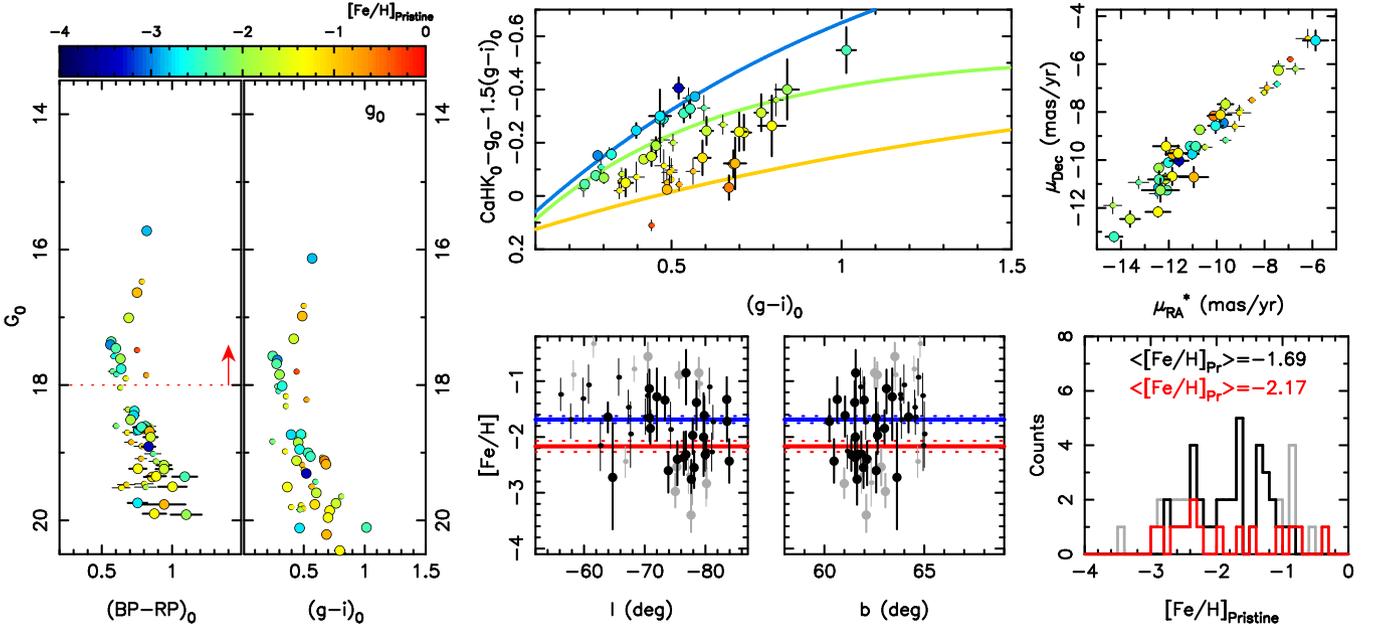}
\caption{Same as Figure~\ref{fig:M5stream} but for the Sylgr stream. Small symbols correspond to stars with a \SF\ significance between $8\sigma$ and $10\sigma$. The red lines and labels in the bottom panels correspond to the $\FeHPrave$ and the MDF obtained when restricting the analysis to only stars with $G_0<18.0$ (red dotted line and arrow in the CMD presented in the left-most panel).}
\label{fig:Sylgr}
\end{center}
\end{figure*}

Sylgr was first discovered by \citet{ibata19} through the application of \SF\ to the \Gaia\ DR2 data. Three stars in common with the SDSS yield an average metallicity of $\langle\FeH\rangle=-2.7$, confirmed with high-resolution spectroscopy for two of them ($\FeH=-2.92\pm0.06$; \citealt{roederer19}). The stream is located at the edge of the \Pris\ footprint but there are still 29 stars in common with the high-significance \SF\ catalogue. We find a mean metallicity that is significantly higher than the literature values, with $\langle\FeH\rangle=-1.69\pm0.06$. The \Pris\ MDF of this structure is fairly wide and, with a closer look at the metallicity values in the bottom-middle panels of Figure~\ref{fig:Sylgr}, it appears that there is a grouping of stars around $\FeHPr\sim-2.5$ while the rest of the sample scatters over the full metallicity range. The CMD of Sylgr shows that these latter stars are preferentially at fainter magnitudes and, contrary to most other streams (\eg Slidr, Figure~\ref{fig:Slidr}), the deeper SDSS CMD does not lead to a tighter main sequence. This is likely the sign that the \SF\ sample of Sylgr members has significant contamination. The presence of some contamination for this stream is also clear from discrepant radial velocities obtained by \citet[][see the top panel of their Figure~3]{ibata21}. Restricting the analysis to only bright stars ($G_0<18.0$) yields a more metal-poor $\FeHPrave=-2.17\pm0.10$ that is in better agreement with but remains more metal-rich than the spectroscopic studies for this stream. This is the metallicity we favor here. In the future, it will be interesting to see whether larger spectroscopic samples confirm the \Pris\ measurement and whether the two spectroscopic metallicities were abnormally low compared to the bulk of the stream's metallicity.

\subsection{Confirmed candidate streams}
The following 4 streams were listed as candidate streams by \citet{ibata21}. We show here that they all have stars with consistent metallicities, even though the MDFs are sometimes contaminated. We therefore confirm that they are all real streams in the MW halo.

\subsubsection{C-9}
\begin{figure*}
\begin{center}
\includegraphics[width=9cm,angle=270]{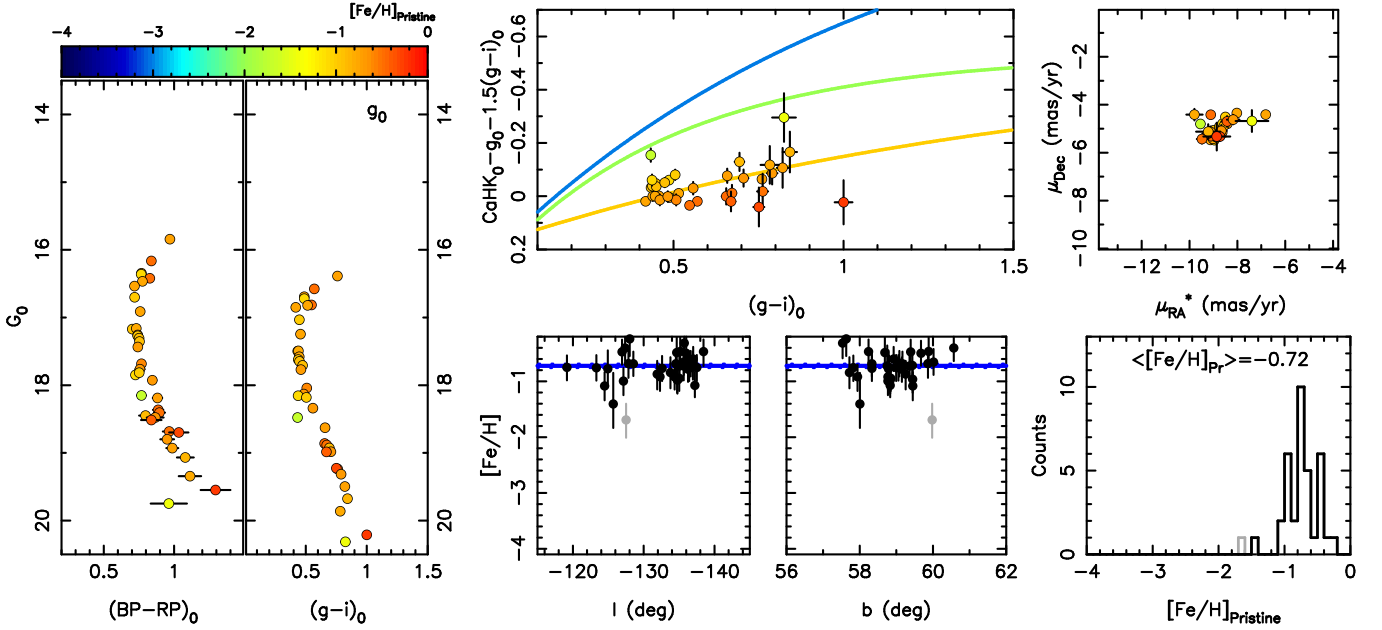}
\caption{Same as Figure~\ref{fig:M5stream} but for the C-9 stream.\label{fig:C-9}}
\end{center}
\end{figure*}

This is a well-defined candidate stellar stream located just above Slidr in Figure~\ref{fig:map}. There are 36 stars from the \SF\ list of high-significance members that have a photometric metallicity in \Pris\ and these show that C-9 is one of the most metal-rich stream in the sample, with $\FeHPrave=-0.72\pm0.04$ (Figure~\ref{fig:C-9}). The very clear peak in the MDF unambiguously confirm that C-9 is a real stream.

\subsubsection{C-11}
\begin{figure*}
\begin{center}
\includegraphics[width=9cm,angle=270]{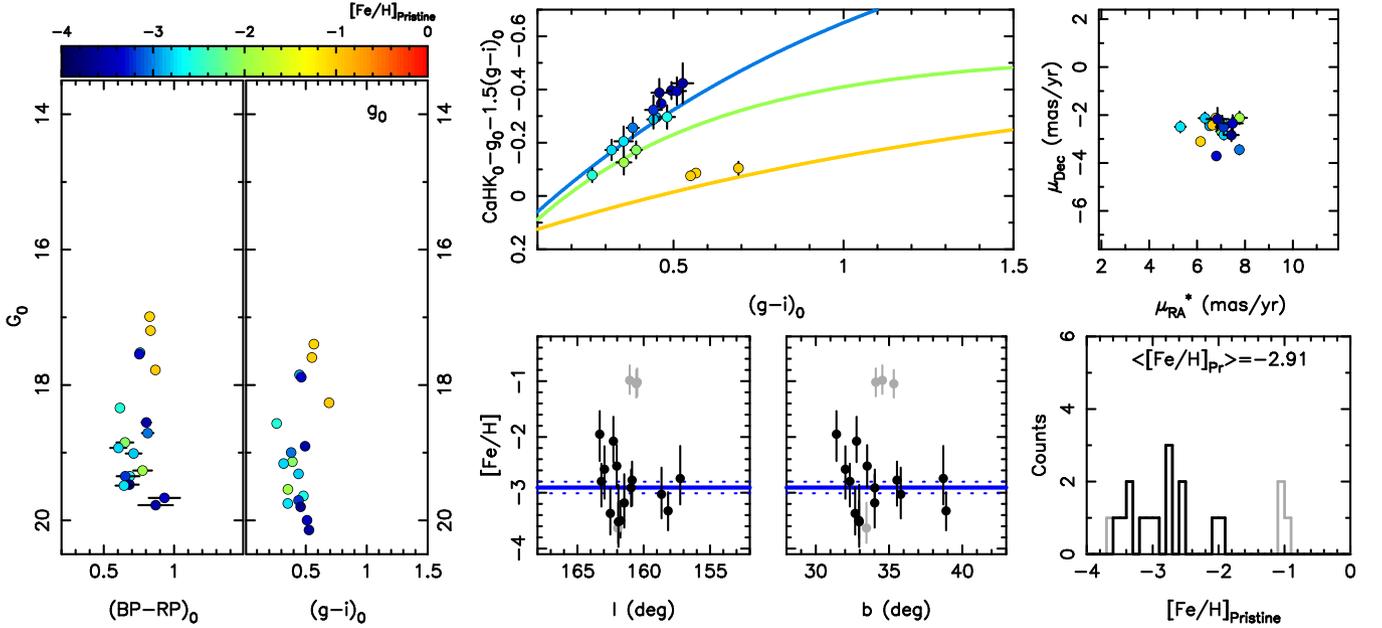}
\caption{Same as Figure~\ref{fig:M5stream} but for the C-11 stream.\label{fig:C-11}}
\end{center}
\end{figure*}

C-11 appears in the \SF\ map at $\sim$8--10\kpc, using \Gaia\ EDR3 data. It is a short stream to the Galactic north of the anticenter and it is almost entirely in \Pris. The picture provided by \Pris\ is that of a very metal-poor stream, for which our $2.5\sigma$-clipping procedure, applied to the list of \SF\ stars with a significance above $10\sigma$, yields $\FeHPrave=-2.91\pm0.10$ from 15 stars (Figure~\ref{fig:C-11}), which is significantly more metal-poor than any known dwarf galaxy or GC.

\subsubsection{C-19}
\begin{figure*}
\begin{center}
\includegraphics[width=9cm,angle=270]{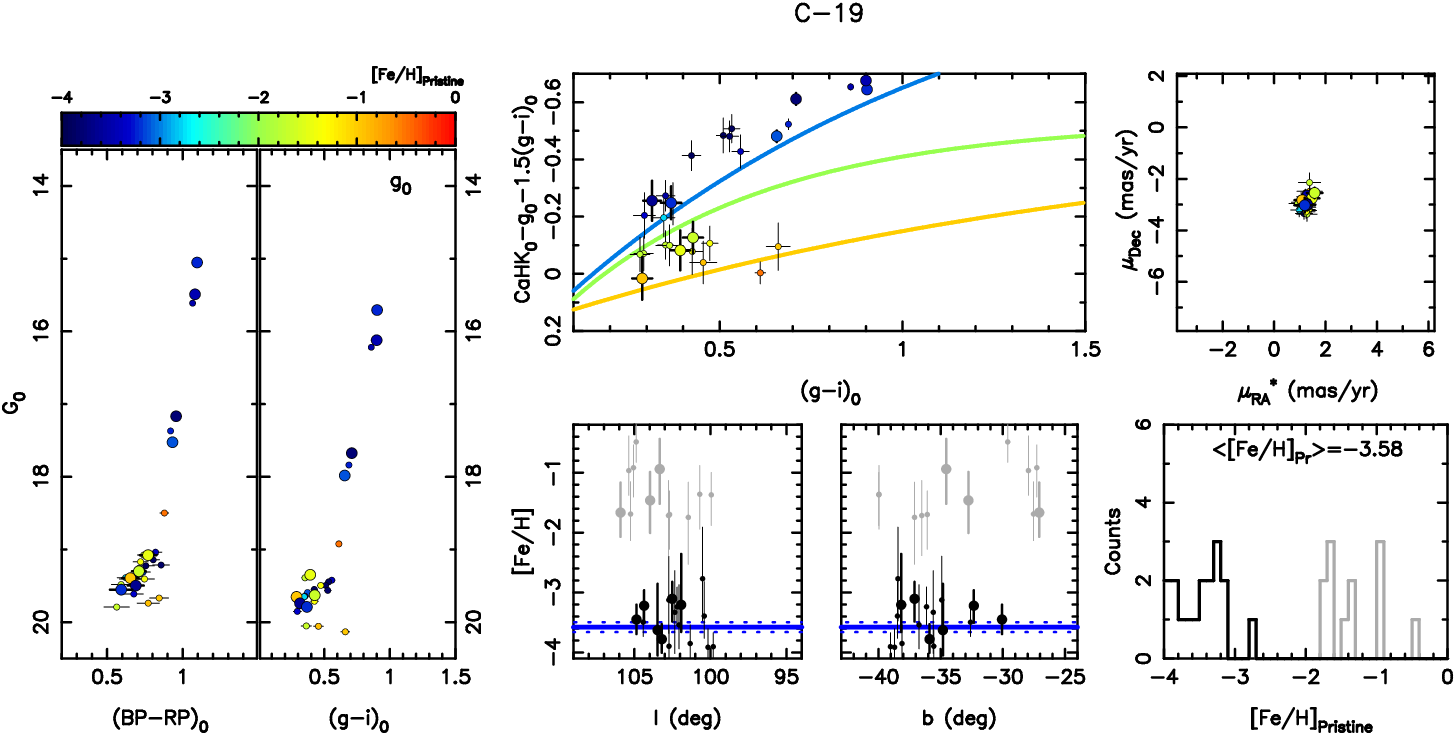}
\caption{Same as Figure~\ref{fig:M5stream} but for the C-19 stream. Small symbols correspond to stars with a \SF\ significance between $8\sigma$ and $10\sigma$.\label{fig:C-19}}
\end{center}
\end{figure*}

C-19 appears in the distant map of \SF\ detections based on the analysis of EDR3 data \citep{ibata21}, in the Southern Galactic cap, and is located almost entirely within the \Pris\ footprint. While the metallicity distribution can, at first glance, look messy, C-19 hosts a very clear distribution of EMP stars according to \Pris. This is made even more obvious from when looking at the \Pris\ color-color space all \SF\ detections above $8\sigma$ (Figure~\ref{fig:C-19}). The presence of more metal-rich stars comes almost entirely from faint stars and we can conclude without ambiguity that, given the vanishingly small probability of randomly finding 17~EMP stars among 29 \SF-selected stars (or 6 out of 7 at the bright end), C-19 unambiguously has the lowest metallicity in the whole sample, with $\FeHPrave=-3.58\pm0.08$.

Follow-up observations of 8~member stars in this structure reveal that it is the most metal-poor structure ever discovered \citep{martin22a}, with $\FeH=-3.38\pm0.06\textrm{ (stat.)} \pm0.20\textrm{ (syst.)}$. These observations further constrain the metallicity dispersion of the system to be consistent with zero and smaller than 0.18\,dex at the 95\% confidence level. Combined with measured difference in the Na abundances, this leads \citet{martin22a} to conclude that C-19 is the remnant of a GC significantly below the observed metallicity floor of GCs, despite being dynamically hot and wider than other known GC streams.

\subsubsection{C-20}
\begin{figure*}
\begin{center}
\includegraphics[width=9cm,angle=270]{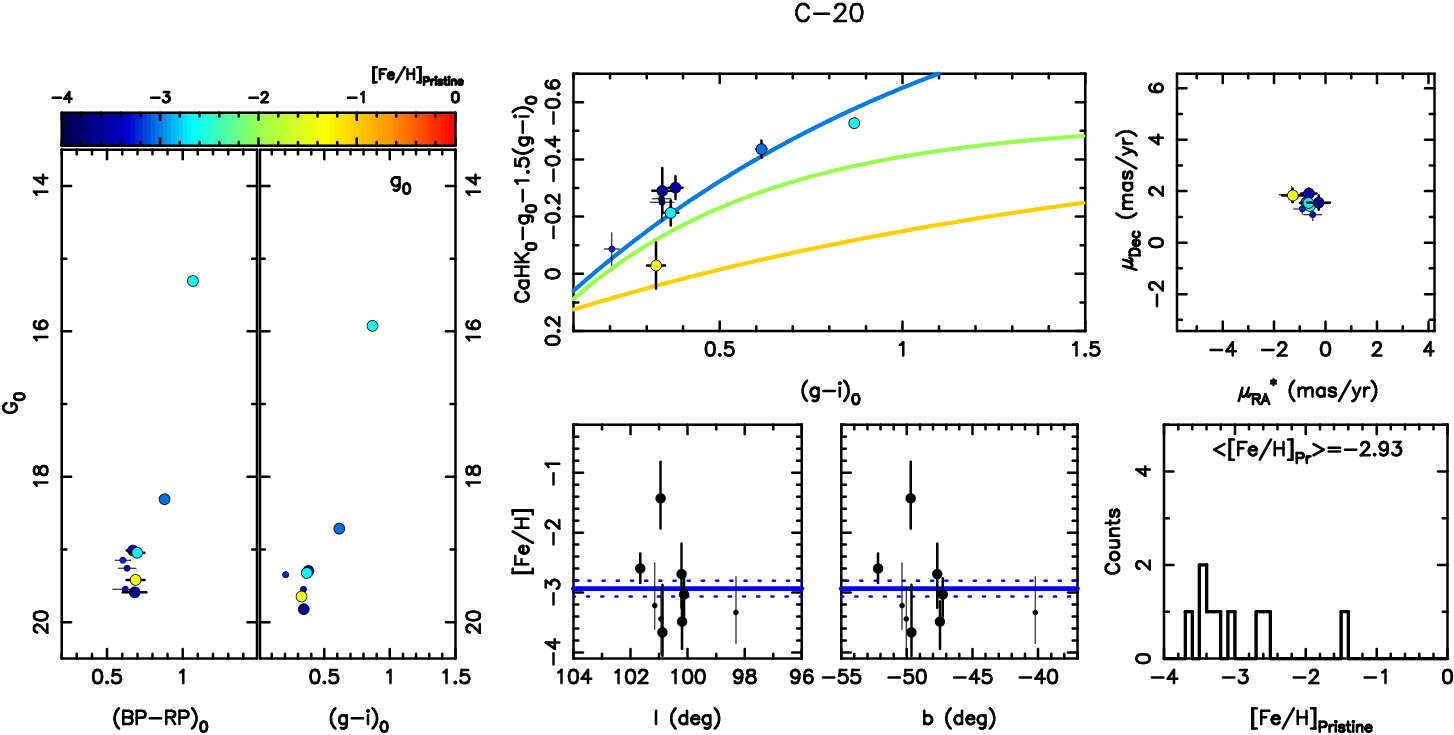}
\caption{Same as Figure~\ref{fig:M5stream} but for the C-20 stream. Small symbols correspond to stars with a \SF\ significance between $8\sigma$ and $10\sigma$.\label{fig:C-20}}
\end{center}
\end{figure*}

C-20 is a distant candidate stream this is visible on the distant map of streams from the \SF\ application to \Gaia\ EDR3 \citep{ibata21}. In Figure~\ref{fig:C-20}, we show that, even though C-20 is sparsely populated, it has a clear metallicity signal in \Pris, which confirms that it is a real stream. It is also the third extremely metal-poor structure in the sample of streams that we are able to track in \Pris, with $\FeHPrave=-2.93\pm0.14$. Added to C-11, C-19, and the Phoenix stream \citep[$\FeH=-2.70\pm0.06$; ][]{wan20}, it is growing increasingly clear that the MW is surrounded by a population of streams that are significantly more metal-poor than its current satellites.

\subsection{New candidates}

The 6~structures mentioned in this sub-section all correspond to a well-defined structure in the phase-space distribution of \SF\ stars detected at more than $8\sigma$ from the processing of EDR3 data presented by \citet{ibata21}, with at least 2~ stars (usually more) detected at $>10\sigma$. The $\Pris$ metallicities confirm that they all display a coherent, often very metal-poor MDF, which we take as a confirmation that they are very likely to be real structures of the MW halo.

\subsubsection{C-21}
\begin{figure*}
\begin{center}
\includegraphics[width=9cm,angle=270]{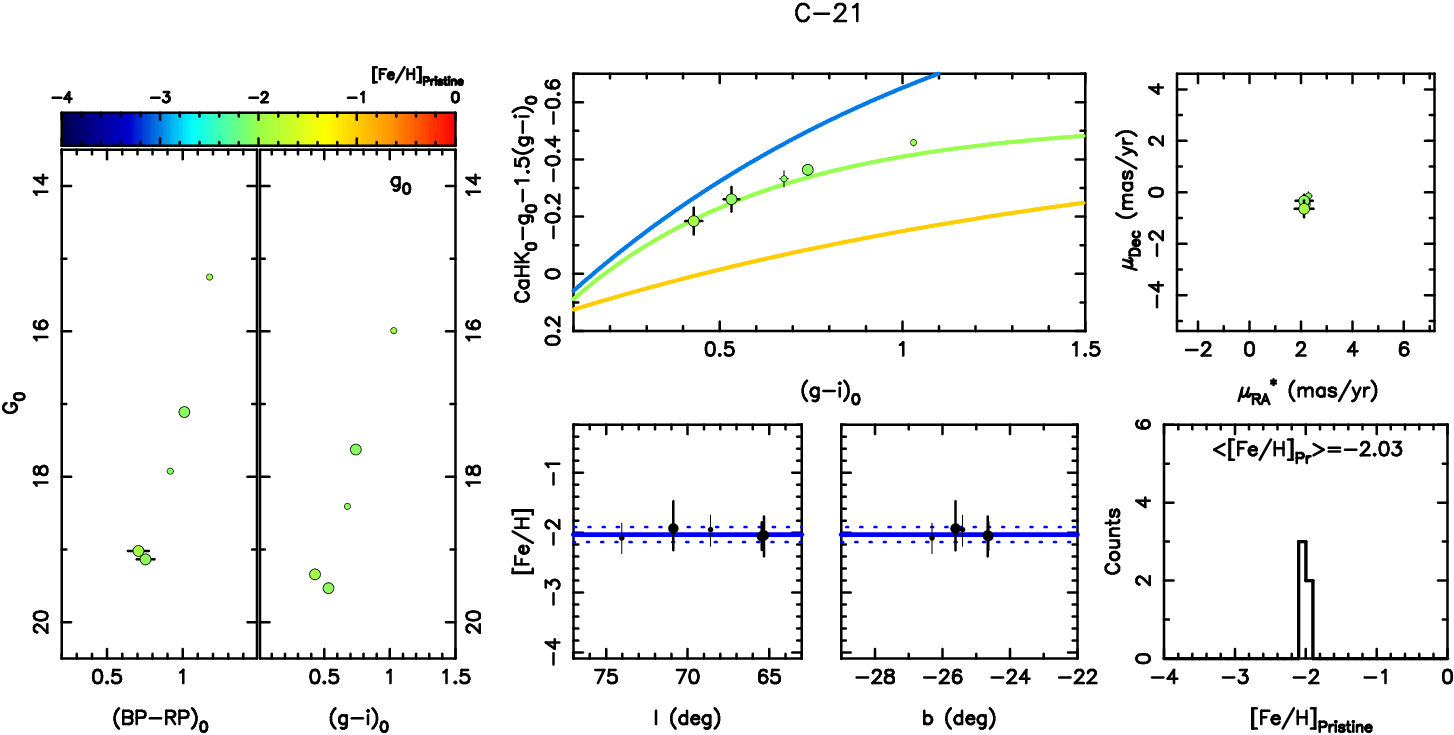}
\caption{Same as Figure~\ref{fig:M5stream} but for the new candidate stream C-21. Small symbols correspond to stars with a \SF\ significance between $8\sigma$ and $10\sigma$.\label{fig:C-21}}
\end{center}
\end{figure*}

C-21 is a grouping of stars that intersect the Phlegeton stream on the sky, but with a very different proper motion and metallicity. C-21 does not have many stars within the \Pris\ footprint but, from those, we find that C-21 has a very narrow MDF with $\FeHPrave=-2.03\pm0.13$ (Figure~\ref{fig:C-21}).

\subsubsection{C-22}
\begin{figure*}
\begin{center}
\includegraphics[width=9cm,angle=270]{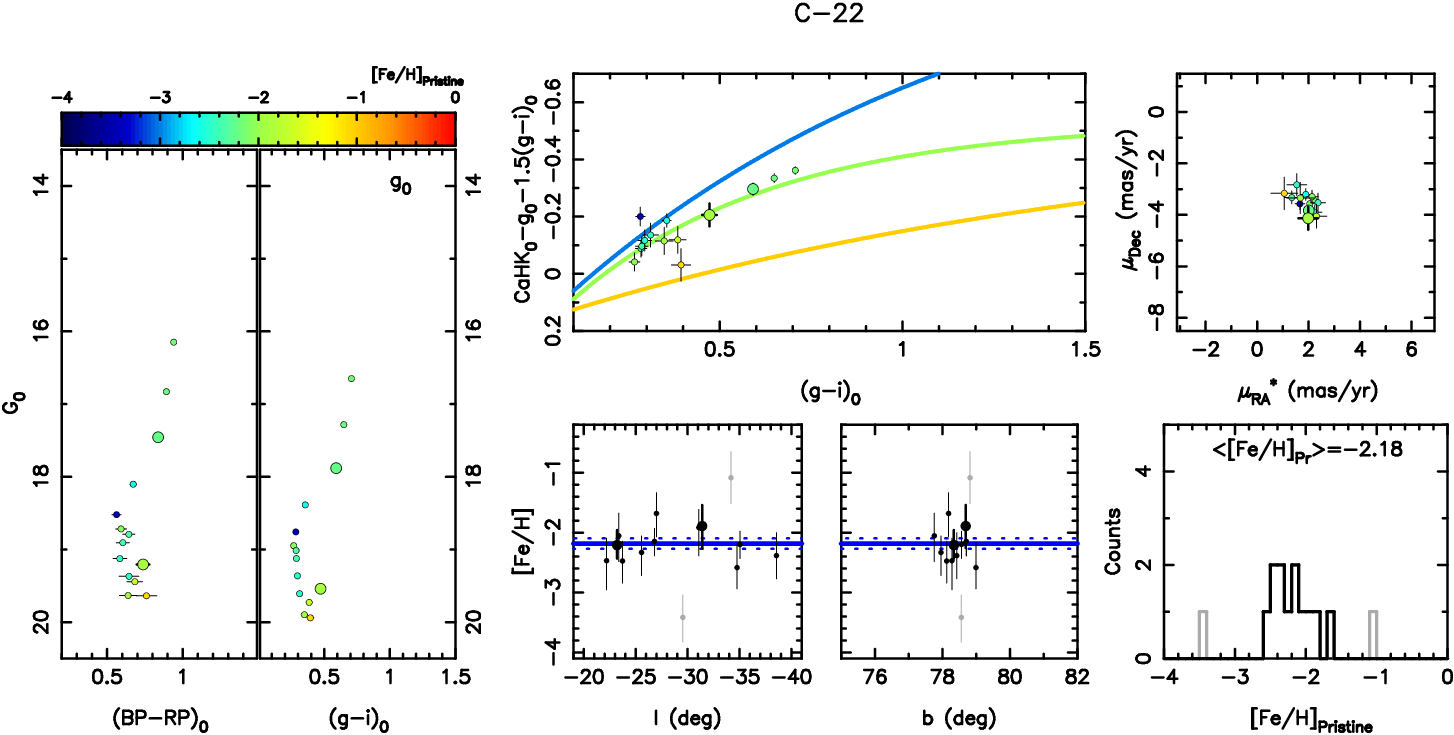}
\caption{Same as Figure~\ref{fig:M5stream} but for the new candidate stream C-22. Small symbols correspond to stars with a \SF\ significance between $8\sigma$ and $10\sigma$.\label{fig:C-22}}
\end{center}
\end{figure*}

C-22 is barely visible in the sample of high-significance \SF\ stars ($>10\sigma$) but becomes well populated once we lower the significance threshold to $8\sigma$. This is possibly because C-22 is located in region with many criss-crossing stream (Figure~\ref{fig:map}; Fj\"orm, Gaia-1, the M5 stream). In total, there are 12 stars in common between the \SF\ list and \Pris\ and they yield an average metallicity of $\FeHPrave=-2.18\pm0.09$ (Figure~\ref{fig:C-22}).

\subsubsection{C-23}
\begin{figure*}
\begin{center}
\includegraphics[width=9cm,angle=270]{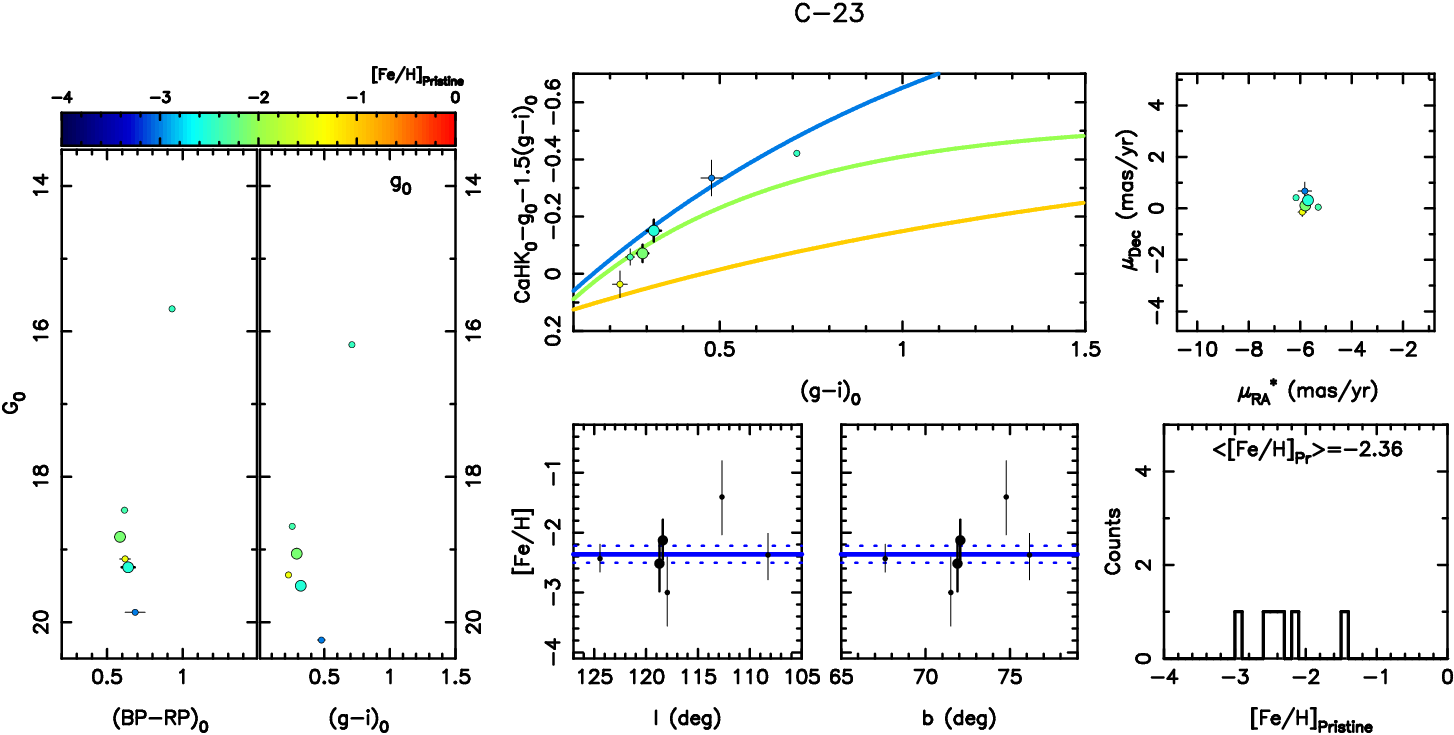}
\caption{Same as Figure~\ref{fig:M5stream} but for the new candidate stream C-23. Small symbols correspond to stars with a \SF\ significance between $8\sigma$ and $10\sigma$.\label{fig:C-23}}
\end{center}
\end{figure*}

This detection of a new stream shows the very limit of what can be achieved when combining the Gaia EDR3 output of \SF\ with a vast external data set like \Pris. Our initial detection of C-23 corresponds to only 2~stars detected as stream-like at $10\sigma$ by \SF. From \Pris, we learn that these two stars that are only $12'$ from each other and share similar proper motions also have statistically identical metallicities (Figure~\ref{fig:C-23}). Extending the search to the sample of $8\sigma$ \SF\ detections confirms that these two stars are but the tip of the iceberg and that this structure appears to be a very metal-poor stream that extends over $\sim10\deg$ ($\FeHPrave=-2.36\pm0.14$).

\subsubsection{C-24}
\begin{figure*}
\begin{center}
\includegraphics[width=9cm,angle=270]{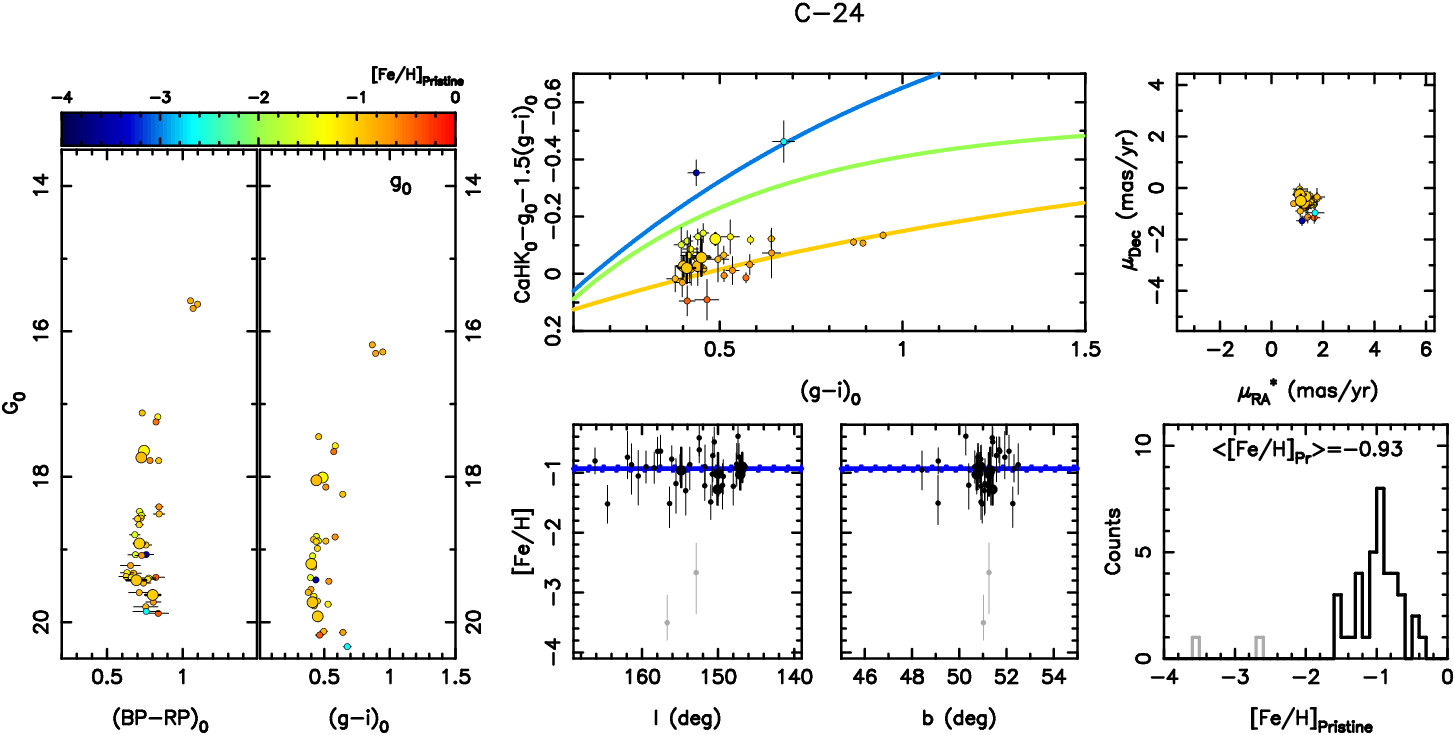}
\caption{Same as Figure~\ref{fig:M5stream} but for the new candidate stream C-24. Small symbols correspond to stars with a \SF\ significance between $8\sigma$ and $10\sigma$.\label{fig:C-24}}
\end{center}
\end{figure*}

Among all the candidate streams, C-24 is an exception in that it is the only one that is not very metal-poor. C-24 is a well populated structure located between the GD-1 and C-11 streams (Figure~\ref{fig:map}). \Pris\ unambiguously shows that the stream is a thin stream of the halo with $\langle\FeHPrave\rangle=-0.93\pm0.05$ (Figure~\ref{fig:C-24}).

\subsubsection{C-25}
\begin{figure*}
\begin{center}
\includegraphics[width=9cm,angle=270]{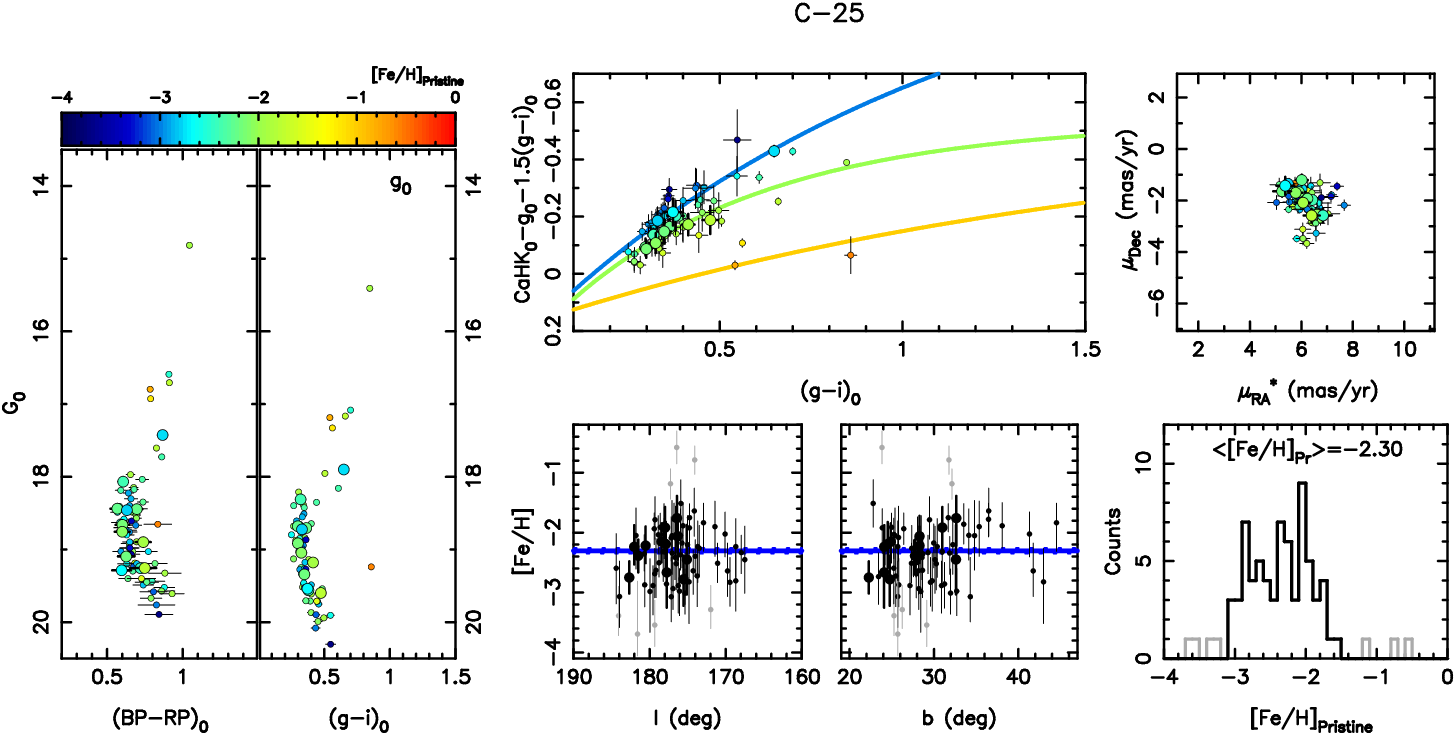}
\caption{Same as Figure~\ref{fig:M5stream} but for the new candidate stream C-25. Small symbols correspond to stars with a \SF\ significance between $8\sigma$ and $10\sigma$.\label{fig:C-25}}
\end{center}
\end{figure*}

Candidate C-25 appears as a diffuse structure at low latitude, in the anticenter direction. Figure~\ref{fig:C-25} shows that the large number of stars picked up by \SF\ as significantly part of a stream consistently show a metallicity close or lower than $\FeHPr=-2.0$, are clumped in proper motion and metallicity, confirming that this is likely a genuine structure with $\FeHPrave=-2.30\pm0.04$. This low metallicity makes it unlikely that it is associated with the ACS or other structures at the edge of the MW in the direction of the anticenter. The region of C-25 is complex in the \SF\ catalogue and there are hints that it could in fact correspond to two structures that overlap on the sky, but at different distances ($\sim6\kpc$ and $\sim11\kpc$ according to the \SF\ orbital solutions). Follow-up observations will be required to confirm whether the group of stars we identify as C-25 correspond to a single system.

\subsubsection{C-26}
\begin{figure*}
\begin{center}
\includegraphics[width=9cm,angle=270]{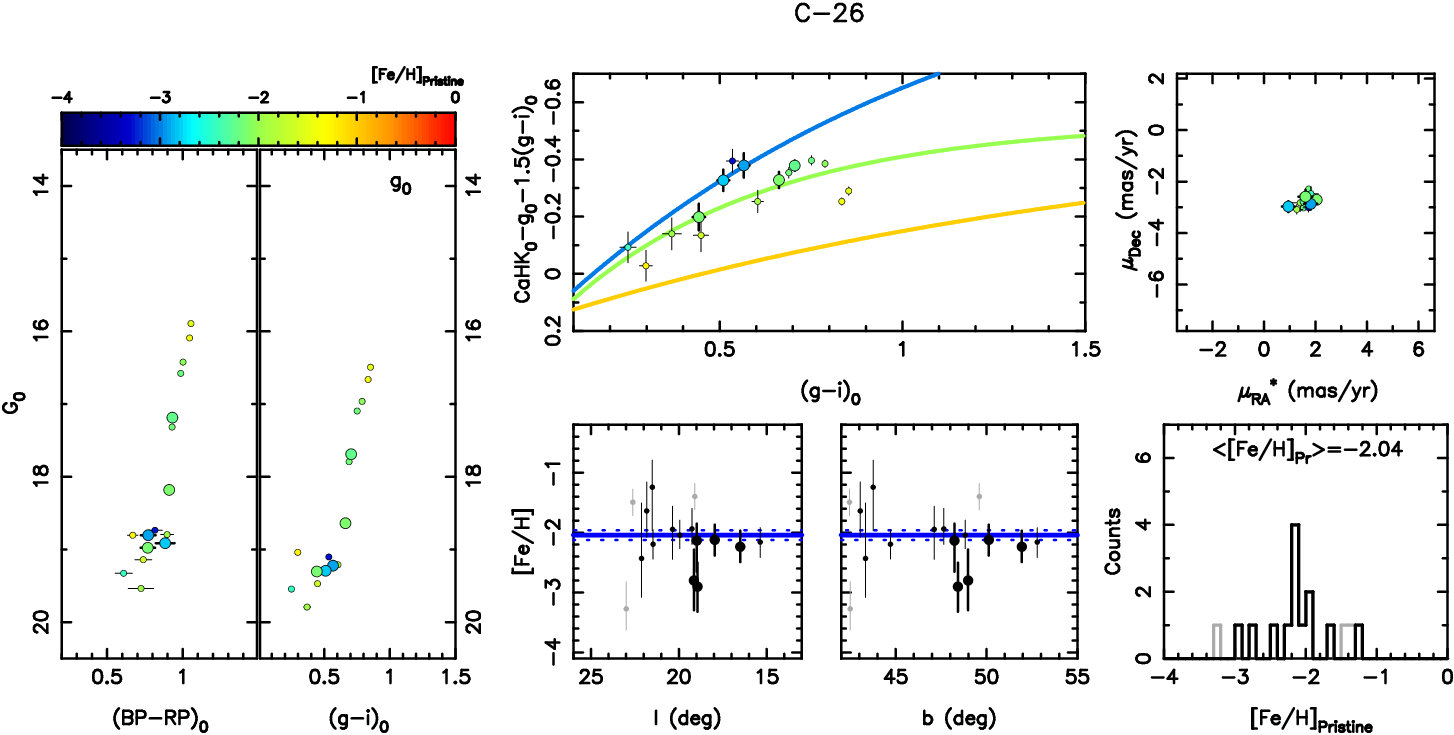}
\caption{Same as Figure~\ref{fig:M5stream} but for the new candidate stream C-26. Small symbols correspond to stars with a \SF\ significance between $8\sigma$ and $10\sigma$.\label{fig:C-26}}
\end{center}
\end{figure*}

This structure is a small but clearly defined streak of stars just north-west of the Pal~5 stream in the distant (10--30\kpc) maps in Figure~7 of \citet{ibata21}, in green/yellow-green in the $\mu_l$/$\mu_b$ panels. This structure is fully within \Pris\ and it has a mean metallicity $\FeHPrave=-2.04\pm0.08$, which shows that C-26 is yet another candidate very metal-poor stream (Figure~\ref{fig:C-26}).

\section{Discussion and conclusions}
In this paper, we show the power of combining the \Gaia\ EDR3 \SF\ sample of stars that are significantly part of coherent stellar structures of the MW halo with the metallicity catalogue generated from the \Pris\ survey. Based on the narrow Ca H\&K filter available on the MegaCam imager at CFHT, the \Pris\ photometry provides accurate metallicities, even in the very metal-poor regime ($\FeH<-2.0$) to extremely metal-poor regimes ($\FeH<-3.0$). This is particularly valuable here as it yields metallicity measurements for all streams, without crippling uncertainties at the very metal-poor end ($\FeH\lta-2.0$). We are therefore able to derive metallicities for 26 stellar streams, including 6 new stream candidates (C-21, C-22, C-23, C-24, C-25, and C-26) that we present here for the first time. The mean metallicities of all those streams are summarize in Table~\ref{table}.

We note that most of the stellar streams unveiled by \Gaia\ in the MW halo have low metallicity, with typically $-3.0\lta\FeHPrave\lta-1.0$. In total, we count 14 streams with $\FeHPrave<-2.0$ (one of which is the stream of M~92, the most metal-poor MW GC), and 3 streams with $\FeHPrave<-2.9$. These results are similar to those reported by \citet{li22} for a mainly different set of streams and using spectroscopic metallicities. While some of these systems could be dwarf galaxy remnants and dwarf galaxies are known to reach lower metallicities than GCs at the faint end \citep[\eg][]{willman12,li18}, the physical width\footnote{The \SF\ run used here and presented in detail by \citet{ibata21} uses a spacial model of Gaussian width $\sigma=50\pc$ and therefore preferentially detects thin streams.} of these streams \citep{ibata21} implies that it is likely most of them result from disrupting clusters whose metallicity is, on the whole, lower than that of known MW clusters. Nowhere is this more striking than for C-19 that we determine to have $\FeHPrave=-3.58\pm0.08$ from 17 stars. A follow-up spectroscopic study confirms that this structure is indeed the most metal-poor structure known to date ($\FeH=-3.38\pm0.06\textrm{ (stat.)}\pm0.20\textrm{ (syst.)}$) and that its progenitor was a GC that was an order of magnitude less metal-enriched than current MW clusters \citep{martin22a}. More metal-poor clusters than observed today must have therefore existed in the past and the observed metallicity floor of GCs \citep{beasley19} is therefore driven by the time evolution of clusters \citep{kruijssen19}. A very low-metallicity GC was recently identified around M31 \citep[$\FeH=-2.91\pm0.04$;][]{larsen20} so the discovery of C-19 unambiguously confirms theories that also the MW hosted more metal-poor GCs in its past.

\begin{figure}
\begin{center}
\includegraphics[width=6cm,angle=270]{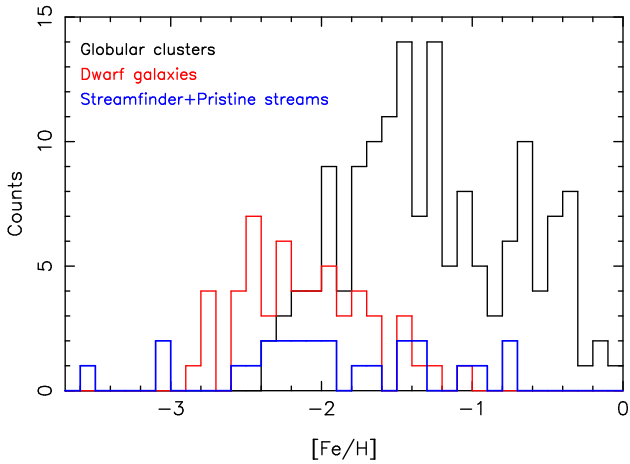}
\caption{Comparison of the metallicities of disitrbution functions of all known MW GCs (black histogram; \citealt{harris96}), dwarf galaxies (red histogram; \citealt{longeard21}), and the streams from the \SF+\Pris\ sample studied in this paper. The streams are systematically more metal-poor than the MW GCs and overlap the metallicities of the dwarf galaxies but also extends it to the extremely metal-poor regime with $\FeH<-3.0$.\label{fig:FeHcomp}}
\end{center}
\end{figure}

Figure~\ref{fig:FeHcomp} presents a comparison of the metallicity distribution functiona of all known MW GCs (black histogram, using metallicity values compiled by \citealt{harris96}, 2010 update), dwarf galaxies (red histogram, using metallicities values compiled by \citealt{longeard21}), and from the \SF+\Pris\ sample (blue histogram). This comparison is na\"ive and assumes that the Pristine footprint limits the number of streams we can study but does not bias their global metallicity properties. Taken at face value, Figure~\ref{fig:FeHcomp} shows that the newly discovered streams that comprise most of the \SF+\Pris\ sample are systematically more metal-poor than the distribution of MW GCs, overlaps the dwarf galaxy metallicities, while also extending it the extremely metal-poor regime. This comparison confirms that stellar streams were made from progenitors with metallicity properties that are seldom or never seen in satellites at present times, especially if most streams come from GCs, as mentioned above.

One key issue for future (spectroscopic) studies will be to systematically determine the metallicity dispersion of the various structures identified by \SF\ and other stream searches. This information is key to constraining the nature of the streams' progenitors since, for faint stellar systems, a dwarf galaxy is expected to exhibit a significant metallicity dispersion while GCs show negligible metallicity dispersions\footnote{Spectroscopic measurements of light-element variations (C, N, O, Na, Al) could also help discriminating between dwarf galaxy and globular cluster progenitors for these streams \citep[\eg][]{ji20}.} \citep[]{willman12,li22}. While it appears from the varying shapes of the MDFs shown in Figures~\ref{fig:M5stream}--\ref{fig:C-26} that the \Pris\ metallicities could hold some of this information, our attempts at determining accurate metallicity dispersions from the photometric metallicities yielded inconclusive results (see Appendix~A, available online as supplementary material). The individual uncertainties on the \Pris\ photometric metallicities translate into weak constraints on the dispersions of the MDFs and, in the case of strong contamination from field stars, especially at the faint end, can yield non-zero dispersions for known GCs. From the \Pris\ data, we nevertheless infer that ACS, C-9, C-24, GD-1, Pal~5, and Phlegeton have metallicity dispersions that are strongly constrained to be small ($\lta0.2$\,dex at the 90\% confidence level). Better constraints on the metallicity dispersions could be gained through low S/N spectroscopy of the \Pris+\SF\ stars, especially for faint stars. Such observations would easily confirm membership through simple radial velocities of likely member stars \citep[\eg][]{ibata21} without requiring the high signal to noise necessary for \FeH\ spectroscopic measurements \citep[see \eg][]{longeard18}.

\begin{figure*}
\begin{center}
\includegraphics[width=5.7cm,angle=270]{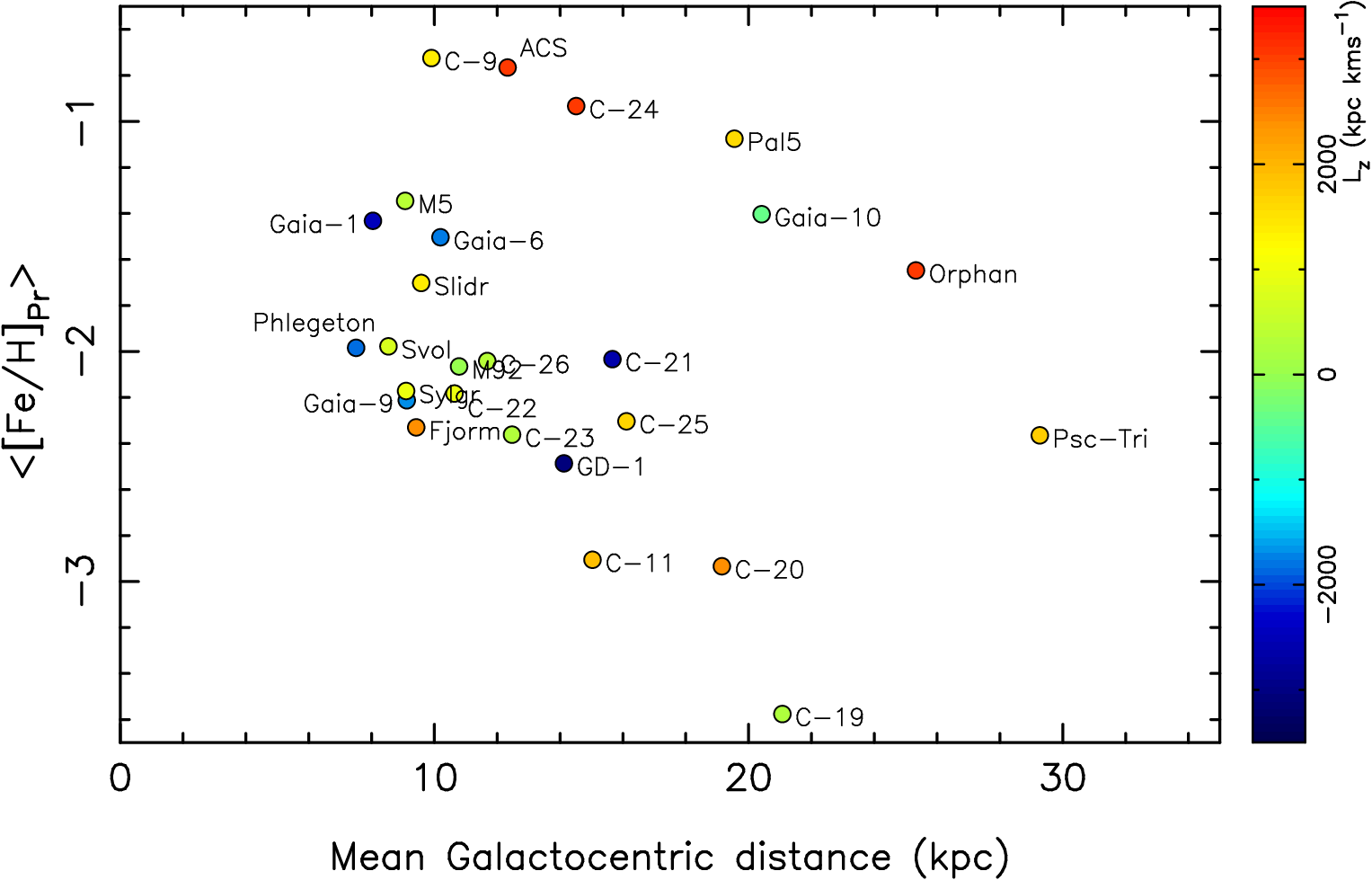}
\includegraphics[width=5.7cm,angle=270]{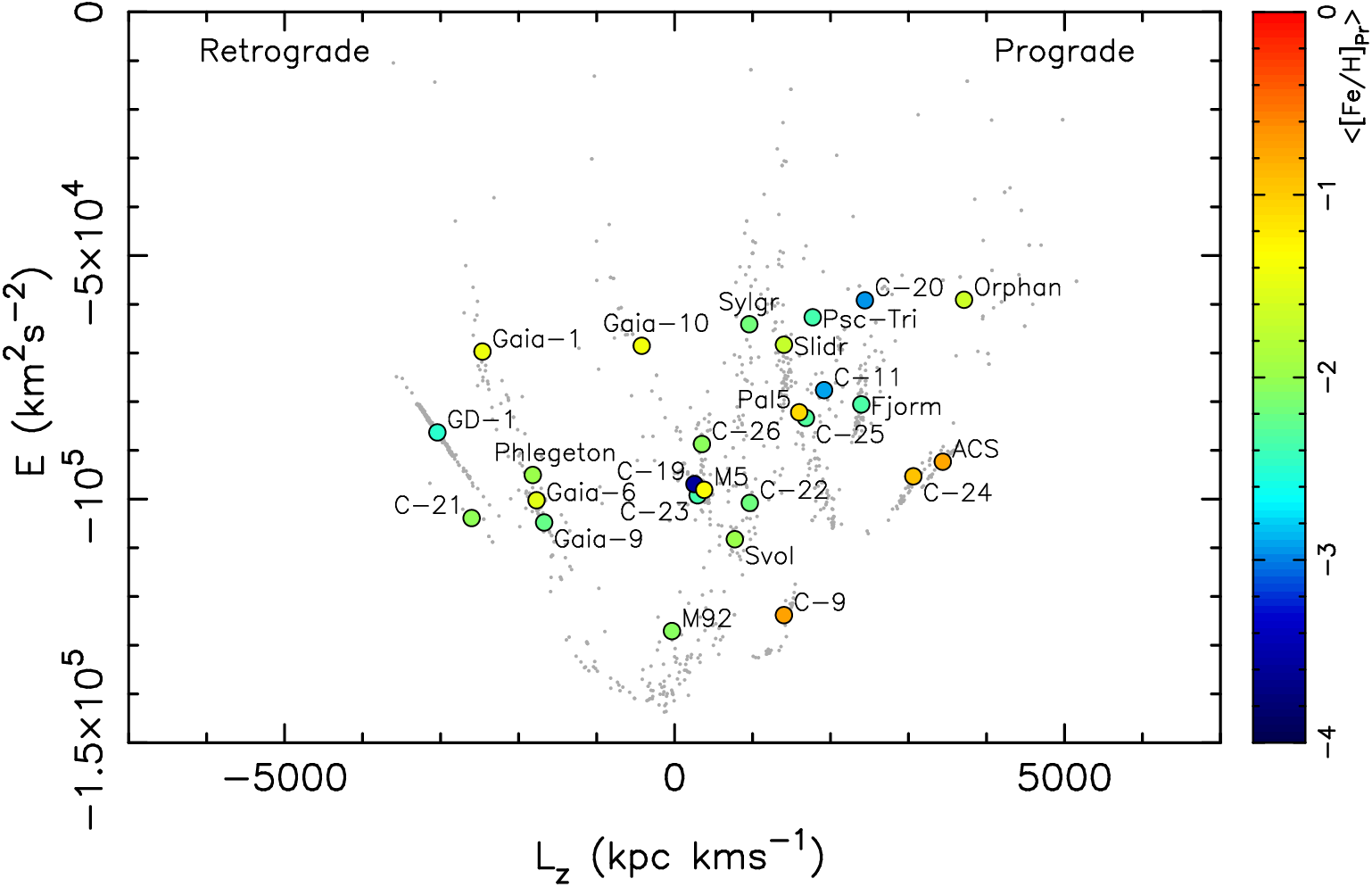}
\caption{Global properties of the 26\, studied streams. \emph{Left panel:} The average \Pris\ metallicity ($\FeHPrave$) of the streams, compared to the mean Galactocentric distance of the stars in a stream's sample, as determined from the \SF\ orbital solution of every individual stream star. The streams are color-coded by their mean angular momentum along the direction perpendicular to the MW plane ($L_z$). \emph{Right panel:} Distribution of stream stars in the energy--angular momentum ($E$,$L_z$) plane. Here as well, each stream is assigned the average $E$ and $L_z$ from the \SF\ orbital solutions of their members.\label{fig:FeHdist}}
\end{center}
\end{figure*}

Figure~\ref{fig:FeHdist} provides a global view of the \SF\ and \Pris\ stream dataset. While this data set is not a complete picture of the nearby MW streams because of the limited \Pris\ footprint\footnote{The window function imparted onto the stream map by the Pristine footprint is, for instance, likely responsible for the absence of streams at distances smaller than $\sim8\kpc$ that can be seen in the left-hand panel of Figure~\ref{fig:FeHdist}. The Pristine survey is mainly tied to the SDSS footprint that does not include lines of sight towards the inner Galaxy and, therefore, misses structures at small Galactocentric distances.} compared to \Gaia, a few intriguing trends can be teased out of this set of 26~stellar streams.

The metallicity range of the streams slides to lower metallicities with increasing Galactocentric distance (left panel). What we report here are the average distances of the stream stars, as determined by \SF, and may not represent the average distances to the orbits themselves. However, most stream stars will spend the majority of their time closer to apocenter (with smaller velocities) rather than pericenter (with larger velocities). With this caveat in mind, it is nevertheless striking that the lowest metallicities streams ($\FeHPrave\lta-2.5$) are all beyond $\sim15\kpc$ and that the metallicity of the most metal-rich streams drops from $\FeHPrave\sim-0.8$ at $\sim10\kpc$ from the Galactic center to $\FeHPrave<-1.4$ beyond 20\kpc. While some nearby structures like ACS may be tied to the MW disk these do not drive the trend and one of the most metal-rich streams, C-9, has an orbit that is distinctly different from a disk orbit. The hierarchical formation of structures could naturally lead to such a trend: high-mass dwarf galaxies accreted during the formation of the MW would spiral in through dynamical friction and deposit their preferentially more metal-rich clusters (and stars) in the inner reaches of the halo whereas lower mass dwarf galaxies would be less affected by dynamical friction and deposit their more metal-poor constituants at larger radii. The lowering metallicity trend with radius would then be made more prominent if some of the closer streams were formed by globular clusters formed in situ since these would naturally be more metal-rich. These ideas are not new for GCs \citep[\eg][]{mackey05} but it is particularly rewarding that we can now start firming up these scenarios with the remnants of past systems. 

Finally, the majority of streams in the combined \SF\ and \Pris\ sample are on prograde orbits around the MW (Figure~\ref{fig:FeHdist} right panel) as has been reported before for different sets of MW streams \citep[\eg][]{bonaca21,li22}. Considering the average \Pris\ metallicities of the streams, there is a hint that the most metal-rich streams are systematically prograde. While this is not surprising for ACS that has been tied to the MW disk, we note that C-24, despite being located at a high Galactic latitude ($b>50\deg$), has a very similar ($L_z$,$E$) and metallicity and may also be disk-related. Detailed studied of the dynamical clustering properties of streams will hopefully continue to shed light on the origin of all those streams \citep[\eg][]{bonaca21} and the metallicity information provided by surveys like \Pris\ can further help characterize their past hosts and progenitors \citep{malhan22}.

Even in an era of increasingly large spectroscopic surveys, those will only ever be able to observe a limited fraction of all MW halo stars. Large-scale surveys of high-quality photometric metallicities like \Pris\ are therefore very powerful to confirm the reality of the faint stellar streams that are being discovered around the MW (and to also clean up samples of member stars for future follow-up spectroscopy). Once the \Pris\ coverage is more homogeneous, we plan on pushing this idea fully by folding the \Pris\ photometric metallicities directly in the calculation of the \SF\ likelihood of a star to be part of a stream. On a closer time-scale, \Gaia's full DR3 release includes stellar parameters, including metallicities, for $\sim200$ million stars \citep{andrae18}. This all-sky sample does not reach the depth reached by \Pris\ but remains very valuable to extend the work presented here to the full MW sky.

The synergetic combination of \SF\ and \Pris\ to discover the MW structures with the most extreme metallicities reveals their past progenitors that have no current counterparts. Together, these two projects contribute to unveiling the complex puzzle of the assembly of the MW halo, bearing out the promises of the \Gaia\ mission.

\section*{Acknowledgements}
We are indebted to the referee, Ting Li, for pointing out that the Gaia-10 stream was in fact a re-discovery of the 300S stream. We gratefully thank the CFHT staff for performing the \Pris\ observations in queue mode, for their reactivity in adapting the schedule, and for answering our questions during the data-reduction process. We are also grateful to the High Performance Computing centre of the Universit\'e de Strasbourg and its staff for a very generous time allocation and for their support over the development of this project.

NFM, RI, ZY, AA, VH, and GK gratefully acknowledge support from the French National Research Agency (ANR) funded project ``Pristine'' (ANR-18-CE31-0017) along with funding from CNRS/INSU through the Programme National Galaxies et Cosmologie and through the CNRS grant PICS07708 and from the European Research Council (ERC) under the European Unions Horizon 2020 research and innovation programme (grant agreement No. 834148). ES acknowledges funding through VIDI grant ``Pushing Galactic Archaeology to its limits''(with project number VI.Vidi.193.093) which is funded by the Dutch Research Council (NWO). MB acknowledges the financial support to this research by INAF, through the Mainstream Grant 1.05.01.86.22 assigned to the project ``Chemo-dynamics of globular clusters: the Gaia revolution'' (P.I. E. Pancino). DA acknowledges support from the ERC Starting Grant NEFERTITI H2020/808240. J.I.G.H. acknowledges financial support from the Spanish Ministry of Science and Innovation (MICINN) project PID2020-117493GB-I00 and also from the Spanish MICINN under 2013 Ram\'on y Cajal program RYC-2013-14875. We benefited from the International Space Science Institute (ISSI) in Bern, CH, thanks to the funding of the Team ``Pristine''. 

This work has made use of data from the European Space Agency (ESA) mission {\it Gaia} (\url{https://www.cosmos.esa.int/gaia}), processed by the {\it Gaia} Data Processing and Analysis Consortium (DPAC, \url{https://www.cosmos.esa.int/web/gaia/dpac/consortium}). Funding for the DPAC has been provided by national institutions, in particular the institutions participating in the {\it Gaia} Multilateral Agreement. 

Based on observations obtained with MegaPrime/MegaCam, a joint project of CFHT and CEA/DAPNIA, at the Canada-France-Hawaii Telescope (CFHT) which is operated by the National Research Council (NRC) of Canada, the Institut National des Sciences de l'Univers of the Centre National de la Recherche Scientifique of France, and the University of Hawaii.

Funding for the Sloan Digital Sky Survey IV has been provided by the Alfred P. Sloan Foundation, the U.S. Department of Energy Office of Science, and the Participating Institutions. SDSS-IV acknowledges support and resources from the Center for High Performance Computing  at the University of Utah. The SDSS website is www.sdss.org. SDSS-IV is managed by the Astrophysical Research Consortium for the Participating Institutions of the SDSS Collaboration including the Brazilian Participation Group, the Carnegie Institution for Science, Carnegie Mellon University, Center for Astrophysics | Harvard \& Smithsonian, the Chilean Participation Group, the French Participation Group, Instituto de Astrof\'isica de Canarias, The Johns Hopkins University, Kavli Institute for the Physics and Mathematics of the Universe (IPMU) / University of Tokyo, the Korean Participation Group, Lawrence Berkeley National Laboratory, Leibniz Institut f\"ur Astrophysik Potsdam (AIP),  Max-Planck-Institut f\"ur Astronomie (MPIA Heidelberg), Max-Planck-Institut f\"ur Astrophysik (MPA Garching), Max-Planck-Institut f\"ur Extraterrestrische Physik (MPE), National Astronomical Observatories of China, New Mexico State University, New York University, University of Notre Dame, Observat\'ario Nacional / MCTI, The Ohio State University, Pennsylvania State University, Shanghai Astronomical Observatory, United Kingdom Participation Group, Universidad Nacional Aut\'onoma de M\'exico, University of Arizona, University of Colorado Boulder, University of Oxford, University of Portsmouth, University of Utah, University of Virginia, University of Washington, University of Wisconsin, Vanderbilt University, and Yale University.

\section*{Data availability statement}
The data underlying this article will be shared on reasonable request to the corresponding author.




\bibliographystyle{mnras}

\begin{thebibliography}{}
\makeatletter
\relax
\def\mn@urlcharsother{\let\do\@makeother \do\$\do\&\do\#\do\^\do\_\do\%\do\~}
\def\mn@doi{\begingroup\mn@urlcharsother \@ifnextchar [ {\mn@doi@}
  {\mn@doi@[]}}
\def\mn@doi@[#1]#2{\def\@tempa{#1}\ifx\@tempa\@empty \href
  {http://dx.doi.org/#2} {doi:#2}\else \href {http://dx.doi.org/#2} {#1}\fi
  \endgroup}
\def\mn@eprint#1#2{\mn@eprint@#1:#2::\@nil}
\def\mn@eprint@arXiv#1{\href {http://arxiv.org/abs/#1} {{\tt arXiv:#1}}}
\def\mn@eprint@dblp#1{\href {http://dblp.uni-trier.de/rec/bibtex/#1.xml}
  {dblp:#1}}
\def\mn@eprint@#1:#2:#3:#4\@nil{\def\@tempa {#1}\def\@tempb {#2}\def\@tempc
  {#3}\ifx \@tempc \@empty \let \@tempc \@tempb \let \@tempb \@tempa \fi \ifx
  \@tempb \@empty \def\@tempb {arXiv}\fi \@ifundefined
  {mn@eprint@\@tempb}{\@tempb:\@tempc}{\expandafter \expandafter \csname
  mn@eprint@\@tempb\endcsname \expandafter{\@tempc}}}

\bibitem[\protect\citeauthoryear{{Aguado} et~al.,}{{Aguado}
  et~al.}{2019}]{aguado19}
{Aguado} D.~S.,  et~al., 2019, \mn@doi [\mnras] {10.1093/mnras/stz2643}, \href
  {https://ui.adsabs.harvard.edu/abs/2019MNRAS.490.2241A} {490, 2241}

\bibitem[\protect\citeauthoryear{{An} \& {Beers}}{{An} \& {Beers}}{2021}]{an21}
{An} D.,  {Beers} T.~C.,  2021, \mn@doi [\apj] {10.3847/1538-4357/abccd2},
  \href {https://ui.adsabs.harvard.edu/abs/2021ApJ...907..101A} {907, 101}

\bibitem[\protect\citeauthoryear{{Andrae} et~al.,}{{Andrae}
  et~al.}{2018}]{andrae18}
{Andrae} R.,  et~al., 2018, \mn@doi [\aap] {10.1051/0004-6361/201732516}, \href
  {https://ui.adsabs.harvard.edu/abs/2018A&A...616A...8A} {616, A8}

\bibitem[\protect\citeauthoryear{{Antoja} et~al.,}{{Antoja}
  et~al.}{2018}]{antoja18}
{Antoja} T.,  et~al., 2018, \mn@doi [\nat] {10.1038/s41586-018-0510-7}, \href
  {https://ui.adsabs.harvard.edu/abs/2018Natur.561..360A} {561, 360}

\bibitem[\protect\citeauthoryear{{Arentsen} et~al.,}{{Arentsen}
  et~al.}{2020}]{arentsen20b}
{Arentsen} A.,  et~al., 2020, \mn@doi [\mnras] {10.1093/mnras/staa1661}, \href
  {https://ui.adsabs.harvard.edu/abs/2020MNRAS.496.4964A} {496, 4964}

\bibitem[\protect\citeauthoryear{{Balbinot}, {Cabrera-Ziri}  \&
  {Lardo}}{{Balbinot} et~al.}{2022}]{balbinot22}
{Balbinot} E.,  {Cabrera-Ziri} I.,   {Lardo} C.,  2022, \mn@doi [\mnras]
  {10.1093/mnras/stac1953}, \href
  {https://ui.adsabs.harvard.edu/abs/2022MNRAS.tmp.1837B} {}

\bibitem[\protect\citeauthoryear{{Banik}, {Bovy}, {Bertone}, {Erkal}  \& {de
  Boer}}{{Banik} et~al.}{2021}]{banik21}
{Banik} N.,  {Bovy} J.,  {Bertone} G.,  {Erkal} D.,   {de Boer} T.~J.~L.,
  2021, \mn@doi [\mnras] {10.1093/mnras/stab210}, \href
  {https://ui.adsabs.harvard.edu/abs/2021MNRAS.502.2364B} {502, 2364}

\bibitem[\protect\citeauthoryear{{Beasley}, {Leaman}, {Gallart}, {Larsen},
  {Battaglia}, {Monelli}  \& {Pedreros}}{{Beasley} et~al.}{2019}]{beasley19}
{Beasley} M.~A.,  {Leaman} R.,  {Gallart} C.,  {Larsen} S.~S.,  {Battaglia} G.,
   {Monelli} M.,   {Pedreros} M.~H.,  2019, \mn@doi [\mnras]
  {10.1093/mnras/stz1349}, \href
  {https://ui.adsabs.harvard.edu/abs/2019MNRAS.487.1986B} {487, 1986}

\bibitem[\protect\citeauthoryear{{Belokurov} et~al.,}{{Belokurov}
  et~al.}{2007}]{belokurov07}
{Belokurov} V.,  et~al., 2007, \mn@doi [\apj] {10.1086/511302}, \href
  {https://ui.adsabs.harvard.edu/abs/2007ApJ...658..337B} {658, 337}

\bibitem[\protect\citeauthoryear{{Belokurov}, {Sanders}, {Fattahi}, {Smith},
  {Deason}, {Evans}  \& {Grand}}{{Belokurov} et~al.}{2020}]{belokurov20}
{Belokurov} V.,  {Sanders} J.~L.,  {Fattahi} A.,  {Smith} M.~C.,  {Deason}
  A.~J.,  {Evans} N.~W.,   {Grand} R. J.~J.,  2020, \mn@doi [\mnras]
  {10.1093/mnras/staa876}, \href
  {https://ui.adsabs.harvard.edu/abs/2020MNRAS.494.3880B} {494, 3880}

\bibitem[\protect\citeauthoryear{{Binney} \& {Tremaine}}{{Binney} \&
  {Tremaine}}{2008}]{binney08}
{Binney} J.,  {Tremaine} S.,  2008, {Galactic Dynamics: Second Edition}.
Princeton University Press

\bibitem[\protect\citeauthoryear{{Blanton} et~al.,}{{Blanton}
  et~al.}{2017}]{blanton17}
{Blanton} M.~R.,  et~al., 2017, \mn@doi [\aj] {10.3847/1538-3881/aa7567}, \href
  {https://ui.adsabs.harvard.edu/abs/2017AJ....154...28B} {154, 28}

\bibitem[\protect\citeauthoryear{{Bonaca}, {Geha}  \& {Kallivayalil}}{{Bonaca}
  et~al.}{2012}]{bonaca12}
{Bonaca} A.,  {Geha} M.,   {Kallivayalil} N.,  2012, \mn@doi [\apjl]
  {10.1088/2041-8205/760/1/L6}, \href
  {http://cdsads.u-strasbg.fr/abs/2012ApJ...760L...6B} {760, L6}

\bibitem[\protect\citeauthoryear{{Bonaca}, {Hogg}, {Price-Whelan}  \&
  {Conroy}}{{Bonaca} et~al.}{2019}]{bonaca19}
{Bonaca} A.,  {Hogg} D.~W.,  {Price-Whelan} A.~M.,   {Conroy} C.,  2019,
  \mn@doi [\apj] {10.3847/1538-4357/ab2873}, \href
  {https://ui.adsabs.harvard.edu/abs/2019ApJ...880...38B} {880, 38}

\bibitem[\protect\citeauthoryear{{Bonaca} et~al.,}{{Bonaca}
  et~al.}{2020a}]{bonaca20b}
{Bonaca} A.,  et~al., 2020a, \mn@doi [\apj] {10.3847/1538-4357/ab5afe}, \href
  {https://ui.adsabs.harvard.edu/abs/2020ApJ...889...70B} {889, 70}

\bibitem[\protect\citeauthoryear{{Bonaca} et~al.,}{{Bonaca}
  et~al.}{2020b}]{bonaca20}
{Bonaca} A.,  et~al., 2020b, \mn@doi [\apjl] {10.3847/2041-8213/ab800c}, \href
  {https://ui.adsabs.harvard.edu/abs/2020ApJ...892L..37B} {892, L37}

\bibitem[\protect\citeauthoryear{{Bonaca} et~al.,}{{Bonaca}
  et~al.}{2021}]{bonaca21}
{Bonaca} A.,  et~al., 2021, \mn@doi [\apjl] {10.3847/2041-8213/abeaa9}, \href
  {https://ui.adsabs.harvard.edu/abs/2021ApJ...909L..26B} {909, L26}

\bibitem[\protect\citeauthoryear{{Boulade} et~al.,}{{Boulade}
  et~al.}{2003}]{boulade03}
{Boulade} O.,  et~al., 2003, in {Iye} M.,  {Moorwood} A. F.~M.,  eds,  Society
  of Photo-Optical Instrumentation Engineers (SPIE) Conference Series Vol.
  4841, Instrument Design and Performance for Optical/Infrared Ground-based
  Telescopes. pp 72--81, \mn@doi{10.1117/12.459890}

\bibitem[\protect\citeauthoryear{{Bovy}, {Bahmanyar}, {Fritz}  \&
  {Kallivayalil}}{{Bovy} et~al.}{2016}]{bovy16}
{Bovy} J.,  {Bahmanyar} A.,  {Fritz} T.~K.,   {Kallivayalil} N.,  2016, \mn@doi
  [\apj] {10.3847/1538-4357/833/1/31}, \href
  {https://ui.adsabs.harvard.edu/abs/2016ApJ...833...31B} {833, 31}

\bibitem[\protect\citeauthoryear{{Carlberg} \& {Grillmair}}{{Carlberg} \&
  {Grillmair}}{2013}]{carlberg13}
{Carlberg} R.~G.,  {Grillmair} C.~J.,  2013, \mn@doi [\apj]
  {10.1088/0004-637X/768/2/171}, \href
  {https://ui.adsabs.harvard.edu/abs/2013ApJ...768..171C} {768, 171}

\bibitem[\protect\citeauthoryear{{Carlberg}, {Grillmair}  \&
  {Hetherington}}{{Carlberg} et~al.}{2012}]{carlberg12}
{Carlberg} R.~G.,  {Grillmair} C.~J.,   {Hetherington} N.,  2012, \mn@doi
  [\apj] {10.1088/0004-637X/760/1/75}, \href
  {https://ui.adsabs.harvard.edu/abs/2012ApJ...760...75C} {760, 75}

\bibitem[\protect\citeauthoryear{{Carretta}, {Bragaglia}, {Gratton}, {D'Orazi}
  \& {Lucatello}}{{Carretta} et~al.}{2009}]{carretta09}
{Carretta} E.,  {Bragaglia} A.,  {Gratton} R.,  {D'Orazi} V.,   {Lucatello} S.,
   2009, \mn@doi [\aap] {10.1051/0004-6361/200913003}, \href
  {https://ui.adsabs.harvard.edu/abs/2009A&A...508..695C} {508, 695}

\bibitem[\protect\citeauthoryear{{Chiti}, {Frebel}, {Mardini}, {Daniel}, {Ou}
  \& {Uvarova}}{{Chiti} et~al.}{2021}]{chiti21}
{Chiti} A.,  {Frebel} A.,  {Mardini} M.~K.,  {Daniel} T.~W.,  {Ou} X.,
  {Uvarova} A.~V.,  2021, \mn@doi [\apjs] {10.3847/1538-4365/abf73d}, \href
  {https://ui.adsabs.harvard.edu/abs/2021ApJS..254...31C} {254, 31}

\bibitem[\protect\citeauthoryear{{Christlieb} et~al.,}{{Christlieb}
  et~al.}{2019}]{christlieb19}
{Christlieb} N.,  et~al., 2019, \mn@doi [The Messenger]
  {10.18727/0722-6691/5121}, \href
  {https://ui.adsabs.harvard.edu/abs/2019Msngr.175...26C} {175, 26}

\bibitem[\protect\citeauthoryear{{Dalton} et~al.,}{{Dalton}
  et~al.}{2018}]{dalton18}
{Dalton} G.,  et~al., 2018, in {Evans} C.~J.,  {Simard} L.,   {Takami} H.,
  eds,  Society of Photo-Optical Instrumentation Engineers (SPIE) Conference
  Series Vol. 10702, Ground-based and Airborne Instrumentation for Astronomy
  VII. p. 107021B, \mn@doi{10.1117/12.2312031}

\bibitem[\protect\citeauthoryear{{Erkal} et~al.,}{{Erkal}
  et~al.}{2019}]{erkal19}
{Erkal} D.,  et~al., 2019, \mn@doi [\mnras] {10.1093/mnras/stz1371}, \href
  {https://ui.adsabs.harvard.edu/abs/2019MNRAS.487.2685E} {487, 2685}

\bibitem[\protect\citeauthoryear{{Fern{\'a}ndez-Alvar}, {Tissera}, {Carigi},
  {Schuster}, {Beers}  \& {Belokurov}}{{Fern{\'a}ndez-Alvar}
  et~al.}{2019}]{fernandez-alvar19}
{Fern{\'a}ndez-Alvar} E.,  {Tissera} P.~B.,  {Carigi} L.,  {Schuster} W.~J.,
  {Beers} T.~C.,   {Belokurov} V.~A.,  2019, \mn@doi [\mnras]
  {10.1093/mnras/stz443}, \href
  {https://ui.adsabs.harvard.edu/abs/2019MNRAS.485.1745F} {485, 1745}

\bibitem[\protect\citeauthoryear{{Fern{\'a}ndez-Alvar}
  et~al.,}{{Fern{\'a}ndez-Alvar} et~al.}{2021}]{fernandez-alvar21}
{Fern{\'a}ndez-Alvar} E.,  et~al., 2021, \mn@doi [\mnras]
  {10.1093/mnras/stab2617}, \href
  {https://ui.adsabs.harvard.edu/abs/2021MNRAS.508.1509F} {508, 1509}

\bibitem[\protect\citeauthoryear{{Freeman} \& {Bland-Hawthorn}}{{Freeman} \&
  {Bland-Hawthorn}}{2002}]{freeman02}
{Freeman} K.,  {Bland-Hawthorn} J.,  2002, \mn@doi [\araa]
  {10.1146/annurev.astro.40.060401.093840}, \href
  {http://cdsads.u-strasbg.fr/abs/2002ARA%26A..40..487F} {40, 487}

\bibitem[\protect\citeauthoryear{{Gaia Collaboration} et~al.,}{{Gaia
  Collaboration} et~al.}{2016}]{Gaia16a}
{Gaia Collaboration} et~al., 2016, \mn@doi [\aap]
  {10.1051/0004-6361/201629272}, \href
  {https://ui.adsabs.harvard.edu/abs/2016A&A...595A...1G} {595, A1}

\bibitem[\protect\citeauthoryear{{Gaia Collaboration} et~al.,}{{Gaia
  Collaboration} et~al.}{2018}]{Gaia18a}
{Gaia Collaboration} et~al., 2018, \mn@doi [\aap]
  {10.1051/0004-6361/201833051}, \href
  {https://ui.adsabs.harvard.edu/abs/2018A&A...616A...1G} {616, A1}

\bibitem[\protect\citeauthoryear{{Gaia Collaboration} et~al.,}{{Gaia
  Collaboration} et~al.}{2021}]{Gaia21a}
{Gaia Collaboration} et~al., 2021, \mn@doi [\aap]
  {10.1051/0004-6361/202039657}, \href
  {https://ui.adsabs.harvard.edu/abs/2021A&A...649A...1G} {649, A1}

\bibitem[\protect\citeauthoryear{{Grillmair}}{{Grillmair}}{2009}]{grillmair09}
{Grillmair} C.~J.,  2009, \mn@doi [\apj] {10.1088/0004-637X/693/2/1118}, \href
  {http://cdsads.u-strasbg.fr/abs/2009ApJ...693.1118G} {693, 1118}

\bibitem[\protect\citeauthoryear{{Grillmair}}{{Grillmair}}{2019}]{grillmair19}
{Grillmair} C.~J.,  2019, \mn@doi [\apj] {10.3847/1538-4357/ab441d}, \href
  {https://ui.adsabs.harvard.edu/abs/2019ApJ...884..174G} {884, 174}

\bibitem[\protect\citeauthoryear{{Grillmair} \& {Dionatos}}{{Grillmair} \&
  {Dionatos}}{2006a}]{grillmair06}
{Grillmair} C.~J.,  {Dionatos} O.,  2006a, \mn@doi [\apjl] {10.1086/503744},
  \href {http://adsabs.harvard.edu/abs/2006ApJ...641L..37G} {641, L37}

\bibitem[\protect\citeauthoryear{{Grillmair} \& {Dionatos}}{{Grillmair} \&
  {Dionatos}}{2006b}]{grillmair06b}
{Grillmair} C.~J.,  {Dionatos} O.,  2006b, \mn@doi [\apjl] {10.1086/505111},
  \href {https://ui.adsabs.harvard.edu/abs/2006ApJ...643L..17G} {643, L17}

\bibitem[\protect\citeauthoryear{{Harris}}{{Harris}}{1996}]{harris96}
{Harris} W.~E.,  1996, \mn@doi [\aj] {10.1086/118116}, \href
  {http://cdsads.u-strasbg.fr/abs/1996AJ....112.1487H} {112, 1487}

\bibitem[\protect\citeauthoryear{{Haywood}, {Di Matteo}, {Lehnert}, {Snaith},
  {Khoperskov}  \& {G{\'o}mez}}{{Haywood} et~al.}{2018}]{haywood18}
{Haywood} M.,  {Di Matteo} P.,  {Lehnert} M.~D.,  {Snaith} O.,  {Khoperskov}
  S.,   {G{\'o}mez} A.,  2018, \mn@doi [\apj] {10.3847/1538-4357/aad235}, \href
  {https://ui.adsabs.harvard.edu/abs/2018ApJ...863..113H} {863, 113}

\bibitem[\protect\citeauthoryear{{Helmi}, {Babusiaux}, {Koppelman}, {Massari},
  {Veljanoski}  \& {Brown}}{{Helmi} et~al.}{2018}]{helmi18}
{Helmi} A.,  {Babusiaux} C.,  {Koppelman} H.~H.,  {Massari} D.,  {Veljanoski}
  J.,   {Brown} A. G.~A.,  2018, \mn@doi [\nat] {10.1038/s41586-018-0625-x},
  \href {https://ui.adsabs.harvard.edu/abs/2018Natur.563...85H} {563, 85}

\bibitem[\protect\citeauthoryear{{Helmi} et~al.,}{{Helmi}
  et~al.}{2019}]{helmi19}
{Helmi} A.,  et~al., 2019, \mn@doi [The Messenger] {10.18727/0722-6691/5120},
  \href {https://ui.adsabs.harvard.edu/abs/2019Msngr.175...23H} {175, 23}

\bibitem[\protect\citeauthoryear{{Hogg} et~al.,}{{Hogg} et~al.}{2016}]{hogg16}
{Hogg} D.~W.,  et~al., 2016, \mn@doi [\apj] {10.3847/1538-4357/833/2/262},
  \href {https://ui.adsabs.harvard.edu/abs/2016ApJ...833..262H} {833, 262}

\bibitem[\protect\citeauthoryear{{Ibata}, {Lewis}  \& {Martin}}{{Ibata}
  et~al.}{2016}]{ibata16}
{Ibata} R.~A.,  {Lewis} G.~F.,   {Martin} N.~F.,  2016, \mn@doi [\apj]
  {10.3847/0004-637X/819/1/1}, \href
  {https://ui.adsabs.harvard.edu/abs/2016ApJ...819....1I} {819, 1}

\bibitem[\protect\citeauthoryear{{Ibata} et~al.,}{{Ibata}
  et~al.}{2017}]{ibata17}
{Ibata} R.~A.,  et~al., 2017, \mn@doi [\apj] {10.3847/1538-4357/aa8562}, \href
  {https://ui.adsabs.harvard.edu/abs/2017ApJ...848..129I} {848, 129}

\bibitem[\protect\citeauthoryear{{Ibata}, {Malhan}, {Martin}  \&
  {Starkenburg}}{{Ibata} et~al.}{2018}]{ibata18}
{Ibata} R.~A.,  {Malhan} K.,  {Martin} N.~F.,   {Starkenburg} E.,  2018,
  \mn@doi [\apj] {10.3847/1538-4357/aadba3}, \href
  {https://ui.adsabs.harvard.edu/abs/2018ApJ...865...85I} {865, 85}

\bibitem[\protect\citeauthoryear{{Ibata}, {Malhan}  \& {Martin}}{{Ibata}
  et~al.}{2019}]{ibata19}
{Ibata} R.~A.,  {Malhan} K.,   {Martin} N.~F.,  2019, \mn@doi [\apj]
  {10.3847/1538-4357/ab0080}, \href
  {https://ui.adsabs.harvard.edu/abs/2019ApJ...872..152I} {872, 152}

\bibitem[\protect\citeauthoryear{{Ibata}, {Thomas}, {Famaey}, {Malhan},
  {Martin}  \& {Monari}}{{Ibata} et~al.}{2020}]{ibata20}
{Ibata} R.,  {Thomas} G.,  {Famaey} B.,  {Malhan} K.,  {Martin} N.,   {Monari}
  G.,  2020, \mn@doi [\apj] {10.3847/1538-4357/ab7303}, \href
  {https://ui.adsabs.harvard.edu/abs/2020ApJ...891..161I} {891, 161}

\bibitem[\protect\citeauthoryear{{Ibata} et~al.,}{{Ibata}
  et~al.}{2021}]{ibata21}
{Ibata} R.,  et~al., 2021, \mn@doi [\apj] {10.3847/1538-4357/abfcc2}, \href
  {https://ui.adsabs.harvard.edu/abs/2021ApJ...914..123I} {914, 123}

\bibitem[\protect\citeauthoryear{{Ishigaki}, {Hwang}, {Chiba}  \&
  {Aoki}}{{Ishigaki} et~al.}{2016}]{ishigaki16}
{Ishigaki} M.~N.,  {Hwang} N.,  {Chiba} M.,   {Aoki} W.,  2016, \mn@doi [\apj]
  {10.3847/0004-637X/823/2/157}, \href
  {https://ui.adsabs.harvard.edu/abs/2016ApJ...823..157I} {823, 157}

\bibitem[\protect\citeauthoryear{{Ivezi{\'c}} et~al.,}{{Ivezi{\'c}}
  et~al.}{2008}]{ivezic08}
{Ivezi{\'c}} {\v{Z}}.,  et~al., 2008, \mn@doi [\apj] {10.1086/589678}, \href
  {https://ui.adsabs.harvard.edu/abs/2008ApJ...684..287I} {684, 287}

\bibitem[\protect\citeauthoryear{{Ji} et~al.,}{{Ji} et~al.}{2020}]{ji20}
{Ji} A.~P.,  et~al., 2020, \mn@doi [\aj] {10.3847/1538-3881/abacb6}, \href
  {https://ui.adsabs.harvard.edu/abs/2020AJ....160..181J} {160, 181}

\bibitem[\protect\citeauthoryear{{Johnston}, {Sigurdsson}  \&
  {Hernquist}}{{Johnston} et~al.}{1999}]{johnston99}
{Johnston} K.~V.,  {Sigurdsson} S.,   {Hernquist} L.,  1999, \mn@doi [\mnras]
  {10.1046/j.1365-8711.1999.02200.x}, \href
  {http://adsabs.harvard.edu/abs/1999MNRAS.302..771J} {302, 771}

\bibitem[\protect\citeauthoryear{{Koch} \& {C{\^o}t{\'e}}}{{Koch} \&
  {C{\^o}t{\'e}}}{2017}]{koch17}
{Koch} A.,  {C{\^o}t{\'e}} P.,  2017, \mn@doi [\aap]
  {10.1051/0004-6361/201629872}, \href
  {https://ui.adsabs.harvard.edu/abs/2017A&A...601A..41K} {601, A41}

\bibitem[\protect\citeauthoryear{{Koposov}, {Rix}  \& {Hogg}}{{Koposov}
  et~al.}{2010}]{koposov10}
{Koposov} S.~E.,  {Rix} H.-W.,   {Hogg} D.~W.,  2010, \mn@doi [\apj]
  {10.1088/0004-637X/712/1/260}, \href
  {https://ui.adsabs.harvard.edu/abs/2010ApJ...712..260K} {712, 260}

\bibitem[\protect\citeauthoryear{{Koposov} et~al.,}{{Koposov}
  et~al.}{2019}]{koposov19}
{Koposov} S.~E.,  et~al., 2019, \mn@doi [\mnras] {10.1093/mnras/stz457}, \href
  {https://ui.adsabs.harvard.edu/abs/2019MNRAS.485.4726K} {485, 4726}

\bibitem[\protect\citeauthoryear{{Kruijssen}}{{Kruijssen}}{2019}]{kruijssen19}
{Kruijssen} J.~M.~D.,  2019, \mn@doi [\mnras] {10.1093/mnrasl/slz052}, \href
  {https://ui.adsabs.harvard.edu/abs/2019MNRAS.486L..20K} {486, L20}

\bibitem[\protect\citeauthoryear{{Laporte}, {Belokurov}, {Koposov}, {Smith}  \&
  {Hill}}{{Laporte} et~al.}{2020}]{laporte20}
{Laporte} C. F.~P.,  {Belokurov} V.,  {Koposov} S.~E.,  {Smith} M.~C.,   {Hill}
  V.,  2020, \mn@doi [\mnras] {10.1093/mnrasl/slz167}, \href
  {https://ui.adsabs.harvard.edu/abs/2020MNRAS.492L..61L} {492, L61}

\bibitem[\protect\citeauthoryear{{Larsen}, {Romanowsky}, {Brodie}  \&
  {Wasserman}}{{Larsen} et~al.}{2020}]{larsen20}
{Larsen} S.~S.,  {Romanowsky} A.~J.,  {Brodie} J.~P.,   {Wasserman} A.,  2020,
  \mn@doi [Science] {10.1126/science.abb1970}, \href
  {https://ui.adsabs.harvard.edu/abs/2020Sci...370..970L} {370, 970}

\bibitem[\protect\citeauthoryear{{Li} et~al.,}{{Li} et~al.}{2018}]{li18}
{Li} T.~S.,  et~al., 2018, \mn@doi [\apj] {10.3847/1538-4357/aadf91}, \href
  {https://ui.adsabs.harvard.edu/abs/2018ApJ...866...22L} {866, 22}

\bibitem[\protect\citeauthoryear{{Li} et~al.,}{{Li} et~al.}{2022}]{li22}
{Li} T.~S.,  et~al., 2022, \mn@doi [\apj] {10.3847/1538-4357/ac46d3}, \href
  {https://ui.adsabs.harvard.edu/abs/2022ApJ...928...30L} {928, 30}

\bibitem[\protect\citeauthoryear{{Longeard} et~al.,}{{Longeard}
  et~al.}{2018}]{longeard18}
{Longeard} N.,  et~al., 2018, \mn@doi [\mnras] {10.1093/mnras/sty1986}, \href
  {https://ui.adsabs.harvard.edu/abs/2018MNRAS.480.2609L} {480, 2609}

\bibitem[\protect\citeauthoryear{{Longeard} et~al.,}{{Longeard}
  et~al.}{2021}]{longeard21}
{Longeard} N.,  et~al., 2021, arXiv e-prints, \href
  {https://ui.adsabs.harvard.edu/abs/2021arXiv210710849L} {p. arXiv:2107.10849}

\bibitem[\protect\citeauthoryear{{Mackey} \& {van den Bergh}}{{Mackey} \& {van
  den Bergh}}{2005}]{mackey05}
{Mackey} A.~D.,  {van den Bergh} S.,  2005, \mn@doi [\mnras]
  {10.1111/j.1365-2966.2005.09080.x}, \href
  {http://adsabs.harvard.edu/abs/2005MNRAS.360..631M} {360, 631}

\bibitem[\protect\citeauthoryear{{Majewski} et~al.,}{{Majewski}
  et~al.}{2017}]{majewski17}
{Majewski} S.~R.,  et~al., 2017, \mn@doi [\aj] {10.3847/1538-3881/aa784d},
  \href {https://ui.adsabs.harvard.edu/abs/2017AJ....154...94M} {154, 94}

\bibitem[\protect\citeauthoryear{{Malhan} \& {Ibata}}{{Malhan} \&
  {Ibata}}{2019}]{malhan19c}
{Malhan} K.,  {Ibata} R.~A.,  2019, \mn@doi [\mnras] {10.1093/mnras/stz1035},
  \href {https://ui.adsabs.harvard.edu/abs/2019MNRAS.486.2995M} {486, 2995}

\bibitem[\protect\citeauthoryear{{Malhan}, {Ibata}  \& {Martin}}{{Malhan}
  et~al.}{2018}]{malhan18}
{Malhan} K.,  {Ibata} R.~A.,   {Martin} N.~F.,  2018, \mn@doi [\mnras]
  {10.1093/mnras/sty2474}, \href
  {https://ui.adsabs.harvard.edu/abs/2018MNRAS.481.3442M} {481, 3442}

\bibitem[\protect\citeauthoryear{{Malhan}, {Ibata}, {Carlberg}, {Valluri}  \&
  {Freese}}{{Malhan} et~al.}{2019a}]{malhan19b}
{Malhan} K.,  {Ibata} R.~A.,  {Carlberg} R.~G.,  {Valluri} M.,   {Freese} K.,
  2019a, \mn@doi [\apj] {10.3847/1538-4357/ab2e07}, \href
  {https://ui.adsabs.harvard.edu/abs/2019ApJ...881..106M} {881, 106}

\bibitem[\protect\citeauthoryear{{Malhan}, {Ibata}, {Carlberg}, {Bellazzini},
  {Famaey}  \& {Martin}}{{Malhan} et~al.}{2019b}]{malhan19a}
{Malhan} K.,  {Ibata} R.~A.,  {Carlberg} R.~G.,  {Bellazzini} M.,  {Famaey} B.,
    {Martin} N.~F.,  2019b, \mn@doi [\apjl] {10.3847/2041-8213/ab530e}, \href
  {https://ui.adsabs.harvard.edu/abs/2019ApJ...886L...7M} {886, L7}

\bibitem[\protect\citeauthoryear{{Malhan} et~al.,}{{Malhan}
  et~al.}{2022}]{malhan22}
{Malhan} K.,  et~al., 2022, \mn@doi [\apj] {10.3847/1538-4357/ac4d2a}, \href
  {https://ui.adsabs.harvard.edu/abs/2022ApJ...926..107M} {926, 107}

\bibitem[\protect\citeauthoryear{{Martell} et~al.,}{{Martell}
  et~al.}{2017}]{martell17}
{Martell} S.~L.,  et~al., 2017, \mn@doi [\mnras] {10.1093/mnras/stw2835}, \href
  {https://ui.adsabs.harvard.edu/abs/2017MNRAS.465.3203M} {465, 3203}

\bibitem[\protect\citeauthoryear{{Martin}, {Carlin}, {Newberg}  \&
  {Grillmair}}{{Martin} et~al.}{2013}]{martinc13}
{Martin} C.,  {Carlin} J.~L.,  {Newberg} H.~J.,   {Grillmair} C.,  2013,
  \mn@doi [\apjl] {10.1088/2041-8205/765/2/L39}, \href
  {http://cdsads.u-strasbg.fr/abs/2013ApJ...765L..39M} {765, L39}

\bibitem[\protect\citeauthoryear{{Martin}, {Collins}, {Longeard}  \&
  {Tollerud}}{{Martin} et~al.}{2018}]{martin18}
{Martin} N.~F.,  {Collins} M. L.~M.,  {Longeard} N.,   {Tollerud} E.,  2018,
  \mn@doi [\apjl] {10.3847/2041-8213/aac216}, \href
  {https://ui.adsabs.harvard.edu/abs/2018ApJ...859L...5M} {859, L5}

\bibitem[\protect\citeauthoryear{{Martin} et~al.,}{{Martin}
  et~al.}{2022}]{martin22a}
{Martin} N.~F.,  et~al., 2022, \mn@doi [\nat] {10.1038/s41586-021-04162-2},
  \href {https://ui.adsabs.harvard.edu/abs/2022Natur.601...45M} {601, 45}

\bibitem[\protect\citeauthoryear{{Myeong}, {Evans}, {Belokurov}, {Sanders}  \&
  {Koposov}}{{Myeong} et~al.}{2018}]{myeong18}
{Myeong} G.~C.,  {Evans} N.~W.,  {Belokurov} V.,  {Sanders} J.~L.,   {Koposov}
  S.~E.,  2018, \mn@doi [\mnras] {10.1093/mnras/sty1403}, \href
  {https://ui.adsabs.harvard.edu/abs/2018MNRAS.478.5449M} {478, 5449}

\bibitem[\protect\citeauthoryear{{Myeong}, {Vasiliev}, {Iorio}, {Evans}  \&
  {Belokurov}}{{Myeong} et~al.}{2019}]{myeong19}
{Myeong} G.~C.,  {Vasiliev} E.,  {Iorio} G.,  {Evans} N.~W.,   {Belokurov} V.,
  2019, \mn@doi [\mnras] {10.1093/mnras/stz1770}, \href
  {https://ui.adsabs.harvard.edu/abs/2019MNRAS.488.1235M} {488, 1235}

\bibitem[\protect\citeauthoryear{{Odenkirchen} et~al.,}{{Odenkirchen}
  et~al.}{2001}]{odenkirchen01a}
{Odenkirchen} M.,  et~al., 2001, \mn@doi [\apjl] {10.1086/319095}, \href
  {http://adsabs.harvard.edu/abs/2001ApJ...548L.165O} {548, L165}

\bibitem[\protect\citeauthoryear{{Onken} et~al.,}{{Onken}
  et~al.}{2019}]{onken19}
{Onken} C.~A.,  et~al., 2019, \mn@doi [\pasa] {10.1017/pasa.2019.27}, \href
  {https://ui.adsabs.harvard.edu/abs/2019PASA...36...33O} {36, e033}

\bibitem[\protect\citeauthoryear{{Palau} \& {Miralda-Escud{\'e}}}{{Palau} \&
  {Miralda-Escud{\'e}}}{2019}]{palau19}
{Palau} C.~G.,  {Miralda-Escud{\'e}} J.,  2019, \mn@doi [\mnras]
  {10.1093/mnras/stz1790}, \href
  {https://ui.adsabs.harvard.edu/abs/2019MNRAS.488.1535P} {488, 1535}

\bibitem[\protect\citeauthoryear{{Pearson}, {Price-Whelan}  \&
  {Johnston}}{{Pearson} et~al.}{2017}]{pearson17}
{Pearson} S.,  {Price-Whelan} A.~M.,   {Johnston} K.~V.,  2017, \mn@doi [Nature
  Astronomy] {10.1038/s41550-017-0220-3}, \href
  {https://ui.adsabs.harvard.edu/abs/2017NatAs...1..633P} {1, 633}

\bibitem[\protect\citeauthoryear{{Price-Whelan} \& {Bonaca}}{{Price-Whelan} \&
  {Bonaca}}{2018}]{price-whelan18}
{Price-Whelan} A.~M.,  {Bonaca} A.,  2018, \mn@doi [\apjl]
  {10.3847/2041-8213/aad7b5}, \href
  {https://ui.adsabs.harvard.edu/abs/2018ApJ...863L..20P} {863, L20}

\bibitem[\protect\citeauthoryear{{Ramos}, {Antoja}, {Mateu}, {Anders},
  {Laporte}, {Carballo-Bello}, {Famaey}  \& {Ibata}}{{Ramos}
  et~al.}{2021}]{ramos21}
{Ramos} P.,  {Antoja} T.,  {Mateu} C.,  {Anders} F.,  {Laporte} C.~F.~P.,
  {Carballo-Bello} J.~A.,  {Famaey} B.,   {Ibata} R.,  2021, \mn@doi [\aap]
  {10.1051/0004-6361/202039830}, \href
  {https://ui.adsabs.harvard.edu/abs/2021A&A...646A..99R} {646, A99}

\bibitem[\protect\citeauthoryear{{Recio-Blanco}, {Fern{\'a}ndez-Alvar}, {de
  Laverny}, {Antoja}, {Helmi}  \& {Crida}}{{Recio-Blanco}
  et~al.}{2021}]{recio-blanco21}
{Recio-Blanco} A.,  {Fern{\'a}ndez-Alvar} E.,  {de Laverny} P.,  {Antoja} T.,
  {Helmi} A.,   {Crida} A.,  2021, \mn@doi [\aap]
  {10.1051/0004-6361/202038943}, \href
  {https://ui.adsabs.harvard.edu/abs/2021A&A...648A.108R} {648, A108}

\bibitem[\protect\citeauthoryear{{Rockosi} et~al.,}{{Rockosi}
  et~al.}{2002}]{rockosi02}
{Rockosi} C.~M.,  et~al., 2002, \mn@doi [\aj] {10.1086/340957}, \href
  {http://cdsads.u-strasbg.fr/abs/2002AJ....124..349R} {124, 349}

\bibitem[\protect\citeauthoryear{{Roederer} \& {Gnedin}}{{Roederer} \&
  {Gnedin}}{2019}]{roederer19}
{Roederer} I.~U.,  {Gnedin} O.~Y.,  2019, \mn@doi [\apj]
  {10.3847/1538-4357/ab365c}, \href
  {https://ui.adsabs.harvard.edu/abs/2019ApJ...883...84R} {883, 84}

\bibitem[\protect\citeauthoryear{{Shelton} et~al.,}{{Shelton}
  et~al.}{2021}]{shelton21}
{Shelton} S.,  et~al., 2021, arXiv e-prints, \href
  {https://ui.adsabs.harvard.edu/abs/2021arXiv210207257S} {p. arXiv:2102.07257}

\bibitem[\protect\citeauthoryear{{Simon} et~al.,}{{Simon}
  et~al.}{2011}]{simon11}
{Simon} J.~D.,  et~al., 2011, \mn@doi [\apj] {10.1088/0004-637X/733/1/46},
  \href {http://cdsads.u-strasbg.fr/abs/2011ApJ...733...46S} {733, 46}

\bibitem[\protect\citeauthoryear{{Slater} et~al.,}{{Slater}
  et~al.}{2014}]{slater14}
{Slater} C.~T.,  et~al., 2014, \mn@doi [\apj] {10.1088/0004-637X/791/1/9},
  \href {http://cdsads.u-strasbg.fr/abs/2014ApJ...791....9S} {791, 9}

\bibitem[\protect\citeauthoryear{{Sollima}}{{Sollima}}{2020}]{sollima20}
{Sollima} A.,  2020, \mn@doi [\mnras] {10.1093/mnras/staa1209}, \href
  {https://ui.adsabs.harvard.edu/abs/2020MNRAS.495.2222S} {495, 2222}

\bibitem[\protect\citeauthoryear{{Starkenburg} et~al.,}{{Starkenburg}
  et~al.}{2017}]{starkenburg17b}
{Starkenburg} E.,  et~al., 2017, \mn@doi [\mnras] {10.1093/mnras/stx1068},
  \href {https://ui.adsabs.harvard.edu/abs/2017MNRAS.471.2587S} {471, 2587}

\bibitem[\protect\citeauthoryear{{Steinmetz} et~al.,}{{Steinmetz}
  et~al.}{2006}]{steinmetz06}
{Steinmetz} M.,  et~al., 2006, \mn@doi [\aj] {10.1086/506564}, \href
  {https://ui.adsabs.harvard.edu/abs/2006AJ....132.1645S} {132, 1645}

\bibitem[\protect\citeauthoryear{{Thomas}, {Ibata}, {Famaey}, {Martin}  \&
  {Lewis}}{{Thomas} et~al.}{2016}]{thomas16}
{Thomas} G.~F.,  {Ibata} R.,  {Famaey} B.,  {Martin} N.~F.,   {Lewis} G.~F.,
  2016, \mn@doi [\mnras] {10.1093/mnras/stw1189}, \href
  {https://ui.adsabs.harvard.edu/abs/2016MNRAS.460.2711T} {460, 2711}

\bibitem[\protect\citeauthoryear{{Thomas} et~al.,}{{Thomas}
  et~al.}{2019}]{thomas19}
{Thomas} G.~F.,  et~al., 2019, \mn@doi [\apj] {10.3847/1538-4357/ab4a77}, \href
  {https://ui.adsabs.harvard.edu/abs/2019ApJ...886...10T} {886, 10}

\bibitem[\protect\citeauthoryear{{Thomas} et~al.,}{{Thomas}
  et~al.}{2020}]{thomas20}
{Thomas} G.~F.,  et~al., 2020, \mn@doi [\apj] {10.3847/1538-4357/abb6f7}, \href
  {https://ui.adsabs.harvard.edu/abs/2020ApJ...902...89T} {902, 89}

\bibitem[\protect\citeauthoryear{{Venn}, {Irwin}, {Shetrone}, {Tout}, {Hill}
  \& {Tolstoy}}{{Venn} et~al.}{2004}]{venn04}
{Venn} K.~A.,  {Irwin} M.,  {Shetrone} M.~D.,  {Tout} C.~A.,  {Hill} V.,
  {Tolstoy} E.,  2004, \mn@doi [\aj] {10.1086/422734}, \href
  {https://ui.adsabs.harvard.edu/abs/2004AJ....128.1177V} {128, 1177}

\bibitem[\protect\citeauthoryear{{Venn} et~al.,}{{Venn} et~al.}{2020}]{venn20}
{Venn} K.~A.,  et~al., 2020, \mn@doi [\mnras] {10.1093/mnras/stz3546}, \href
  {https://ui.adsabs.harvard.edu/abs/2020MNRAS.492.3241V} {492, 3241}

\bibitem[\protect\citeauthoryear{{Wan} et~al.,}{{Wan} et~al.}{2020}]{wan20}
{Wan} Z.,  et~al., 2020, \mn@doi [\nat] {10.1038/s41586-020-2483-6}, \href
  {https://ui.adsabs.harvard.edu/abs/2020Natur.583..768W} {583, 768}

\bibitem[\protect\citeauthoryear{{Willman} \& {Strader}}{{Willman} \&
  {Strader}}{2012}]{willman12}
{Willman} B.,  {Strader} J.,  2012, \mn@doi [\aj] {10.1088/0004-6256/144/3/76},
  \href {http://cdsads.u-strasbg.fr/abs/2012AJ....144...76W} {144, 76}

\bibitem[\protect\citeauthoryear{{Yanny} et~al.,}{{Yanny}
  et~al.}{2009}]{yanny09}
{Yanny} B.,  et~al., 2009, \mn@doi [\aj] {10.1088/0004-6256/137/5/4377}, \href
  {https://ui.adsabs.harvard.edu/abs/2009AJ....137.4377Y} {137, 4377}

\bibitem[\protect\citeauthoryear{{Youakim} et~al.,}{{Youakim}
  et~al.}{2017}]{youakim17}
{Youakim} K.,  et~al., 2017, \mn@doi [\mnras] {10.1093/mnras/stx2005}, \href
  {https://ui.adsabs.harvard.edu/abs/2017MNRAS.472.2963Y} {472, 2963}

\bibitem[\protect\citeauthoryear{{Youakim} et~al.,}{{Youakim}
  et~al.}{2020}]{youakim20}
{Youakim} K.,  et~al., 2020, \mn@doi [\mnras] {10.1093/mnras/stz3619}, \href
  {https://ui.adsabs.harvard.edu/abs/2020MNRAS.492.4986Y} {492, 4986}

\bibitem[\protect\citeauthoryear{{Yuan} et~al.,}{{Yuan}
  et~al.}{2020a}]{yuan20a}
{Yuan} Z.,  et~al., 2020a, \mn@doi [\apj] {10.3847/1538-4357/ab6ef7}, \href
  {https://ui.adsabs.harvard.edu/abs/2020ApJ...891...39Y} {891, 39}

\bibitem[\protect\citeauthoryear{{Yuan}, {Chang}, {Beers}  \& {Huang}}{{Yuan}
  et~al.}{2020b}]{yuan20b}
{Yuan} Z.,  {Chang} J.,  {Beers} T.~C.,   {Huang} Y.,  2020b, \mn@doi [\apjl]
  {10.3847/2041-8213/aba49f}, \href
  {https://ui.adsabs.harvard.edu/abs/2020ApJ...898L..37Y} {898, L37}

\bibitem[\protect\citeauthoryear{{Yuan} et~al.,}{{Yuan} et~al.}{2022}]{yuan22a}
{Yuan} Z.,  et~al., 2022, \mn@doi [\apj] {10.3847/1538-4357/ac616f}, \href
  {https://ui.adsabs.harvard.edu/abs/2022ApJ...930..103Y} {930, 103}

\makeatother
\end{thebibliography}



\appendix
\section{Constraints on the metallicity dispersions of the streams}\label{appendix}

\begin{figure*}
\begin{center}
\includegraphics[width=10.5cm,angle=270]{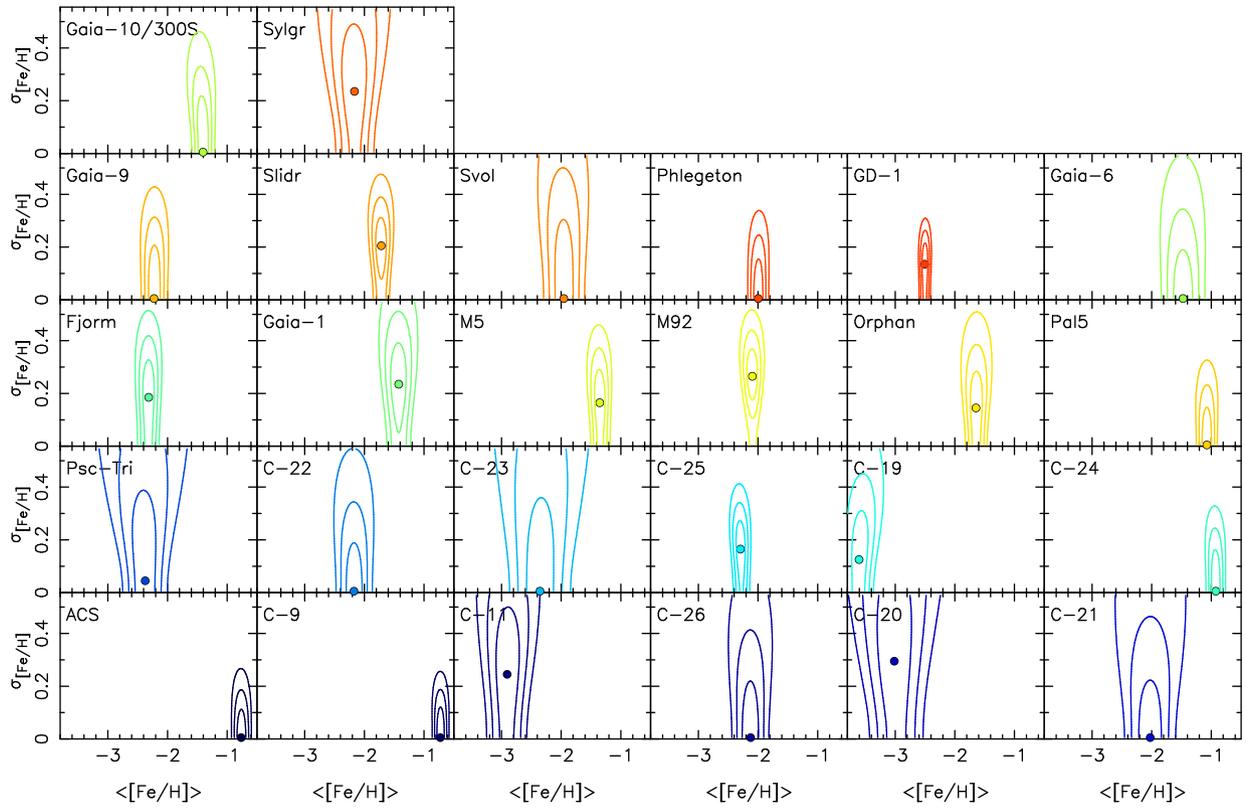}
\caption{PDFs of the mean ($\langle\FeH\rangle$) and dispersion ($\sigma_\mathrm{\FeH}$) of the metallicity distributions shown in Figures~\ref{fig:M5stream}--\ref{fig:C-26} obtained through forward modeling (see text for details). The contours correspond to 2-dimensional Gaussian 1-, 2-, and 3-$\sigma$ uncertainties and are calculated from the peak of the likelihood distribution, represented by the large dot. For ease of reference, the colors are the same as in the bottom two panels of Figure~\ref{fig:map}.\label{fig:dispersions}}
\end{center}
\end{figure*}

We explore the constraints that the \Pris\ data can provide on the metallicity dispersion of the streams by forward modeling a given MDF with a Gaussian distribution, folding in the individual metallicity uncertainties on each data point. The framework we use is the same as presented in \citet{martin18}, except that metallicity data here replaced the velocity data in their equations~1 and~2. The resulting 2-dimensional PDFs are shown in Figure~\ref{fig:dispersions} for all the studied streams. Since this technique does not specifically flag outliers that could correspond to contaminants, the resulting mean metallicities may deviate slightly from those obtained through the $\sigma$-clipping procedure and listed in Table~\ref{table}. The mean metallicities obtained from the two different technique are nevertheless always consistent with each other.

Unfortunately, the large uncertainties of the \Pris\ metallicity and the presence of contamination, especially at the faint end, do not yield very constraining results. We also note that the stream of known GC M92 is non consistent with the absence of a metallicity dispersion. This discrepancy is due to the heavy contamination that was already pointed out in sub-section~\ref{M92} for this stream that is near apocenter. With these caveats in mind, we conservatively use these inferences to confirm those dispersions that are constrained to be strongly compatible with zero, \ie ACS, C-9, C-24, GD-1, Pal~5, and Phlegeton, that all show dispersions $\lta0.2$\,dex at the 90\% confidence level.


\bsp	
\label{lastpage}
\end{document}